\documentclass[a4paper,11pt]{article}
\usepackage[utf8]{inputenc}
\usepackage[english]{babel}
\usepackage{amsmath, amssymb,graphicx}
\usepackage{braket}
\usepackage{yfonts}
\usepackage{caption}
\usepackage{subcaption}
\usepackage{ulem}
\pdfoutput=1
\usepackage{jheppub}
\newcommand{\be}{\begin{equation}}
\newcommand{\ee}{\end{equation}}
\newcommand{\bea}{\begin{eqnarray}}
\newcommand{\eea}{\end{eqnarray}}
\newcommand{\nn}{\nonumber}

\author[a]{Irina Ya. Aref'eva,}
\author[b,c]{Anastasia A. Golubtsova}
\author[d]{and Giuseppe Policastro}

\affiliation[a]{Steklov Mathematical Institute, Russian Academy of Sciences,\\Gubkina str. 8, 119991, Moscow, Russia}
\affiliation[b]{Bogoliubov Laboratory of Theoretical Physics, JINR,\\141980  Dubna, Moscow region, Russia}
\affiliation[c]{Dubna State University,\\Universitetskaya str. 19, Dubna, 141980, Russia}
\affiliation[d]{Laboratoire de Physique Th\'eorique de l'  \'Ecole Normale Sup\'erieure, \\
PSL University, CNRS, Sorbonne Universit\'es, UPMC Univ. Paris 06, \\
 24 rue Lhomond, 75231 Paris Cedex 05, France}

\emailAdd{arefeva@mi.ras.ru}
\emailAdd{golubtsova@theor.jinr.ru}
\emailAdd{policast@lpt.ens.fr}

\title{Exact holographic RG flows and the $A_{1}\times A_{1}$ Toda chain}
\abstract{We construct analytic solutions of Einstein gravity coupled to a dilaton field with a potential given by a sum of two 
exponentials, by rewriting the equations of motion in terms of an integrable Toda chain. 
These solutions can be interpreted as domain walls interpolating between different asymptotics, and as such they can have interesting applications in holography. In some cases, we can construct a solution which
interpolates between an AdS fixed point in the UV limit and a hyperscaling violating boundary in  the IR region.
We also find analytic black brane solutions at finite temperature. We discuss the properties of the solutions and the interpretation in terms of RG flow.}

\begin{document}
\begin{flushright}
LPTENS/18/06
\end{flushright}
\vfil
\maketitle

\flushbottom
\newpage
\section{Introduction}

\label{sec:intro}

The notion of the renormalization group has dominated our thinking about quantum field theories and statistical systems since 
its elaboration by K. Wilson. It has been a paradigm-changing idea, providing a unified and systematic picture to understand the 
dynamics of systems with many degrees of freedoms. Quantum field theories are now not seen as isolated items, but as classes connected 
by the RG flow. The method is very powerful, however in practice one can usually only determine the structure of the  flow in a perturbative expansion around a fixed point that is weakly coupled. Luckily there are many interesting theories for which the program can be implemented, including of course Yang-Mills and QCD that are asymptotically free, the Wilson-Fisher fixed point of $\lambda \phi^4$ in dimension $4-\epsilon$, etc, but strongly-coupled examples are few. 

The holographic correspondence gives a description of a class of strongly-coupled field theories in terms of a dual weakly-coupled gravitational description. The RG flow in this description is geometrized, and corresponds to a gravitational solution with particular asymptotic properties; the holographic direction corresponds to the energy scale, and so the Hamiltonian evolution in this direction can be put in correspondence with the evolution of the system under the change of the RG scale \cite{deBoer:1999tgo}. These solutions are often described as ``domain wall" solutions, as they interpolate between two asymptotic regions, each of them being a solution on its own and corresponding to a given theory.  

This aspect of the holographic correspondence has been intensely explored, but some issues have so far eluded a complete resolution; for instance, the precise relation between the holographic and the Wilsonian scheme \cite{Heemskerk:2010hk,Faulkner:2010jy}, and the fact that Einstein equations are second order while RG equations are first-order; even though the equations can be cast in first-order form using the Hamilton-Jacobi formalism, there seems still to be a mismatch in that the couplings of the field theory are promoted to fluctuating fields in the gravity description; recently an attempt to solve this problem has been made \cite{Lee:2013dln} with the notion of a ``quantum RG flow", that arises from the classical one after integrating out the double-trace operators. This flow may be more complex than the one typically arising in QFT, and these possibilities have only been considered in the last few years \cite{KNP}. 

The matching of RG equations with the gravity equations has been precisely formulated only for the solutions that have the Poincar\'e invariance at the boundary, and correspond to the vacuum of the dual field theory. In these cases, one can show that the equations of motion can be expressed in terms of a superpotential, which is determined by the scalar potential (up to the choice of some integration constants), and 
is in one-to-one correspondence with the beta function of the dual theory. But it is not yet known how to extend this procedure to non-Poincar\'e invariant states, in particular finite temperature/density. 

In this paper we will not be concerned with these conceptual issues. Rather, in the spirit of ``bottom-up" holography, we remark that there are 
some relatively simple models of gravity coupled to a scalar field, that allow for interesting examples of RG flows that can be found analytically. This is due to the fact that the Einstein equations, with the Ansatz that corresponds to domain-walls solutions, reduce to 
dynamical equations that are completely integrable (they can be reduced to the equations of a Toda chain). It is remarkable that such analytic domain-wall solutions can be found not only for the vacuum but also at finite temperature. 
The Toda equations arise when the potential is a sum of exponentials, when the coefficients in the exponent satisfy certain relations. 
The case of a single exponential term had been solved in \cite{ChR}. The Chamblin-Reall black brane solution has found applications to the study of the dynamics of a non-conformal strongly coupled plasma in the hydro regime and beyond \cite{GJP, Betzios:2017dol}. 
However the single exponential case has some limitations: the potential is monotonous, so it does not have a minimum but only run-away solutions; the RG flow in this case does not start from a UV fixed point, and in fact the dual theory is not well-behaved in the UV. Moreover, the single exponential has a definite sign. This precludes the possibility of studying cases in which the potential changes sign along the solution. The general analysis of RG flow with a single scalar, performed in \cite{KNP}, only deals with potentials that are definite negative; the general case of a potential with zeros is less explored.  

In this paper we consider the next simplest case, with a sum of two exponentials. Integrability leaves one of the two exponents unfixed, so we have one free parameter, like in the case of a single exponential. This case already allows to overcome the limitations just mentioned. We will leave the consideration of more complicated potentials for future investigations. Qualitatively, the properties of the potential is that it has a minimum with $V_{min} <0$, it goes from being negative and vanishing at $\phi \to - \infty$ to positive and diverging at $\phi \to + \infty$. 

It turns out that for our choice of the potential, the integrable Toda chain is associated to  the Lie algebra is $A_1 \times A_1$ \cite{LS, Perelomov}, so it is a particularly simple case  and the solutions can be given explicitly in terms of elementary functions. 

The solutions we find depend on a certain number of parameters. One parameter distinguishes between vacuum and nonvacuum solutions. The vacuum solutions are Poincar\'e invariant. Turning on the parameter gives a deformation to non-vacuum solutions, in which Poincar\'e symmetry is broken, and horizons can be formed. One more constant is the "energy" of the solutions (more exactly the energy in the associated Toda chain description) and determines the type of solutions; there are four general classes of solutions -- depending on the type of functions that appear, we will call them the $\sinh$-class, $\sin$-class, linear class and  $\cosh$-class  of solutions (see eqs.\eqref{Ffunc}). 
We will consider mainly the  $\sinh$-class. Two other parameters determine the position of singularities, whose presence requires to split the 
domain on the radial variable in three regions; correspondingly we have "left", "middle" and  "right" solutions. In the left and middle solutions the dilaton interpolates between $+ \infty$ and $- \infty$, whereas the right  solution is bouncing: the dilaton starts at $-\infty$,  goes to a maximal value and then goes back to  $-\infty$. 
It should be noted that the singularities are not just coordinate singularities; the scalar curvature usually diverges at the end points of the branches, except for the right end of the middle branch. \\

 Coming to the domain wall coordinates we explore the solutions in the holographic framework. We define the energy scale A$=e^{\mathcal{A}}$, where $\mathcal{A}$ is the scale factor of the domain wall metric  and the running coupling as $\lambda = e^{\phi}$ through the dilaton $\phi$.
It is interesting that, in spite of the fact that the dilaton has a similar behaviour in the  left and middle solutions, the scale factor  $\mathcal{A}$ has rather different behaviour on these solutions, namely,  on the left solution when the  dilaton varies from $-\infty$ to a special value $\phi _s$,  the energy scale increases, but after the  dilaton passes the special value
 $\phi _s$, the scale factor starts to decrease down. This non-monotonic behavior of $\mathcal{A}$ precludes the possibility of a holographic interpretation of this branch of solutions. 
For the middle solution, when the dilaton increases from $-\infty$ to  $\infty$, the  scale factor decreases from large positive values to large negative values. This behaviour corresponds to the running coupling in an asymptotically UV free theory, so it is of interest from the point of view of possible applications in QCD. In the right solution the scale factor is increasing from minus infinity to plus infinity, but the dilaton bounces back after reaching a maximal value $\phi_{max}$, as already mentioned. These behaviours are illustrated in Fig.~\ref{fig:X'}. 
 
 There is also a special case when the points of singularities coincide. In this case  the singularities are cancelled  and the solution 
 has a smooth AdS boundary in the UV limit and a hyperscaling violating boundary in the IR region. The dilaton supporting this geometry runs from a constant value in the UV to $-\infty$ in the IR. 
 
As we mentioned before there  are other solutions, in particular this one from the linear class describes the opposite flow, from hyperscaling-violating in the UV to AdS in the IR. 
  
 When we turn to non-vacuum solutions, the deformation can give solutions  with a horizon or without.  For the cases having  horizons, the parameter characterizing  the deformation is related to the temperature. It happens that the black brane solutions can be constructed only from those vacuum solutions which are defined for $u \to \pm \infty$.  For these solutions the dilaton potential evaluated  on-shell is bounded from above and therefore the solutions obey the Gubser's criterion \cite{Gubser}. In particular, for the solutions defined for $u \to +\infty$ we change the scaling properties of these solutions in the IR regime varying the temperature, meanwhile the UV behavior does not change. 
 
In accordance with Gubser's criterion, we find finite temperature solutions only for the ``regular" vacuum flow. In particular, we do not have any black brane solution with AdS UV asymptotics, as this is a singular flow. The finite temperature generalization of the vacuum solution with coinciding singularities yields to be just a AdS-Schwarzschild black brane, i.e. it does not flow anymore.

The interpretation of the solutions in terms of RG flows is clarified by using the first order variables $X,Y$ \cite{GKMN} that have a direct relation to the superpotential and the beta function, when a dual theory interpretation is possible. We discuss how our solutions describe flows between different attractor points in the phase space, i.e. the $(\phi, X)$ plane, but in order to recover all possible flows we also have to include  other classes of solutions, namely the linear and $cosh$ solutions. This is illustrated in Figs.~\ref{fig:Xcomparison} and ~\ref{fig:pointP}. In Fig.~\ref{fig:XY} we show  possible flows at non-zero temperature.

The paper is organized as follows. In Sect.~\ref{Sec:setup} we describe the holographic gravity model, the ansatz for the metric and the dilaton that leads to  a special mechanical model, that can be explicitly  integrated. Here (Sect.~\ref{Sect:RG}) we also introduce some general relations for the holographic RG flow. In Sect.~\ref{Sec:vac} we describe in details
 the vacuum solutions and give their interpretation in terms of the holographic RG flow. In Sect.~\ref{Sect:4NV} we study non-vacuum solutions, derive a black brane and study the black brane solution as a holographic RG flow at finite temperature. In the Appendix,   we collect information about 
 curvature invariants for our background  in Sect.~\ref{Sec:Cur}, some formula about the dilaton field in Sect.~\ref{Sec:phi} 
 and  details about the superpotential for vacuum case in Sect.~\ref{App:SuperW}.

\setcounter{equation}{0}
\section{The setup}\label{Sec:setup}

We consider a holographic model with gravity coupled to a dilaton field, and the dilaton potential is  taken as a sum of two exponential functions. As explained in the introduction, the choice of the dilaton potential is motivated by studies of models with only one exponential function in the potential \cite{GJP},\cite{GKMN}.

\subsection{The holographic gravity model}

\begin{figure}[h!]
\centering
\includegraphics[width=7cm]{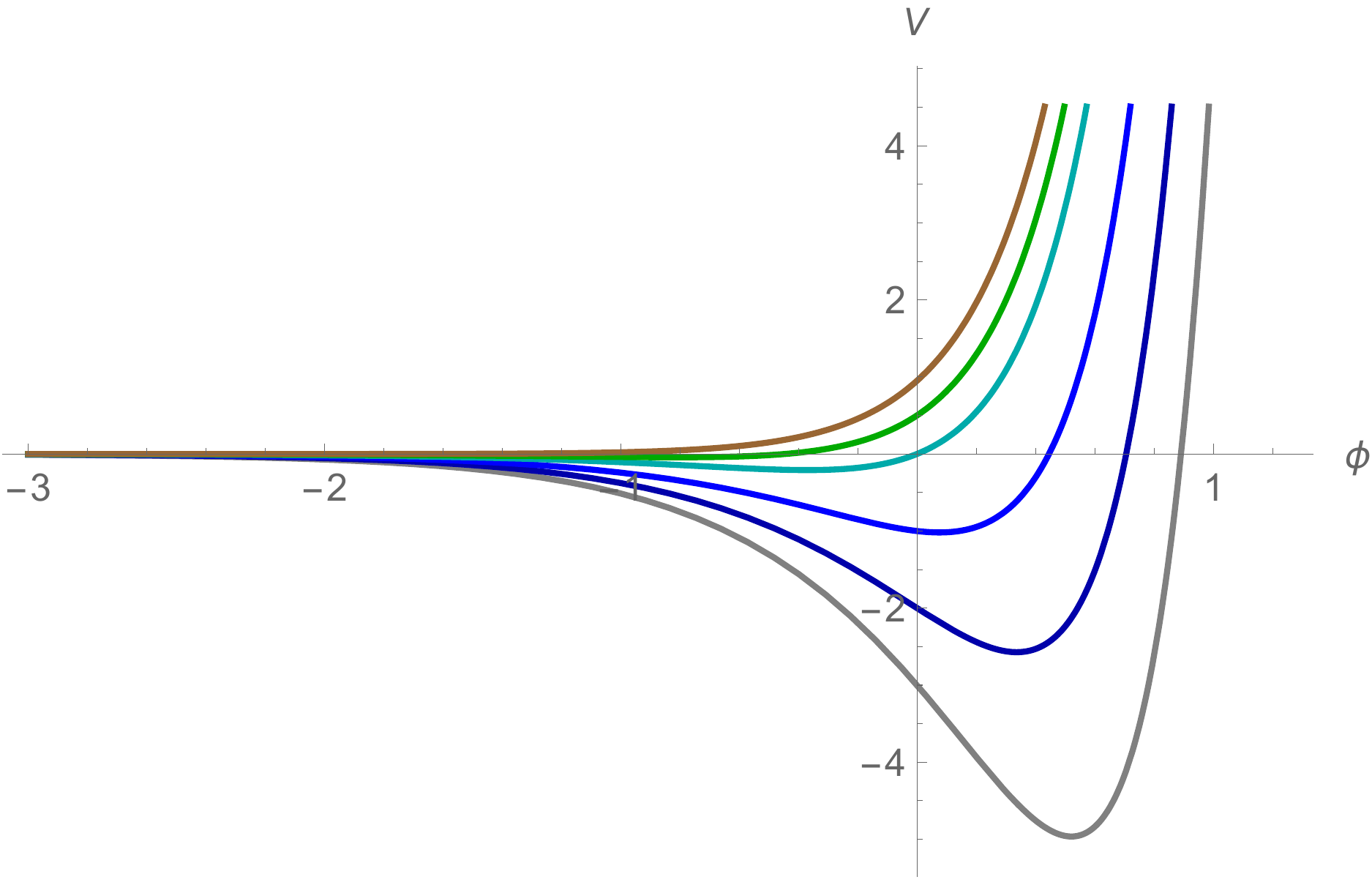} 
\caption{The behaviour of the potential $V(\phi)$ for $C_{1}<0$, $C_{2}>0$.}
\label{2expV}
\end{figure}

The holographic model is governed by an action of  the form
\begin{eqnarray}\label{1.1}
\mathcal{S} = \frac{1}{16\pi G_{5}}\int d^{5}x \sqrt{|g|}\left(R  - \frac{4}{3} \left(\partial \phi\right)^{2}  - V (\phi)\right) + G.H.,
\end{eqnarray}
with the dilaton potential 
\begin{equation}\label{1.1b}
V(\phi) = C_{1}e^{2k_{1}\phi} + C_{2}e^{2k_{2} \phi},
\end{equation}
where $C_{i}$, $k_{i}$ with $i = 1,2$ are constants. We choose the constants $C_{1}$ and $C_{2}$ as follows
\bea\label{1.1b2}
C_{1}<0, \quad C_{2} >0.
\eea
In this case  the potential has a minimum and regions of positive and negative sign, as shown in Fig.~\ref{2expV}.

The model (\ref{1.1})-(\ref{1.1b}) is defined on a 5-dim manifold $M$ equipped with the metric
\begin{eqnarray}\label{1.1c}
ds^{2} = -  e^{2A(u)}dt^{2} + e^{2B(u)}\sum^{3}_{i =1}dy^{2}_{i} + e^{2C(u)}du^{2},
\end{eqnarray}
where $A = A(u)$, $B = B(u)$ and $C = C(u)$ are some smooth functions. We make an ansatz in which these functions, as well as the dilaton,  depend only on the $u$-coordinate. 

The equations of motion which follow from the action (\ref{1.1}) read
\bea\label{EOME}
R_{MN} - \frac{1}{2}g_{MN}R = \frac{4}{3}\left(\partial_{M}\phi \partial_{N}\phi - \frac{1}{2}g_{MN}\partial_{k}\phi\partial^{k}\phi\right) - \frac{1}{2}g_{MN} V(\phi).
\eea
The equation for the scalar field is
\bea\label{EOMD}
\Box \phi =\frac{3}{8}\frac{\partial V}{\partial \phi}.
\eea

Taking into account the relations (\ref{app2})-(\ref{app4}), (\ref{app5}), the Lagrangian can be reduced to the following form, up to total derivative terms : 
\begin{eqnarray}\label{1.5}
  L  =  \frac{1}{2}\Bigl[ - e^{A+3B - C}(6 \dot{ A} \dot{B}+ 6 \dot{ B}^{2} )  +  \frac{4}{3} \dot{\phi}^{2}e^{A+3B - C}  + e^{A+3B + C} V (\phi)\Bigr], 
\end{eqnarray}
where we denote  $\dot{ } \equiv \frac{d}{du}$.
For convenience we redefine the functions as
\bea\label{newdeff}
A(u) = x^{1}(u), \quad B(u) =x^{2}(u), \quad \phi(u) = x^{3}(u), \quad C(u) =x(u),
\eea
 introduce a new variable $x_{0}$
\bea\label{harmc1}
x_{0} = x^{1} + 3x^{2}
\eea
and the so-called lapse function
\begin{equation}
\mathcal{N} = e^{x - x_{0}}.
\end{equation}

In what follows we use the harmonic gauge, i.e.
\bea\label{harmgauge}
\mathcal{N} = 1, \quad \text{or}\quad x = x_{0}.
\eea

The explicit form of E.O.M. is presented in Appendix~\ref{App:EOM} in the harmonic gauge $C=A+3B$.

\subsection{ Mechanical model}\label{Sec:2MM}
The Lagrangian (\ref{1.5}) with the help of (\ref{newdeff})-(\ref{harmgauge}) can be taken to  the following form (see \cite{AGVD} and refs. therein)

\begin{eqnarray}\label{2.1a}
 L  &=& \frac{1}{2}G_{ij}\dot{x}^{i}\dot{x}^{j} - V (x),\\ \label{2.1p}
 V & = & - \frac{1}{2}\sum^{2}_{s =1}C_{s}e^{2(x^{1} + 3 x^{2}+ k_{s}x^{3})},
\end{eqnarray}

where the minisupermetric $G_{ij}$ on the target space $\mathcal{M}$ reads

\begin{eqnarray}\label{6.2}
\left(G_{ij}\right) = \left(\begin{array}{ccc}
0 & -3 & 0\\
-3 & -6 & 0\\
0 &  0& \frac{4}{3}
\end{array}
\right),\quad 
\left(G^{ij}\right) = \displaystyle{\left(\begin{array}{ccc}
\displaystyle{\frac{2}{3}} & -\displaystyle{\frac{1}{3}} & 0\\
-\displaystyle{\frac{1}{3}} & 0 & 0\\
0 & 0 & \frac{3}{4}\end{array}
\right)}, \quad i,j = 1,2,3.
\end{eqnarray}

The corresponding energy constraint is given by
\begin{eqnarray}\label{2.1d}
E = \frac{1}{2}G_{ij}\dot{x}^{i}\dot{x}^{j} + V = 0.
\end{eqnarray}

However, the system (\ref{2.1a})-(\ref{2.1d}) is still difficult to work with, because of the degrees of freedom are coupled in the potential. 
The idea of the next section is to separate the equations of motion rotating the  system \cite{IM}.

\subsection{Integration of the mechanical model}\label{Sec:2IntMM}

To perform the following calculations it is useful to present the mechanical Lagrangian (\ref{2.1a})  in the form

\begin{eqnarray}\label{4.1}
L = \frac{1}{2}\Braket{\dot{x},\dot{x}} + \frac{C_{1}}{2}e^{\Braket{V,x}} + \frac{C_{2}}{2}e^{\Braket{W,x}},
\end{eqnarray}
where $V$ and $W$ are some vectors on the target space (we suppose that the original basis is $(e_{1},e_{2}, e_{3})$), the brackets denote a scalar product on the target space $\mathcal{M}$  with the metric (\ref{6.2}).

Then the components of $V$ and $W$ are defined by
\begin{eqnarray}
\label{Vij}
V^{1} &=& -\frac{2}{3},\quad  V^{2} = -\frac{2}{3},\quad V^{3} = \frac{3}{2} k_{1}, \\
W^{1} &=& -\frac{2}{3},\quad W^{2} = -\frac{2}{3},\quad W^{3} =\frac{3}{2} k_{2}.
\end{eqnarray}

The method we will use applies generally to systems with interactions of the form (\ref{4.1}), with arbitrarily many exponentials \cite{IM,IMS}. It turns out that the system is integrable if the vectors $V$ and $W$ can be identified with the root vectors of a Lie algebra (for more details see \cite{LS,Perelomov,IM}). Then the system becomes a Toda chain. In our case, we can see that the scalar products of the vectors are: 
\bea\label{VWlen}
\Braket{V,V} = 3\left(k^{2}_{1} - \frac{16}{9}\right), \quad \Braket{W,W} = 3\left(k^{2}_{2} - \frac{16}{9}\right),\quad \Braket{V, W} = 3\left(k_{1}k_{2} - \frac{16}{9}\right).
\eea
These correspond to the roots of a Lie algebra only if  $V$ and $W$ are orthogonal, in which case we have the $A_1 \times A_1$ Toda model. 
The integrability requirement puts a restriction on the dilaton couplings: 
 \bea\label{SPWVO}
\Braket{V,W} = 0 , \quad  k_{1}k_{2} = \frac{16}{9}.
\eea

We also suppose that $V$ is a timelike vector and $W$ is a spacelike one (notice that the symmetry between $k_1$ and $k_2$ is broken by the choice of the signs of the coefficients in the potential).  We rename 
\bea\label{kconstr1}
k_1 = k, \quad k_{2} = \frac{16}{9k},
\eea
and we have the condition 
\bea\label{kconstr2}
 0<k < 4/3.
\eea
We can find a basis of orthonormal vectors with respect to the metric $G_{ij}$: 
\bea
\Braket{e^{'}_{i},e^{'}_{j}} = \eta_{ij}, \quad \text{with}\quad (\eta_{ij}) = \text{diag}\left(-1,1,1\right), \quad i,j=1,2,3; 
\eea
one can choose the new basis vectors as follows
\bea
e^{'}_{1} = \frac{V}{||V||}, \quad e^{'}_{2} = \frac{W}{||W||}.
\eea
The new basis is related to the old one by a Lorentz transformation:  
\bea\label{newb2}
e^{'}_{j} = \sum^{3}_{i =1}S^{i}_{j}e_{i}, \quad 
\sum^{3}_{k,l=1}G_{kl}S^{k}_{i}S^{l}_{j} = \eta_{ij},
\eea
and the coordinates transform as 
\bea\label{nLorenz}
 x^{i} = \sum^{3}_{j =1}S^{i}_{j}X^{j}, \quad X^{i} = \eta_{ii}\Braket{e^{'}_{i},x}.
\eea

The Lagrangian and the energy constraint (\ref{2.1a})-(\ref{2.1d}) in the new basis take the form
\bea\label{LAGROR}
L& = &\frac{1}{2}\sum^{3}_{i,j =1}\eta_{ij}\dot{X}^{i}\dot{X}^{j} + \frac{C_{1}}{2}\exp\left[\eta_{11}|\Braket{V,V}|^{1/2}X^{1}\right] + \frac{C_{2}}{2}\exp\left[\eta_{22}|\Braket{W,W}|^{1/2}X^{2}\right], \nonumber\\ \,\\\label{ECC2}
E_{0} &=& \frac{1}{2}\sum^{3}_{i,j =1}\eta_{ij}\dot{X}^{i}\dot{X}^{j} - \frac{C_{1}}{2}\exp \left[\eta_{11}|\Braket{V,V}|^{1/2}X^{1}\right] - \frac{C_{2}}{2}\exp\left[\eta_{22}|\Braket{W,W}|^{1/2}X^{2}\right]. \nonumber\\
\eea

From (\ref{LAGROR})-(\ref{ECC2})  we see that mechanical variables are decoupled, so  
the equations of motion following from the Lagrangian (\ref{LAGROR}) are 
\bea\label{EOMOR}
\ddot{X}^{s} &=& |\Braket{R_{s},R_{s}}|^{1/2}\frac{C_{s}}{2}\exp\left[\eta_{ss}|\Braket{R_{s},R_{s}}|^{1/2}X^{s}\right], \quad\\ \label{EOMOR3}
\ddot{X}^{3} &= &0,
\eea
where $s=1,2$ and we introduce a notation for the scalar product
\bea\label{RR}
\Braket{R_{1},R_{1}} = \Braket{V,V}, \qquad \Braket{R_{2},R_{2}}=\Braket{W,W}.
\eea

The eqs. (\ref{EOMOR}) with $s = 1,2$ are two decoupled Liouville equations. The solution can be given explicitly: 
\bea\label{xsmalXb}
X^{s} = -\eta_{ss}|\Braket{R_{s},R_{s}}|^{-1/2}\ln{\left(F^{2}_{s}(u - u_{0s})\right)},
\eea
with the functions
\bea\label{Ffunc}
F_{s}(u-u_{0s})=
\left\{
\begin{aligned}
   \sqrt{\frac{|C_{s}|}{2|E_{s}|}} & \sinh\left[\sqrt{\frac{\left|E_{s}\Braket{R_{s},R_{s}}\right|}{2}}(u - u_{0s})\right],
&  \text{if}\quad \eta_{ss}C_{s}>0,\quad \eta_{ss}E_{s}>0,     \\
\sqrt{\frac{|C_{s}|}{2|E_{s}|}} & \sin\left[\sqrt{\frac{\left|E_{s}\Braket{R_{s},R_{s}}\right|}{2}}(u - u_{0s})\right],
& \text{if} \quad \eta_{ss}C_{s}>0,\quad \eta_{ss}E_{s}<0,\\
  &\sqrt{\frac{|\Braket{R_{s},R_{s}}C_{s}|}{4}}(u-u_{0s}),
& \text{if}\quad \eta_{ss}C_{s}>0, \quad E_{s} = 0,\\
 \sqrt{\frac{|C_{s}|}{2|E_{s}|}} & \cosh\left[\sqrt{\frac{\left|E_{s}\Braket{R_{s},R_{s}}\right|}{2}}(u - u_{0s})\right], &
\text{if} \quad \eta_{ss}C_{s}<0,\quad \eta_{ss}E_{s}>0,\\ 
\end{aligned}
\right.
\eea
where $u_{0s}$, $E_{s}$ are constants of integration, $s=1,2$. The $E_s$ are the conserved energies of the two decoupled Liouville modes. 
In our case, with $C_{1}< 0$ and $C_{2} >0$, we have $\eta_{ss}  C_s > 0$, 
so the relevant solutions are given by the first three lines.  
We will analyze in detail the $\sinh$ solutions, which are the most interesting from the RG point of view. 

The solutions to eq. (\ref{EOMOR3}) is 
\bea
X^{3} = p^{3} u+ q^{3},
\eea
with constants of integration  $p^{3}$, $q^{3}$.

Having solved the system, we can go back to the original variables applying the transformations (\ref{nLorenz}) with components
\bea
S^{i}_{1} = \frac{V^{i}}{|\Braket{V,V}|^{1/2}}, \quad S^{i}_{2} = \frac{W^{i}}{\Braket{W,W}^{1/2}},
\eea

and obtain 
\bea \label{xsmalsol1}
\exp{x^{1}}  =\left[F_{1}(u -u_{01})\right]^{-\frac{2V^{1}}{\Braket{V,V}}}\left[F_{2}(u -u_{02})\right]^{-\frac{2W^{1}}{\Braket{W,W}}}e^{\alpha^{1}u+\beta^{1}}, \\
\exp{x^{2}} =\left[F_{1}(u -u_{01})\right]^{-\frac{2V^{2}}{\Braket{V,V}}}\left[F_{2}(u -u_{02})\right]^{-\frac{2W^{2}}{\Braket{W,W}}}e^{\alpha^{2}u+\beta^{2}}, \\ \label{xsmallsol3}
\exp{x^{3}} =\left[F_{1}(u -u_{01})\right]^{-\frac{2V^{3}}{\Braket{V,V}}}\left[F_{2}(u -u_{02})\right]^{-\frac{2W^{3}}{\Braket{W,W}}}e^{\alpha^{3}u+\beta^{3}},
\eea
with the parameters $\alpha^{i}$, $\beta^{i}$ defined using (\ref{newb2})
\bea
\alpha^{i} = S^{i}_{3}p^{3}, \quad \beta^{i} = S^{i}_{3}q^{3}.
\eea
The parameters $\alpha^{i}$ satisfy the following conditions
\bea\label{alphaVW}
\Braket{\alpha,V } = 0, \quad \Braket{\alpha,W} = 0,
\eea
which imply 
\be\label{alpha123}
\alpha^{3} = 0, \quad \alpha^{1} =  -3\alpha^{2}\,.
\ee

The constants $E_{1}$, $E_{2}$ and $\alpha^{i}$  are related by the constraint (\ref{ECC2}) which reads
\bea\label{mechee}
E_{1} + E_{2} + \frac{1}{2}\Braket{\alpha, \alpha} = E_{0} \,. 
\eea

\subsection{The exact solutions in the harmonic gauge}

Keeping in mind that the variables $x^{i}$ with (\ref{xsmalsol1})-(\ref{xsmallsol3}), (\ref{alpha123})  are  redefined metric functions $A$, $B$ and the dilaton $\phi$ (\ref{newdeff})-(\ref{harmc1}), and taking the gauge condition $C = A + 3B$ (\ref{harmgauge}) into account, we 
can write down the metric coefficients of (\ref{1.1c})
\bea\label{FABC.1nv}
e^{A} &=& F^{\frac{4}{9k^{2} -16}}_{1}F^{\frac{9k^{2}}{4(16-9k^{2})}}_{2} e^{\alpha^{1}u},\\ \label{FABC.2}
e^{B} &= &F^{\frac{4}{9k^{2} -16}}_{1}F^{\frac{9k^{2}}{4(16-9k^{2})}}_{2}e^{-\frac{\alpha^{1}}{3}u},\\ \label{FABC.3}
e^{C}& =& F^{\frac{16}{9k^{2}-16}}_{1}F^{\frac{9k^{2}}{16-9k^{2}}}_{2},
\eea
where for simplicity we put $\beta^{i}$, $i = 1,2,3$, to zero  (without loss of generality, as this corresponds to a rescaling of the coordinates).  
For the dilaton we have 
\bea\label{FD.4nv}
\phi =  -\frac{9k}{9k^{2}-16}\log{F_{1}} +\frac{9k}{9k^{2}-16}\log{F_{2}},
\eea
moreover the dilaton equation (\ref{EOMD}) requires to take $E_{0} = 0$ in (\ref{mechee}), i.e. 
\bea\label{E1E2c3}
E_{1} + E_{2} + \frac{2(\alpha^{1})^{2}}{3} = 0, 
\eea
with $E_{1} < 0$, $E_{2}>0$ and the value of $E_{1}$ is bounded as $|E_{1}| = E_{2} + \frac{2(\alpha^{1})^{2}}{3}$.\\
 
Then the functions  $F_{1}$ and $F_{2}$ in (\ref{FABC.1nv})-(\ref{FD.4nv}) are given by
\bea\label{nC1pC2.F1E1n}
  F_{1} &= &\sqrt{\left|\frac{C_{1}}{2E_{1}}\right|}\sinh(\mu_1\,|u-u_{01}|),\, \mu_1= \sqrt{\left|\frac{3E_{1}}{2}\left(k^{2}-\frac{16}{9}\right)\right|},\\
\label{nC1pC2.F2E2p}
F_{2}& =& \sqrt{\left|\frac{C_{2}}{2E_{2}}\right|}\sinh(\mu_2\,|u-u_{02}|),\,  \mu_2 = \sqrt{\left|\frac{3E_{2}}{2}\left(\left(\frac{16}{9}\right)^{2}\frac{1}{k^{2}} - \frac{16}{9}\right)\right|},
\eea
where  $0< k <4/3$, $C_{1}$ and $C_{2}$ are given by (\ref{1.1b})-(\ref{1.1b2}).
The generic form of the metric that solves EOM  following from (\ref{1.1})  reads
\bea\label{MABCnv}
ds^{2}=  F^{\frac{8}{9k^{2} -16}}_{1}F^{\frac{9k^{2}}{2(16-9k^{2})}}_{2} \left( - e^{2\alpha^{1}u}dt^{2} + e^{-\frac{2\alpha^{1}}{3}u}d\vec{y}^{~2} \right) +
F^{\frac{32}{9k^{2}-16}}_{1}F^{\frac{18k^{2}}{16-9k^{2}}}_{2}du^{2},\nonumber\\
\eea
where $\vec{y} = (y_{1},y_{2},y_{3})$.\\

The dilaton potential evaluated on the solution becomes 
\bea\label{VDIL}
V = C_{1}e^{2k\phi} + C_{2}e^{32\phi/(9k)}= C_{1}\left(\frac{F_{2}}{F_{1}}\right)^{\frac{18k^{2}}{9k^{2}16}} + C_{2}\left(\frac{F_{2}}{F_{1}}\right)^{\frac{32}{9k^{2}-16}}.
\eea
$$\,$$
We note that we have two more solutions for our choice of the potential (\ref{1.1b}) with $C_{1}<0$ and $C_{2}>0$ governed by $F_{1}$ and $F_{2}$ from the second and third branches of (\ref{Ffunc}). The solutions differ by the range of integrating constants, i.e.
for $E_{1}>0$ and $E_{2}< 0$ (opposite signes to (\ref{nC1pC2.F1E1n})-(\ref{nC1pC2.F2E2p})), we have
\bea\label{F1sin1}
F_{1} &= &\sqrt{\left|\frac{C_{1}}{2E_{1}}\right|}\sin(\mu_1\,|u-u_{01}|),\quad  \mu_1 = \sqrt{\left|\frac{3E_{1}}{2}(k^{2}-\frac{16}{9})\right|},\\
\label{F1sin2}
F_{2}& =& \sqrt{\left|\frac{C_{2}}{2E_{2}}\right|}\sin(\mu_2\,|u-u_{02}|),\quad  \mu_2 = 
\sqrt{\left|\frac{3E_{2}}{2}\left(\left(\frac{16}{9}\right)^{2}\frac{1}{k^{2}} - \frac{16}{9}\right)\right|},
\eea
while taking $E_{1} = E_{2} =0$ one has 
\bea \label{1lin}
F_{1} &= &\sqrt{\frac{3}{4}\left(k^{2} - \frac{16}{9}\right)C_{1}}(u-u_{01}), \quad
F_{2} = \sqrt{\frac{3}{4}\left(\left(\frac{16}{9k}\right)^{2} - \frac{16}{9}\right)C_{2}}(u-u_{02}).
\eea

In the forth line of (\ref{Ffunc}) we observe  cosh-solutions for the functions $F_{1}$ and $F_{2}$
\bea\label{cosh1}
F_{1}& =&\sqrt{\left|\frac{C_{1}}{2E_{1}}\right|}\cosh(\mu_1\,|u-u_{01}|), \quad  \mu_1 = \sqrt{\left|\frac{3E_{1}}{2}(k^{2}-\frac{16}{9})\right|},\\
\label{cosh2}
F_{2} &=&\sqrt{\left|\frac{C_{2}}{2E_{2}}\right|}\cosh(\mu_2\,|u-u_{02}|),\quad  \mu_2 = \sqrt{\left|\frac{3E_{2}}{2}\left(\left(\frac{16}{9}\right)^{2}\frac{1}{k^{2}} - \frac{16}{9}\right)\right|},
\eea
that corresponds to the dilaton potential with $C_{1}>0$, $C_{2}<0$ and integration constants $E_{1}<0$, $E_{2}>0$.

\subsection{Solutions as RG flows}\label{Sect:RG}
Let's briefly discuss the gravity solutions as holographic RG flows. It is useful to come to so-called domain wall coordinates.  
In this coordinates a general form of the non-vacuum solutions is 
\bea\label{DW}
ds^{2} = \frac{dw^{2}}{f(w)} + e^{2\mathcal{A}(w)}\left(-f(w)dt^2+\delta_{ij}dx^{i}dx^{j}\right),
\eea
which covers the vacuum case with $f(w) =1$. 
Both temperature  and vacuum  solutions are characterized by the scale factor $e^{\mathcal{A}(w)}$, 
that measures the field theory energy scale,  the blackening function $f(w)$ and by a scalar field profile $\phi(w)$
\bea\label{lambdaphi}
\lambda  = e^{\phi},
\eea
which is interpreted as the running coupling.

If we define 
\bea\label{XX}
X &=& \frac{1}{3 \lambda}\frac{d\lambda}{d \mathcal{A}}
\\
\label{Y}
Y(\phi)&=&\frac14\,\frac{g'}{\mathcal{A}'}, \quad  g = \log f,
\eea
we get the system of first order differential equations
\bea\label{dXphi}
\frac{dX}{d\phi}& =& -\frac{4}{3}\left(1-X^{2}+Y\right)\left(1+ \frac{3}{8X}\frac{d\log V}{d\phi}\right),\\ \label{dYphi}
\frac{dY}{d\phi} &= &-\frac{4}{3}\left(1 - X^{2} + Y\right)\frac{Y}{X},
\eea

For the vacuum case  we get
\bea\label{dXphi0}
\frac{dX}{d\phi}& =& -\frac{4}{3}\left(1-X^{2}\right)\left(1+ \frac{3}{8X}\frac{d\log V}{d\phi}\right).
\eea

\newpage
\section{Vacuum solutions}\label{Sec:vac}
\subsection{The metric and the dilaton for vacuum exact solutions}

The vacuum solutions are those that preserve the Poincar\'e invariance at the boundary. As can be seen from (\ref{MABCnv}), this requires that   $\alpha^{1} = \alpha^{2} = 0$, so in this case the two Liouville energies have to be: 
\bea\label{vc-cond}
|E_{1}| = |E_{2}|,
\eea
with the opposite signs $E_{1}<0$, $E_{2}>0$.\\
Owing to (\ref{vc-cond}) the metric (\ref{MABCnv}) takes the form
\bea\label{MABC.pC}
ds^{2}=  F^{\frac{8}{9k^{2} -16}}_{1}F^{\frac{9k^{2}}{2(16-9k^{2})}}_{2} ( - dt^{2}+ d\vec{y}^{2} ) +
F^{\frac{32}{9k^{2}-16}}_{1}F^{\frac{18k^{2}}{16-9k^{2}}}_{2}du^{2} \,,
\eea
and the dilaton is given by 
 \be \label{dil-u012}
 \phi =\frac{9k}{9k^{2}-16}\log\left(\sqrt{\frac{C_2}{|C_1|}}\frac{\sinh(\mu_2\, |u - u_{02}| )}{\sinh(\mu_1\,  |u- u_{01}|)}\right) \,.
 \ee
In the vacuum case, due to (\ref{vc-cond}) the following relation holds: 
\be
\frac{\mu_2}{\mu_1} = \frac{4}{3 k} > 1 \,.
\ee

From the form of the functions $F_{1}$ and $F_{2}$  given by (\ref{nC1pC2.F1E1n})-(\ref{nC1pC2.F2E2p}) we can see that the solution will have coordinate singularities at the points $u_{01},u_{02}$.
We have to consider three coordinate charts.
Let us take   $\bf u_{02}<u_{01}$. Then the charts are
\bea\nn
\mbox{{\bf 3-branch solution}}\,\,\,\,\,\,\,\,\,\,&&\\
\mbox{left:} &&\,\,\,\,u<u_{02}\label{chartl}\\
\mbox{middle:} &&\,\,\,\,u_{02}<u<u_{01}\label{chart}\\
\mbox{right:} &&\,\,\,\,u>u_{01}\label{chartr}
\eea
The  degenerate case with 
$u_{01}= u_{02} = u_{0}$ requires only two charts:
\bea\nn
\mbox{{\bf 2-branch solution}}\,\,\,\,\,\,\,\,\,\,&&\\
\mbox{left:} &&\,\,\,\,u<u_{0}\\ \label{chart1}
\mbox{right:} &&\,\,\,\,u>u_{0}.\label{chart2}
\eea

\begin{figure}[h!]
\centering
 \includegraphics[width=4cm]{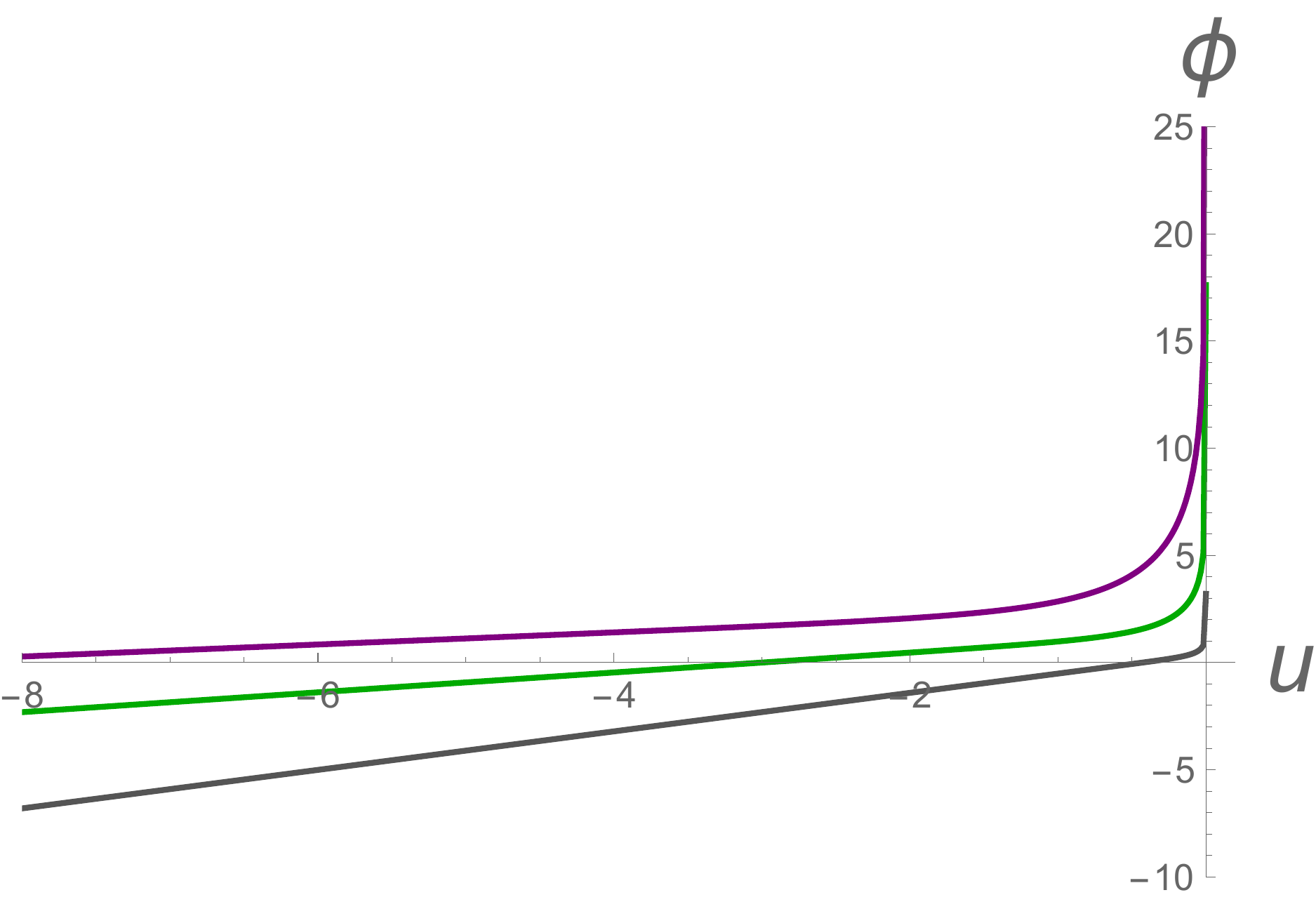}A$\,\,$
 \includegraphics[width=4 cm]{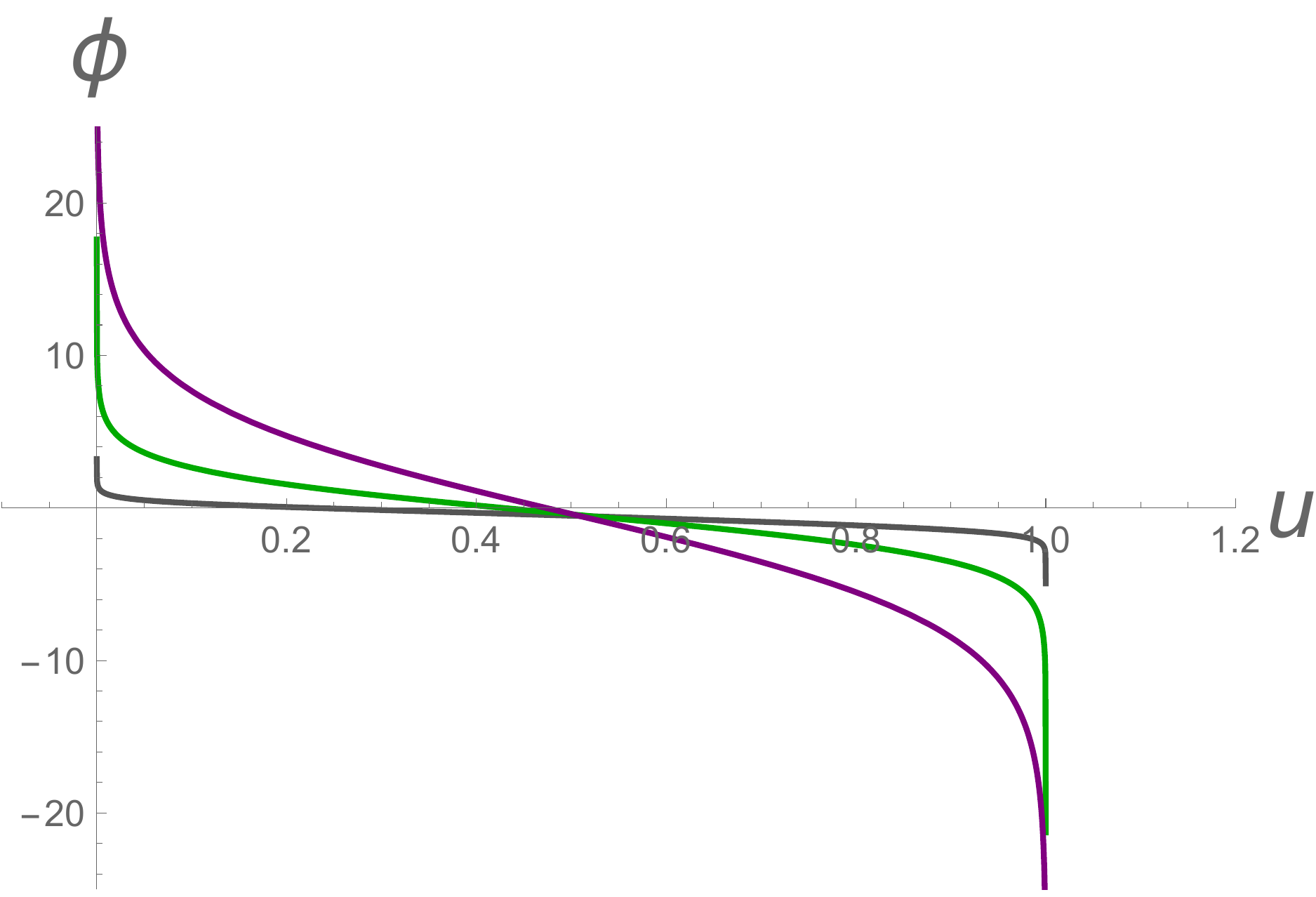}B$\,\,$
  \includegraphics[width=4 cm]{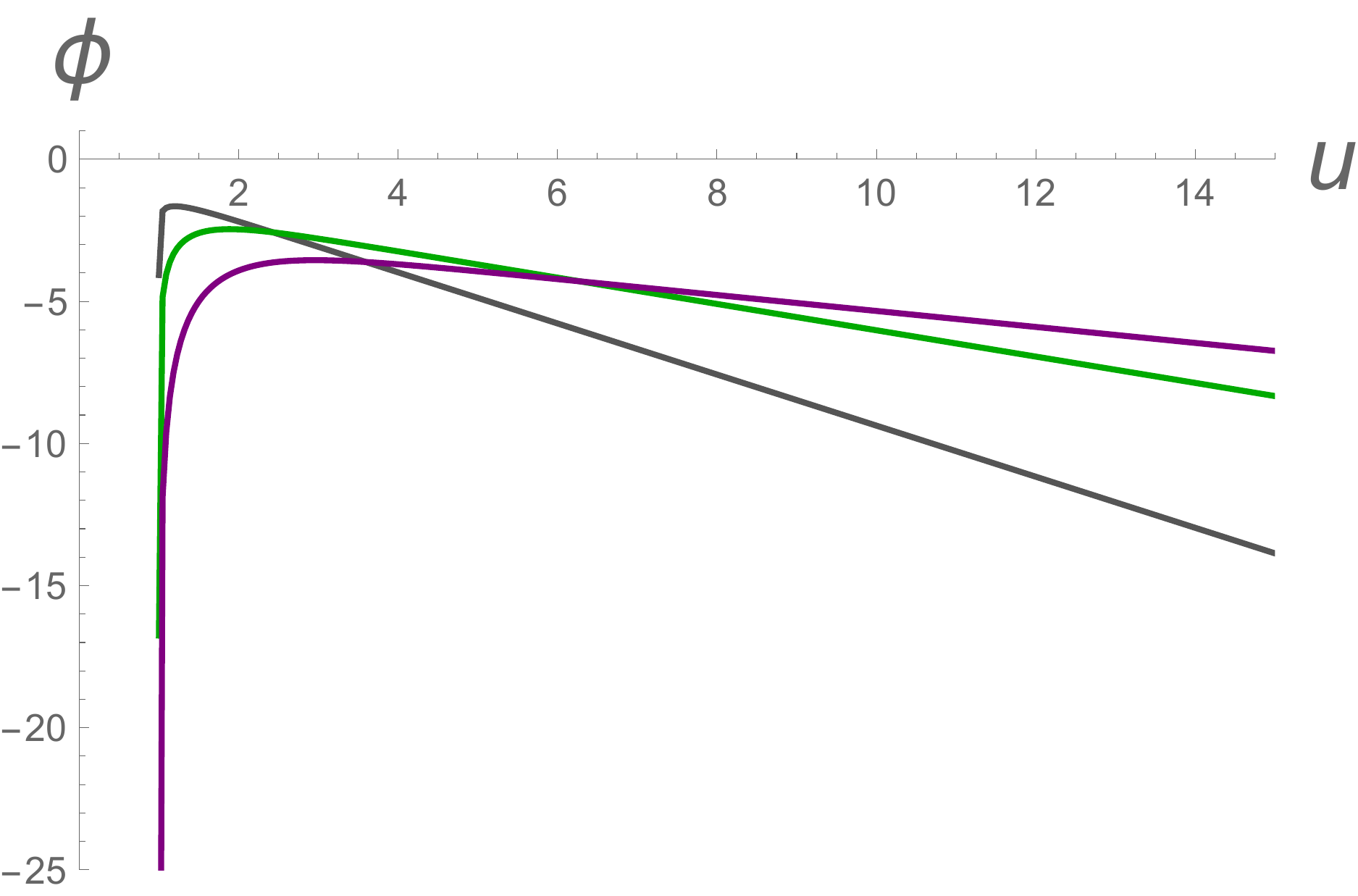}C$\,\,$\\
  \includegraphics[width=2cm]{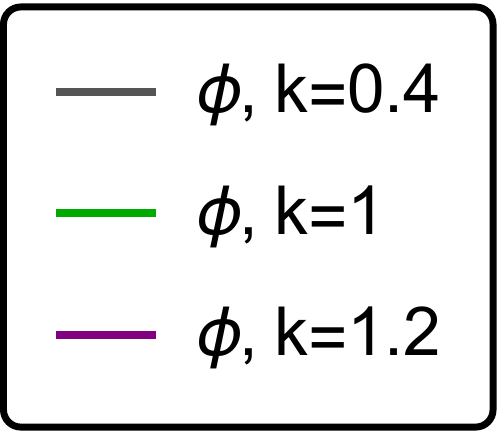}
   \caption{The dilaton solutions (\ref{dil-u012}) as functions of $u$, separately for each type of the solutions: A) the dilaton for $u<u_{02}$, $u_{02} =0$ B) the dilaton for $u_{02} < u< u_{01}$, $u_{01} =1$, $u_{02} = 0$ C) the dilaton for $u> u_{01}$, $u_{01} = 1$. For all $E_{1}= E_{2} = -1$, $C_{1} =- C_{2} = -1$, $k =0.4,1,1.2$. 
   %See also MATHEMATICA file {\bf DIL-POT-z-ALL.nb}
   }
 \label{fig:DILz}
\end{figure}

In Fig.~\ref{fig:DILz} we plot  the solutions for the dilaton $\phi$   (\ref{dil-u012})   for each choice of the coordinates (\ref{chartl})-(\ref{chartr}).

From   Fig.~\ref{fig:DILz} we see that for $u \to u_{01} \pm \epsilon$ the dilaton tends to $-\infty$, while for $u \to u_{02} \pm \epsilon$ the dilaton asymptotes to $+\infty$.
 It can be verified by direct calculations. \\
 
 \subsection{Asymptotics of the metric and the dilaton}
 We present here the asymptotics of the dilaton (\ref{dil-u012}) and the metric (\ref{MABC.pC})  with  (\ref{chartl})-(\ref{chartr}) near the boundaries.\\
$\,$\\
{\bf The left solution} with $u< u_{02}$
\begin{itemize}
\item
at $u\to - \infty$: 

\bea
\label{23z2}
ds^{2} & \sim &z^{2/3}\left(-dt^{2} + dy^{2}_{1} + dy^{2}_{2} + dy^{2}_{3} + dz^{2}\right), \\
\label{leftphiv-1}
\phi  & \sim & \frac{9k}{16-9k^2} (\mu_2-\mu _1)\,u  \sim \log z \to -\infty ,
\eea
where we use a new coordinate $z \sim \frac{4+3k}{3\mu_{1}} e^{\frac{3\mu_{1}u}{4+3k}}$. 
In Appendix~\ref{App:VacR} the scalar curvature of the left solution for both limits is presented. For $u\to -\infty$ the scalar curvature of the left solution (\ref{leftsc1})  tends to $+\infty$.

 \item at $u\to u_{02}-\epsilon$ : 
 \bea
 \label{asymleft2}
ds^{2} & \sim &  z^{\frac{18k^{2}}{64 - 9k^{2}}}\left(-dt^{2} + dy^{2}_{1}+ dy^{2}_{2} + dy^{2}_{3} +dz^{2}\right), \\
 \label{leftphi}
 \phi  & \sim  & - \frac{36 k}{64 - 9 k^2} \log z \to+\infty \,,
\eea
with the radial coordinate defined by
\bea\label{zls}
z \sim\frac{64-9k^{2}}{4(16 - 9k^{2})}(u- u_{02})^{\frac{64-9k^{2}}{4(16 - 9k^{2})}}.
\eea

From the relation for the scalar curvature (\ref{scu02}) with $u\to u_{02} - \epsilon$ it can be seen that the solution has a non-removable singularity at the point $u_{02}$.
\end{itemize}

{\bf The middle solution} with $u_{02}<u<u_{01}$
\begin{itemize}
\item
at $u\to u_{02}+\epsilon$ the asymptotics of the dilaton and the metric are the same as in (\ref{asymleft2})-(\ref{leftphi}). As in the case of the left solution, one can see from the scalar curvature (\ref{scu02}) that the middle solution  has also a  singularity at the point $u_{02}$;

\item at $u\to u_{01}-\epsilon$ : 
\bea \label{LMS.1}
ds^{2} & \sim& z^{\frac{8}{9k^{2} - 4}}\left(-dt^{2} + dy^{2}_{1} + dy^{2}_{2} + dy^{2}_{3} + dz^{2}\right), \\ \label{dm01}
\phi  &\sim& \frac{9k}{4-9k^{2}}\log z  \to -\infty, \, 
\eea
where the conformal coordinate is
\bea\label{zmsr}
z\sim \frac{16-9k^{2}}{9k^{2} - 4}(u - u_{01})^{\frac{4-9k^2}{16 -9k^2}}.
\eea
Notice that as $u \to u_{01}$, the conformal coordinate $z \to 0$ if $0<k<2/3$ and $z \to \infty$ if $2/3 < k < 4/3$. 
These asymptotics are the same as for the solutions in a single exponential potential \cite{ChR}.  
It is worth nothing the scalar curvature of the middle solution (\ref{middleR01}) has a regular behaviour at $u_{01}$.
\end{itemize}

{\bf The right solution} with $u > u_{01}$
\begin{itemize}
\item at $u\to u_{01}+\epsilon$ the asymptotics are as in (\ref{LMS.1})-(\ref{dm01}).
\item at $u\to + \infty$ :  
\bea 
\label{23z1}
ds^{2} & \sim& z^{2/3}\left(-dt^{2} + dy^{2}_{1} + dy^{2}_{2} + dy^{2}_{3} + dz^{2}\right), \\
\label{rightphiv-2}
\phi  & \sim&  \log z \to -\infty, \,
\eea
where $z$ is defined by $z \sim -\frac{4+3k}{3\mu_{1}} e^{-\frac{3\mu_{1}u}{4+3k}}$. The asymptotics are the same as in (\ref{23z2})-(\ref{leftphiv-1}). Even though the dilaton goes to $ - \infty$, these are different than the asymptotics for a single exponential. 
The scalar curvature (\ref{sccrs2}) goes to $+\infty$ with $u\to +\infty$.

\end{itemize}

Now let us turn the discussion to the dilaton potential, that can be written on solutions as 
\bea
V= 
%C_{1}e^{2k_{1}\phi}+C_{2}e^{2k_{2}\phi} =]] 
%
C_{1}\left(\frac{F_{2}}{F_{1}}\right)^{\frac{18k^{2}}{9k^{2}-16}} + C_{2}\left(\frac{F_{2}}{F_{1}}\right)^{\frac{32}{9k^{2}-16}}.
\eea

\begin{figure}[h!]
\centering \includegraphics[width=4cm]{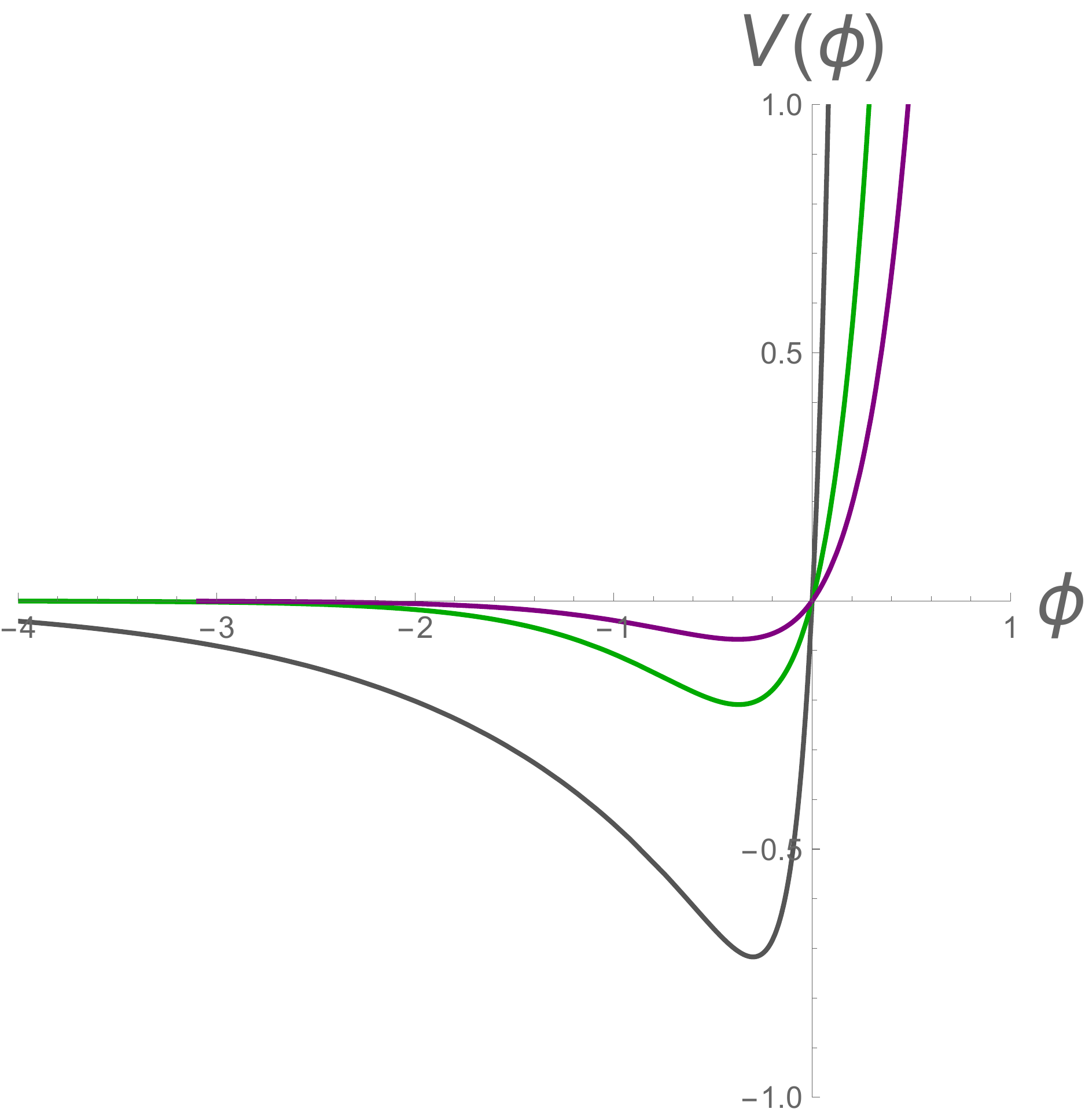}A$\,$
 \includegraphics[width=4cm]{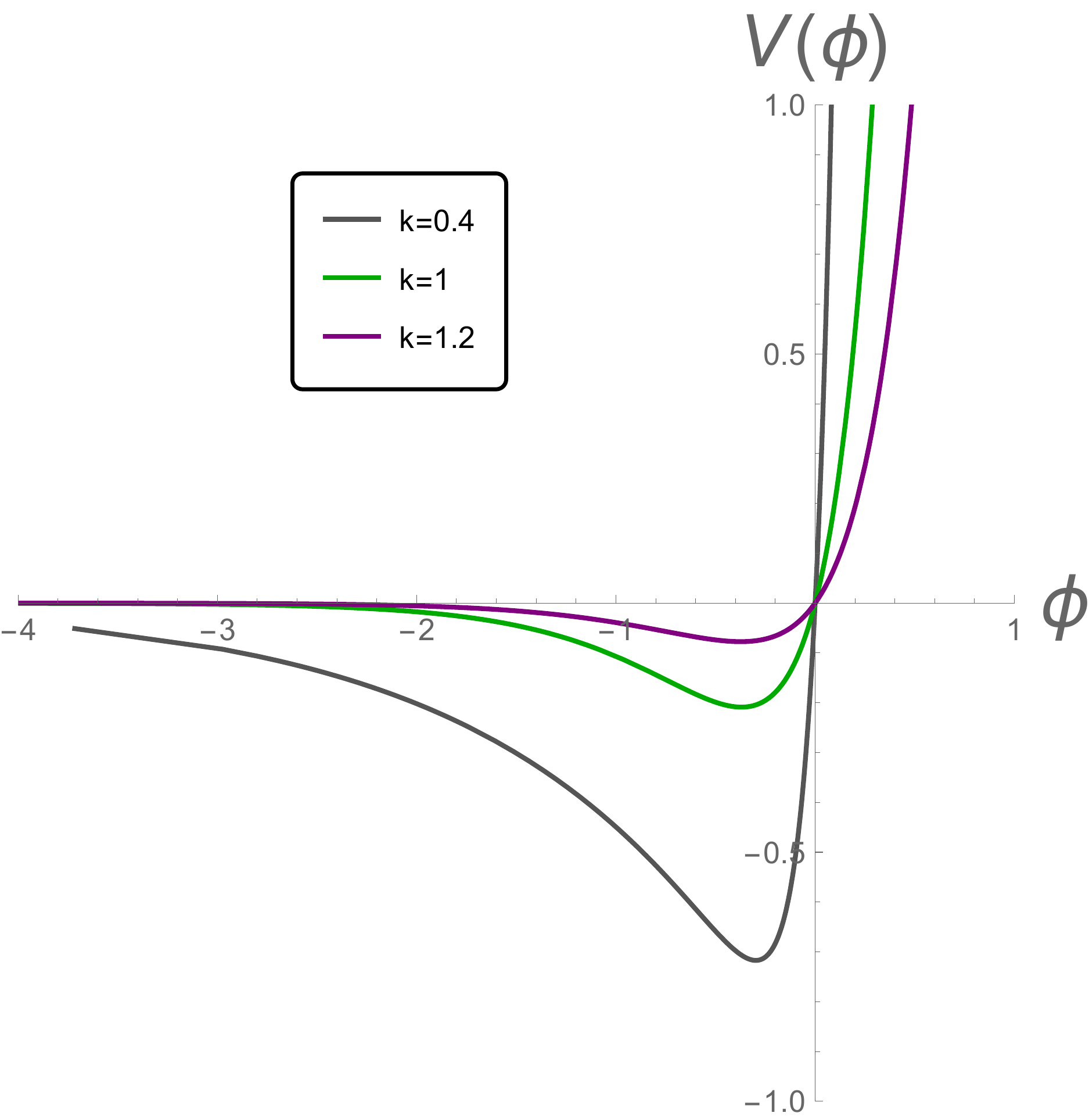}B$\,$
  \includegraphics[width=4cm]{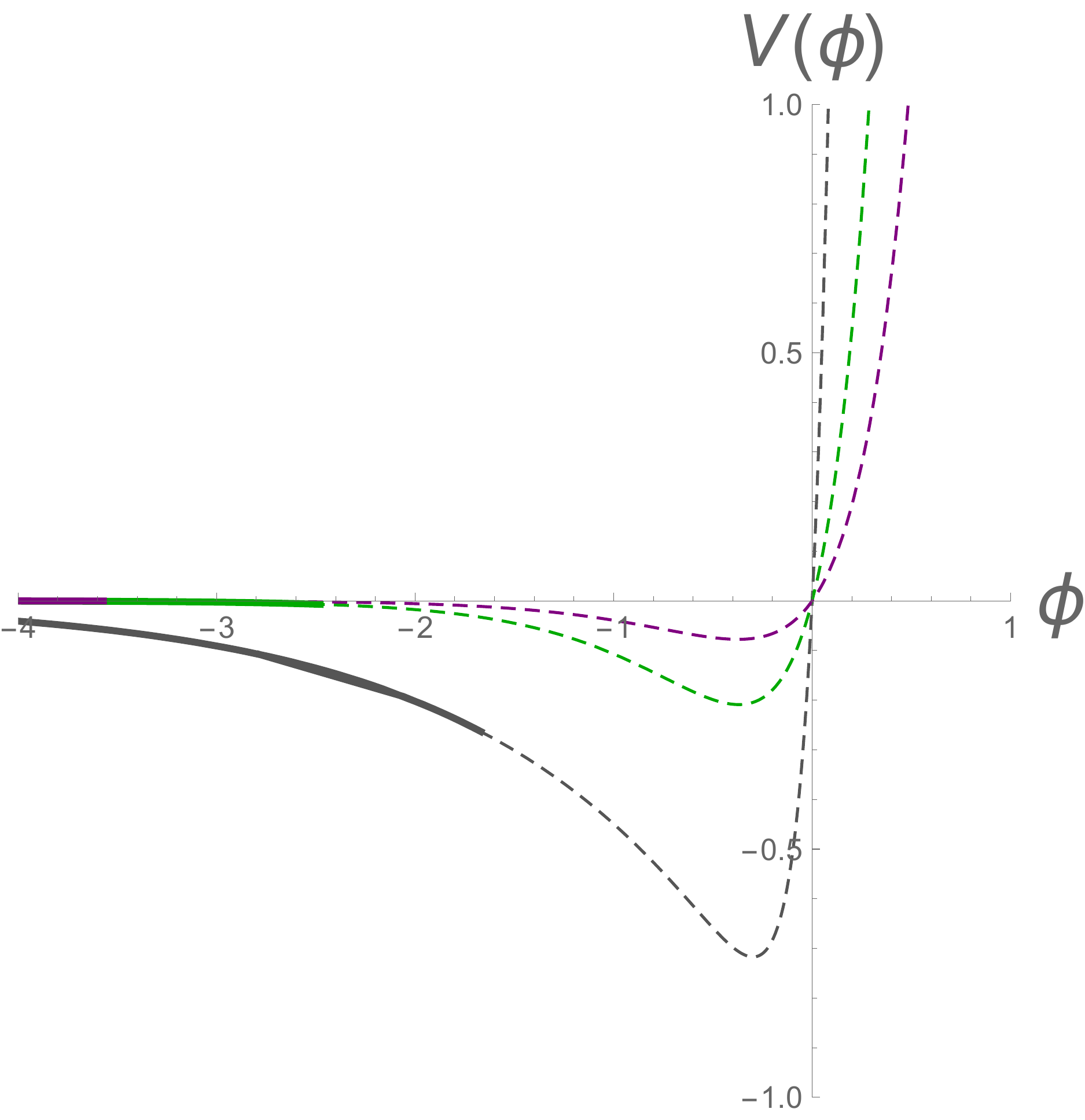}C
 \caption{The dilaton potential plotted on solutions for $\phi$. A) The potential in the  left-solution varies from  $V(\phi_{1}) = 0$ at $\phi_{1} = - \infty$, reaches its minimum and then goes to $V(\phi_{2})= + \infty$  at  $\phi_2(u_{02} - \epsilon) = +\infty$   (thick lines).   B) The potential on the  middle solution varies from $V(\phi_{2}) = + \infty$ at $\phi_{2}=\phi(u_{02} +\epsilon) = +\infty$ then reaches the minimal value  $V_{min}<0$ and  goes to $V(\phi_{1}) = 0$ at $\phi_{1}= \phi(u_{01} - \epsilon) = -\infty$. C) The potential of the right solution varies from $V (\phi_{1})= 0$ with $\phi_{1} = \phi(u_{01}+\epsilon) = - \infty$  to $V_{s}<0$, $V_{s} = V(\phi)$ at  the point $ \phi =\phi_{s}$.  When $\phi_{s}$ goes back to $-\infty$ the potential goes from $V_{s} = V(\phi_{s})$ back to zero. 
 }
 \label{fig:DIL-POT1}
\end{figure}

 In Fig.~\ref{fig:DIL-POT1} we plot the dilaton potential on the solutions for $\phi$. From Fig.~\ref{fig:DIL-POT1} C) one can see that  for the solutions defined for $u > u_{01}$ there is a turning point  $V_{s} = V(\phi)$ with $ \phi =\phi_{s}$, where the potential stops and goes back to zero.

One can find the stop points of the potentials analytically. The stop point $u_s$ can be calculated as follows.  The dilaton should obey
\be
\phi'=0,
\ee
this gives rise to
\bea
\frac{F'_{1}}{F_{1}}=\frac{F'_{2}}{F_{2}}
\eea

and we get 
\be\label{sol}
\frac{\tanh|\mu_1(u_s-u_{01})|}{\tanh|\mu_2(u_s-u_{02})|}=\frac{\mu_1}{\mu_2}.
\ee

%In  Fig.~\ref{fig:DIL-POT-s} we observe that the extremal point of $\phi(u)$ coincides 
%with the  extremal point of $V(u)$, $V'_u=V'_\phi\,\phi'_u$.
%Note that the last does not coincide with an extremal point of $V(\phi)$. 

%\begin{figure}[h!]
%\centering
 %\includegraphics[width=5.5cm]{3branch/Vphi3br.pdf}$\,\,\,\,\,$
  %\includegraphics[width=2.5cm]{3branch/Vphi3brleg.pdf}$\,\,\,\,\,$
 %\caption{The rescaled potential and dilaton as  functions of $u$ in the right solution, $u>u_{01}$. We see that $V'_u|_{u=u_s}=0$
% corresponds to $\phi'_u|_{u=u_s}=0$.}
 %\label{fig:DIL-POT-s}
%\end{figure}
 
\begin{figure}[h!]
\centering
 \includegraphics[width= 9cm]{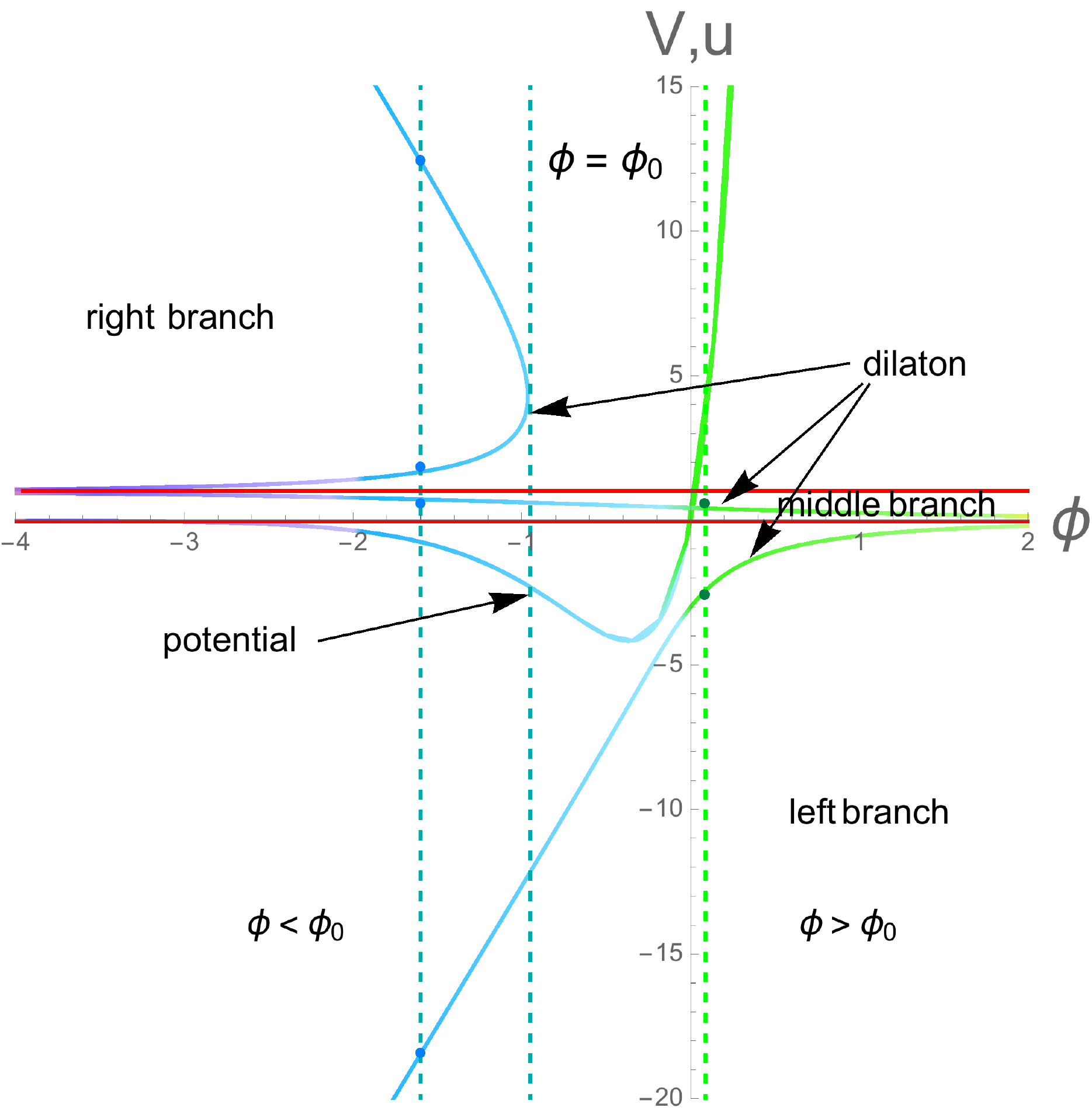}
 \caption{The dilaton potential  $V=V(\phi)$ on the vacuum solutions  $\phi=\phi(u)$ with (\ref{chartl})-(\ref{chartr}) and plots that indicate which values of $u$ 
 correspond to given $\phi$, i.e. $u=u(\phi)$. The function $u(\phi)$ differs for each branch of the solutions, and moreover this function is double-valued at the right branch.  The different values of $u$ corresponding to the same $\phi$ are indicated by points at the vertical lines.}
 \label{fig:DIL-POT1phi}
\end{figure}

 Note, that the value of the potential at the turning point  $V_{s} = V(\phi_{s})$ (where $\phi_s = \phi(u_{s})$) doesn't coincide with the extremal point of the potential.  The  extremal point of $V(\phi)$ can also be computed analytically. Using 
\be
V'_\phi=2k_{1}C_{1}e^{2k_{1}\phi}+2k_{2}C_{2}e^{2k_{2}\phi}, \ee
and taking into account that  $C_1C_2<0$, we can find 
\bea
\phi_c =
\frac{9k}{(16-9k^2)}\log \frac{3k}{4}+\frac{9k}{2(16-9k^2)}\log \Big|\frac{C_1}{C_2}\Big|.
\label{phic}\eea
$$\,$$
We show the behaviour of the potential as a function of  $\phi$ and the dilaton as a function of $u$ on the same plot in Fig.~\ref{fig:DIL-POT1phi}. In this picture we draw the dependences for all solutions.

\subsubsection{Special case $u_{01} =u_{02}$, solutions with AdS boundary}

Let us  see  the features of the solution given by (\ref{MABC.pC})-(\ref{dil-u012})  with (\ref{chart2}) and $u_{01}=u_{02} = u_{0}$.

\begin{figure}[h!]
\centering
 \includegraphics[width=6cm]{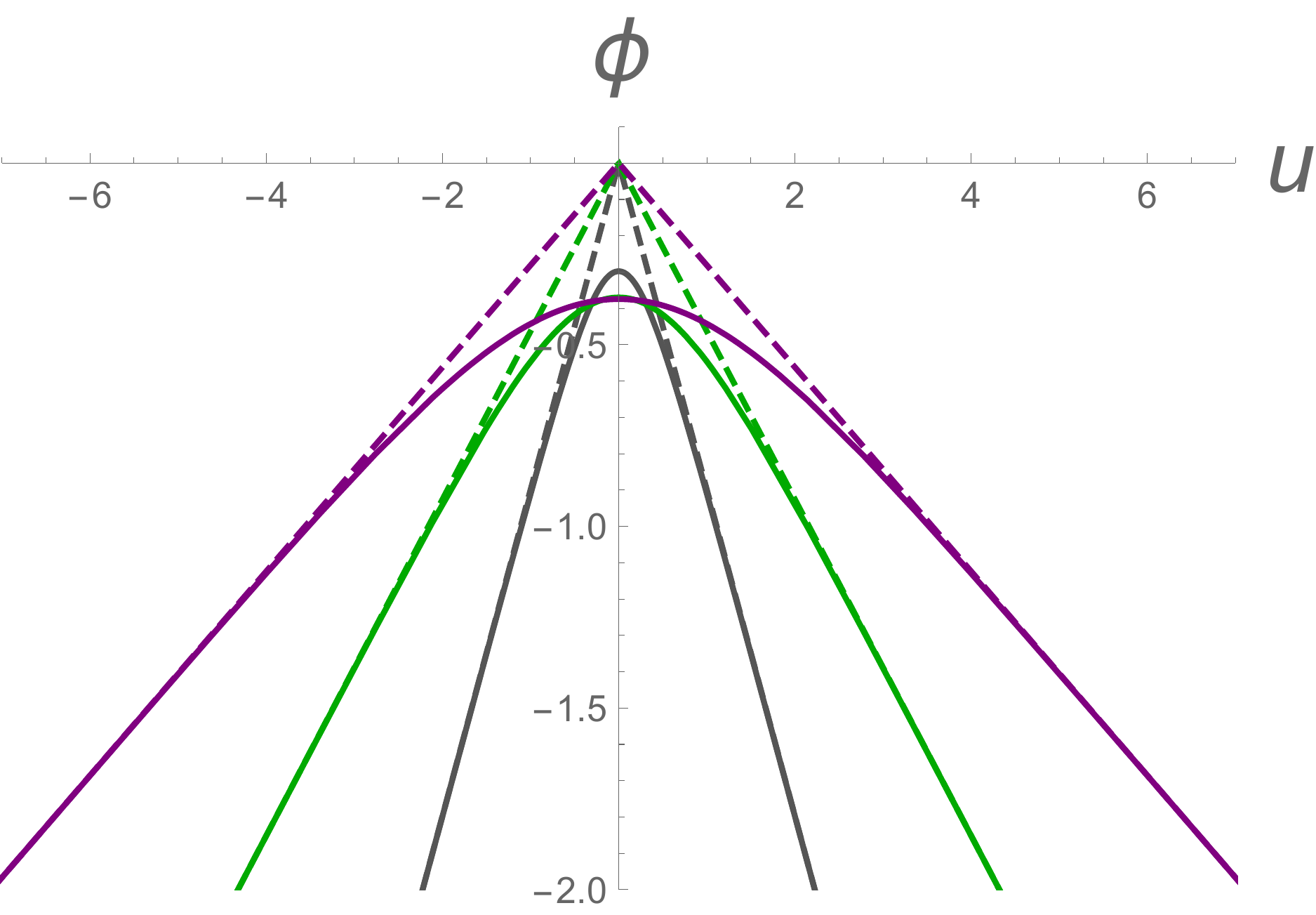}A$\,\,\,$
  \includegraphics[width=2cm]{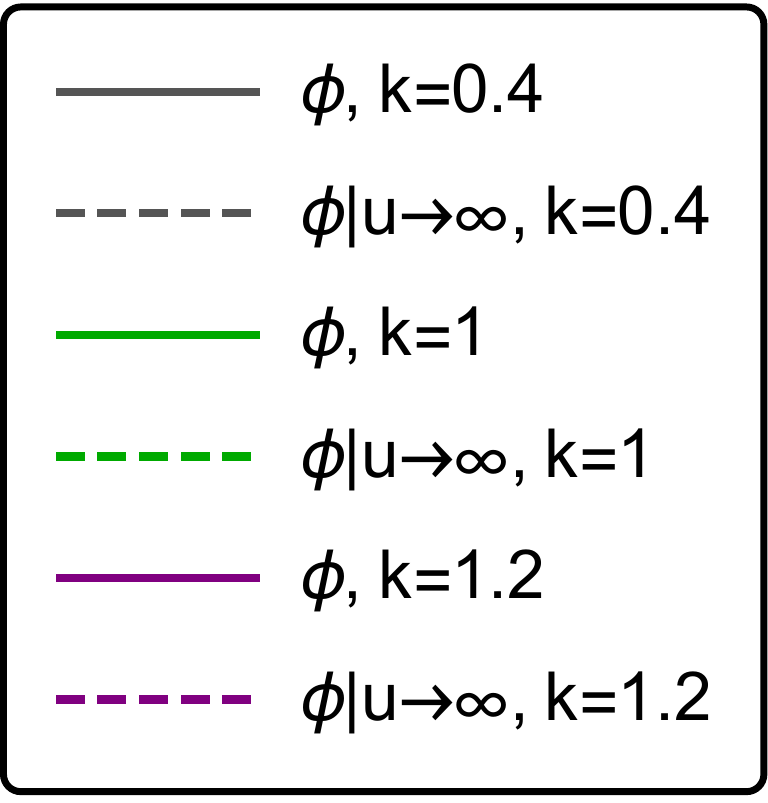}
  \includegraphics[width=4.5cm]{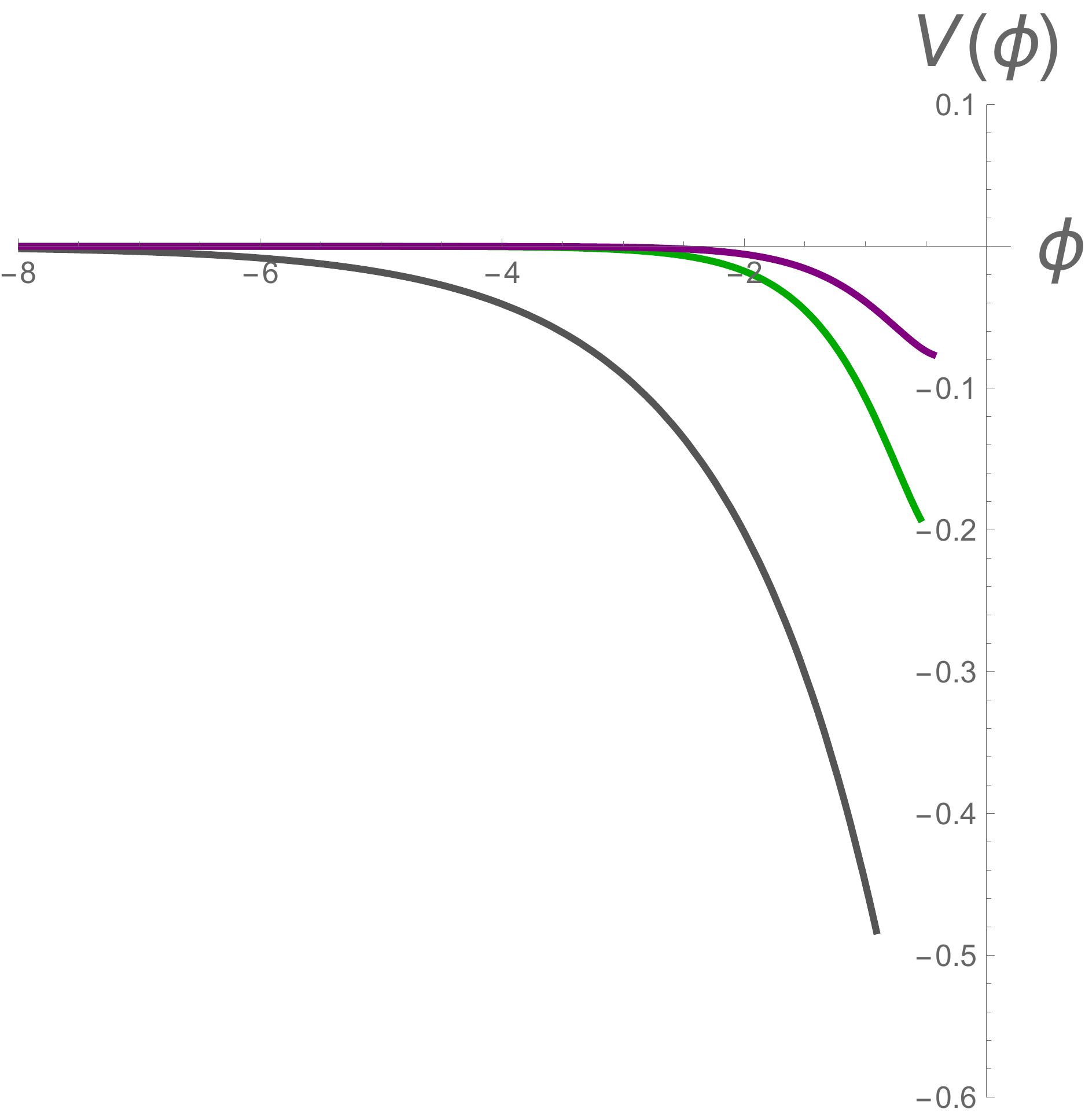}B
   \caption{ A) The behaviour of the dilaton (solid lines) and its asymptotics at infinity (dashed lines) for $u_{01} = u_{02} =0$, $C_1=-C_2= -1$, $E_{1} = - E_{2} = -1$ and different values of $k$. From bottom to top  $k=0.4, 1, 1.2$. B) The dilaton potential as a function of $\phi$ for $u > 0$.}
 \label{fig:EC}
\end{figure}

 In Fig.~\ref{fig:EC} the behavior of the dilaton (\ref{dil-u012}) with $u_{01} =u_{02} = u_0$ is shown. From this picture one can see that the dilaton tends to $-\infty$ as $u \to \pm \infty$.

As for the previous  solutions we present the asymptotics  in the conformal coordinates.\\

$\bullet$ In the limit with $u\rightarrow \pm \infty$  
\bea\label{23z3}
ds^{2} &\sim& z^{2/3}(-dt^{2} + dy^{2}_{1} + dy^{2}_{2} + dy^{2}_{3} + dz^{2}),\\ \label{asymp2b}
\phi& \sim&\frac{9k}{9k^2-16}(\mu_2-\mu _1)\,u \sim \log z \to -\infty ,
\eea
where the conformal radial coordinate is given by $z  \sim \mp \frac{4+3k}{3\mu_{1}}e^{\mp \frac{3\mu_{1}u}{4+3k}}$. 
So we come to the same asymptotics  as for the left  (\ref{23z2}) and right  (\ref{23z1}) solutions with $u \to \pm \infty$ for the 3-branch case. 
The scalar curvature has also a common behaviour with the right and left solutions, namely it goes to infinity (\ref{sclimitinf}) with $u \to + \infty$ .\\

$\bullet$ For $u\to u_{0}$ one gets has the following form of the metric (\ref{MABC.pC}) 
\bea\label{5dAdS}
ds^{2} \sim \frac{1}{z^{2}}(- dt^{2} + dy^{2}_{1} +dy^{2}_{2}+dy^{2}_{3} + dz^{2}),
 \eea
where the conformal "radial" coordinate  is defined  as $z =  4(u - u_{0})^{1/4}$ and $z\to 0$ as $u\to u_{0}$.
In  (\ref{5dAdS}) one can easily recognize the 5d AdS metric supported by  the constant dilaton 
\bea\label{phi2b0}
\phi|_{u\to u_0}\sim \frac{9k}{(16-9k^2)}\log \frac{3k}{4}+\frac{9k}{2(16-9k^2)}\log \Big|\frac{C_1}{C_2}\Big|,
\eea
which coincides with the minimum of the potential (\ref{phic}).  As expected the scalar curvature of this solution with $u\to u_0$ has a constant value (\ref{sclimit0}).
 
We note that the Fig.~\ref{fig:EC} is an agreement with the calculations of the asymptotics for the dilaton.  The  potential of the dilaton  as a function of $\phi$ with $u_{01} = u_{02} = u_0$ is presented in Fig.~\ref{fig:EC} {\bf B)}. From this plot we observe the existence of the turning point of the two-branch solution.
The equation for the stop point (\ref{sol}) in this case becomes 
\be
\frac{\tanh(\mu_1\,(u_s-u_0))}{\tanh(\mu_2\,(u_s-u_0))}=\frac{\mu_1}{\mu_2},
\ee
and has a solution 
\be
u_s=u_0.
\ee
As already observed, in this case we find that the stop point coincides with the extremal point of the potential, 
$
\phi_s=\phi_c=\phi(u_0)
$.

\subsection{RG flow  for vacuum solutions}

\subsubsection{Details of RG flow  for vacuum solutions}
 
The domain wall form (\ref{DW}) of the vacuum solutions  looks
\bea\label{VSDWC}
ds^{2}&=& e^{2{\cal A}} \Big[-dt^{2} +dy^{2}_{1}+dy^{2}_{2} +dy^{2}_{3}\Big]+ dw^{2}.
\eea

To come to (\ref{VSDWC}) we use the change of variables  for  (\ref{MABC.pC})-(\ref{dil-u012}) 
\bea
dw &= &F^{\frac{16}{9k^{2}-16}}_{1}F^{\frac{9k^{2}}{16-9k^{2}}}_{2}du.
\eea
We can represent the scale factor of the domain wall (\ref{VSDWC}) as follows
\bea\label{scaleE-dw}
{\cal A}&=&\frac{4}{9k^2-16}\,\log F_1+
\frac{9k^2}{4(16-9k^2)}\,\log F_2,
\eea
so the energy scale  A is
\bea\label{scaleE-dw2}
{\rm  A}\equiv e^{\mathcal{A}} = F^{\frac{4}{9k^{2} -16}}_{1}F^{\frac{9k^{2}}{4(16-9k^{2})}}_{2}.
\eea
The running coupling  is defined through the dilaton  (\ref{dil-u012}) reads
\bea\label{lambdaphig}
\lambda  = \left(\frac{F_{2}}{F_{1}}\right)^{\frac{9k}{9k^{2} -16}}.
\eea

\begin{figure}[h!]
\centering
 \includegraphics[width=4.5cm]{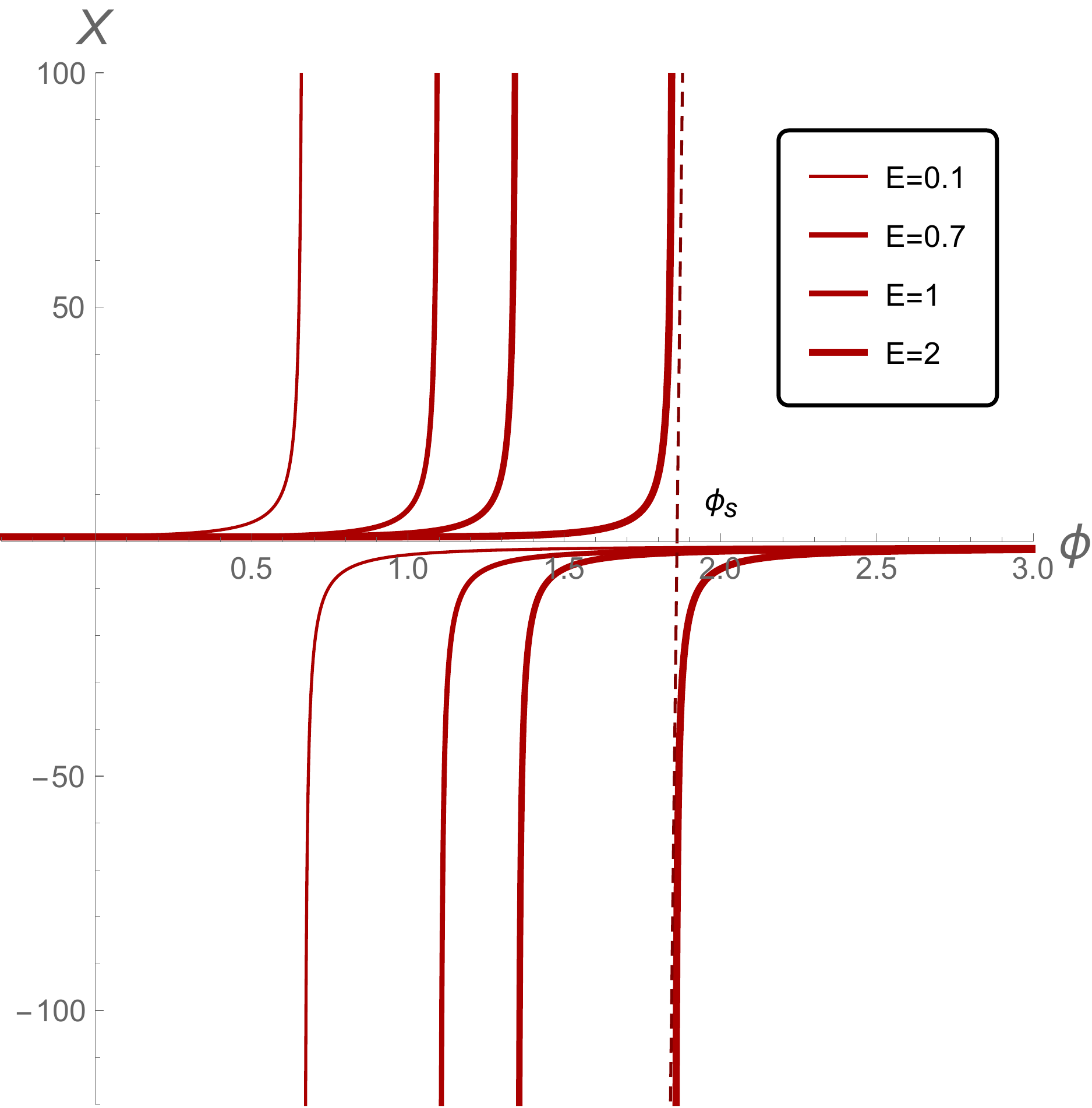}$\,\,\,$
     \includegraphics[width=4.5cm]{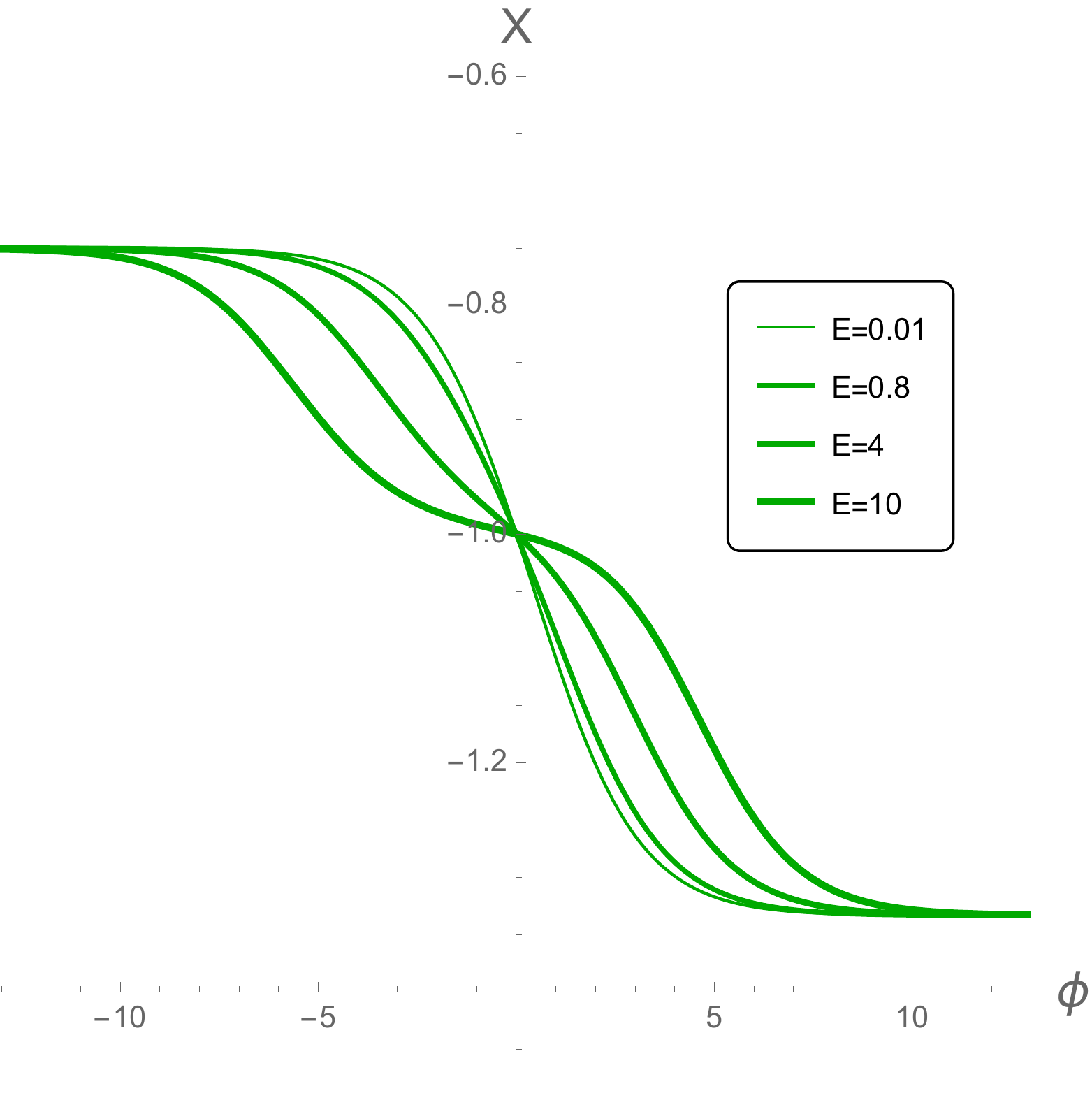}$\,\,\,$
    \includegraphics[width=4.5 cm]{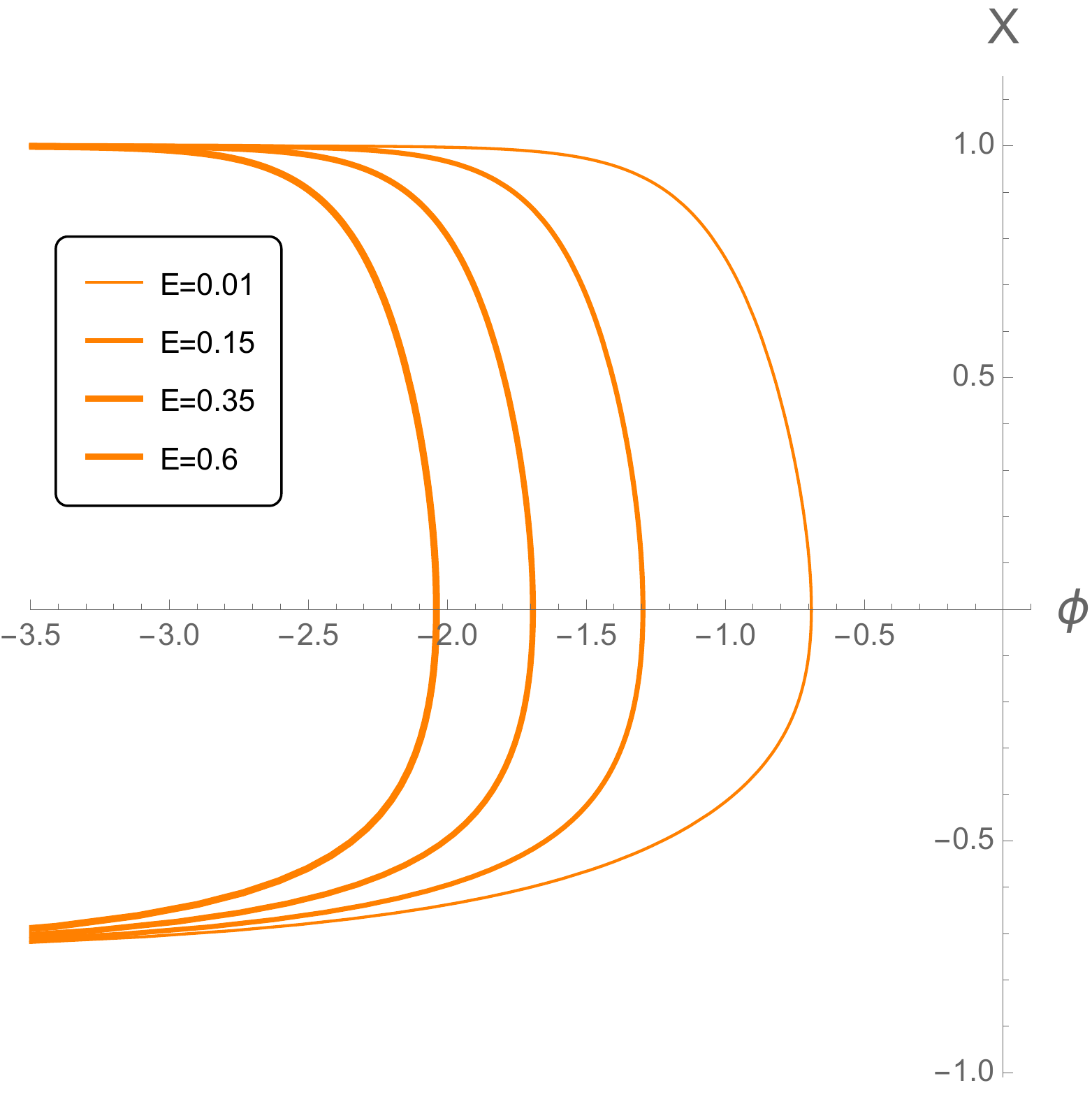}\\
    A$\,\,\,\,\,\,\,\,\,\,\,\,\,\,\,\,\,\,\,\,\,\,\,\,\,\,\,\,\,\,\,\,\,\,\,\,\,\,\,\,\,\,\,\,\,\,\,\,\,\,\,\,\,\,\,\,\,\,\,\,\,\,\,\,\,\,\,\,\,\,\,\,\,\,\,\,\,\,\,\,\,\,$B$\,\,\,\,\,\,\,\,\,\,\,\,\,\,\,\,\,\,\,\,\,\,\,\,\,\,\,\,\,\,\,\,\,\,\,\,\,\,\,\,\,\,\,\,\,\,\,\,\,\,\,\,\,\,\,\,\,\,\,\,\,\,\,\,\,\,\,\,\,\,\,\,\,\,\,\,\,\,\,\,\,\,$C
    \caption{ The behaviour of the $X$-function with the dependence on the dilaton plotted using the solutions for $\mathcal{A}$ and $\phi$: {\bf A)} the left branch with $u_{02}>u$, {\bf B)} the middle branch $u_{02}<u<u_{01}$; {\bf C)} the right branch $u>u_{01}$. For  all plots $u_{01} =0$, $u_{02}=-1$, $k=1$, $C_{1} = -2$, $C_{2}=2$, different curves on the same plot correspond to the different values of $|E_{1}| =|E_{2}|$, labeled as $E$ on the legends.}
    %{\bf Math.file: Reproducing-X-17-02.nb+VacuumRGLeft.nb} }
 \label{fig:X-repr}
 \end{figure}

The function X (\ref{XX}) is represented as follows
\bea\label{XonSOL}
X= \frac{1}{3}\left(\frac{F_{2}}{F_{1}}\right)^{\frac{9k}{16 - 9k^{2}}}\frac{ \lambda'}{ \mathcal{A}'},
\eea
with $\lambda$ given by (\ref{lambdaphig}), $\mathcal{A}$ -- by (\ref{scaleE-dw}).
It is useful to explore the evolution of the rescaled $\beta$-function $X$ (\ref{XonSOL}) as a function of $log$ of the running  coupling (\ref{lambdaphig}), which is the dilaton $\phi$. 
In Fig.~\ref{fig:X-repr} we present the parametric plots of the behaviour of $X$ (\ref{XonSOL})  using  solutions for $\phi$ (\ref{FD.4nv}) and $\mathcal{A}$ (\ref{scaleE-dw}) with the dependence on the parameter $u$ (\ref{chartl})-(\ref{chartr}).  
In these plots we fix the shape of the potential putting $C_{1} =- C_{2} = -2$ and $k= 1$ while we vary the constants $|E_{1}| = |E_{2}|$ (labeled as $E$ in Fig.~\ref{fig:X-repr}). 
From Pic. \ref{fig:X-repr} {\bf A)} and {\bf C)} we see that  the holographic $\beta$-functions at zero temperature constructed on the left $u<u_{02}$ and right solutions $u >u_{01}$ can take both negative and positive values. As for the case {\bf B)}, which corresponds to the middle solution, $X$  is always negative.  

\begin{figure}[h!]
 \centering 
 \includegraphics[width=4.5cm]{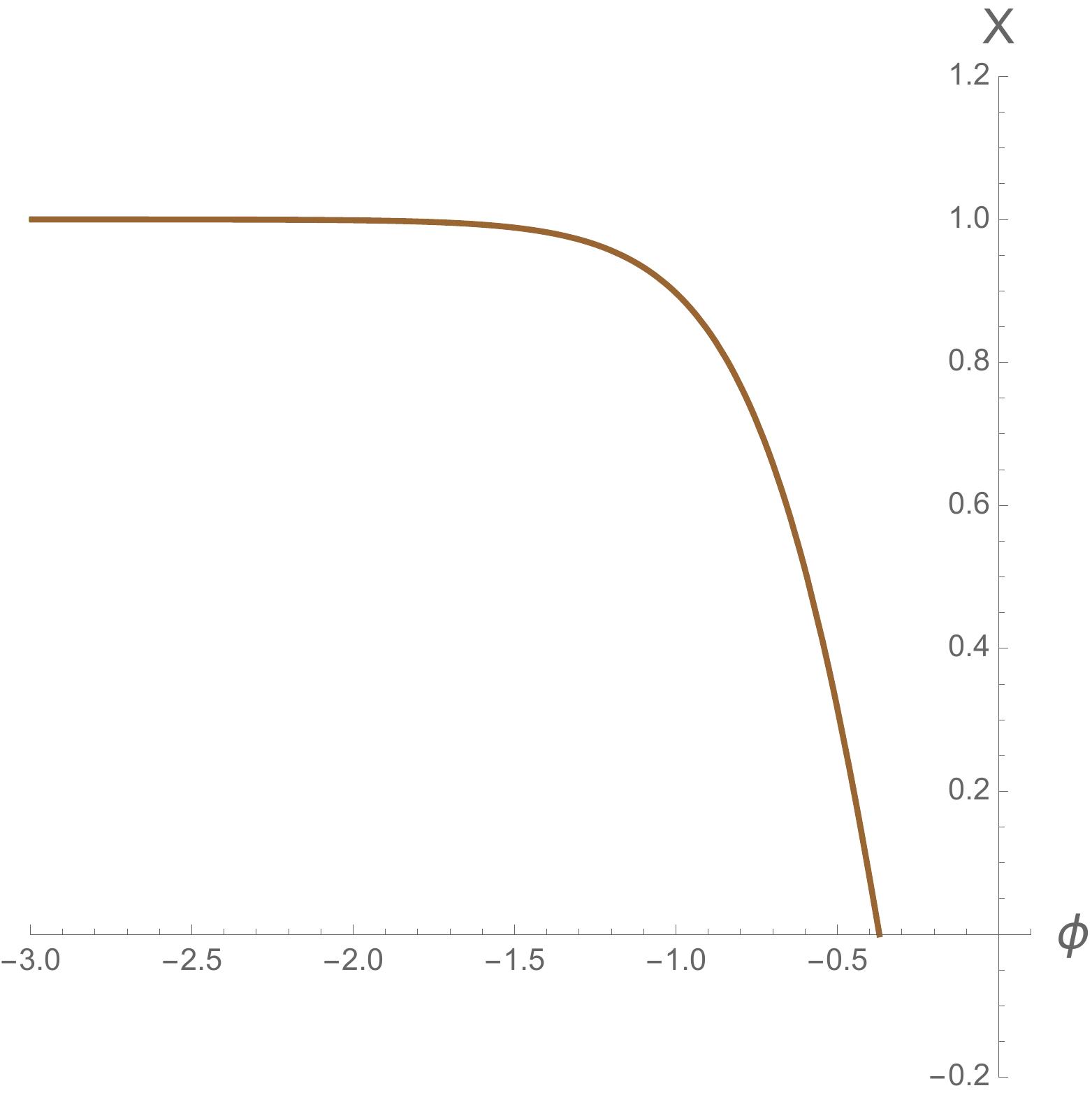}$\,\,\,$
  \includegraphics[width=1.5cm]{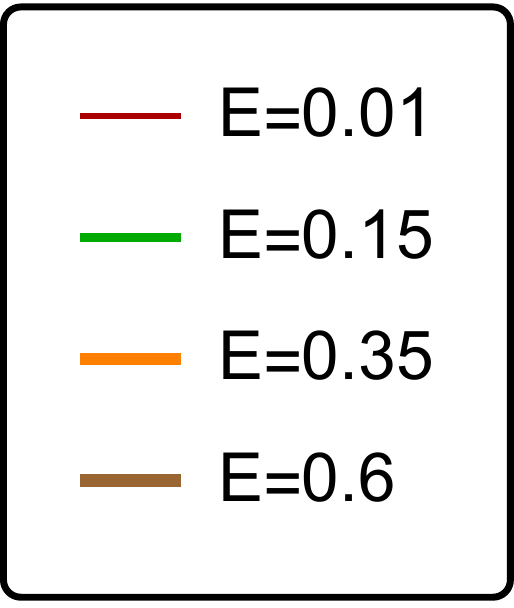}
 \caption{The $X(\phi)$ function for the dilaton solution with $u_{01}=u_{02}=0$,  the potential fixed as $C_{1} = -2$, $C_{2} =2$, $k=1$. For all range of values of $|E_{1}| =|E_{2}|$, labeled as $E$ on the legend, the curves of $X$ coincide.}
 \label{fig:Xphi0}
 \end{figure}

\begin{figure}[h!]
\centering
 \includegraphics[width=4.5cm]{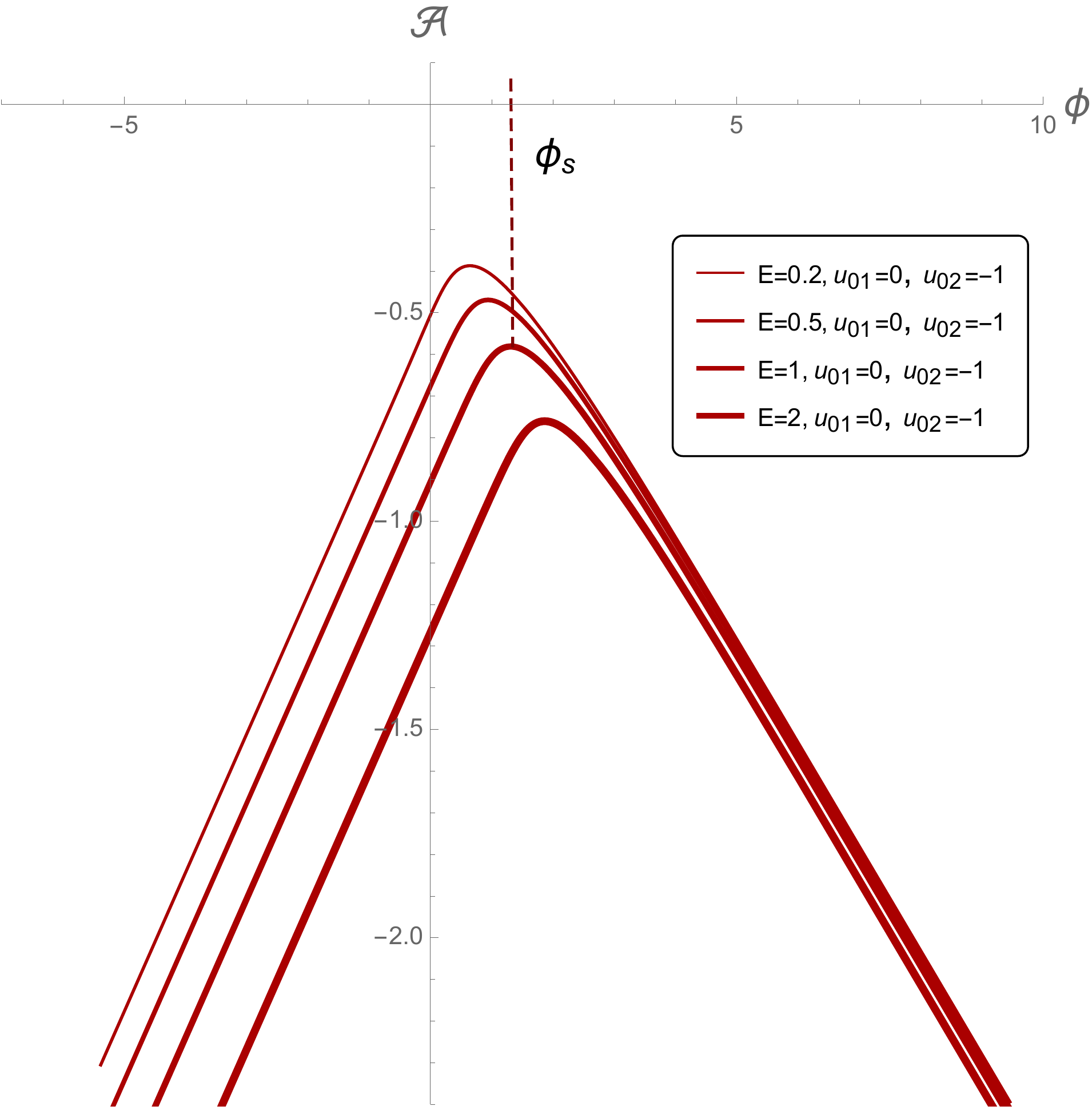}A $\,\,\,$
  \includegraphics[width=4.5cm]{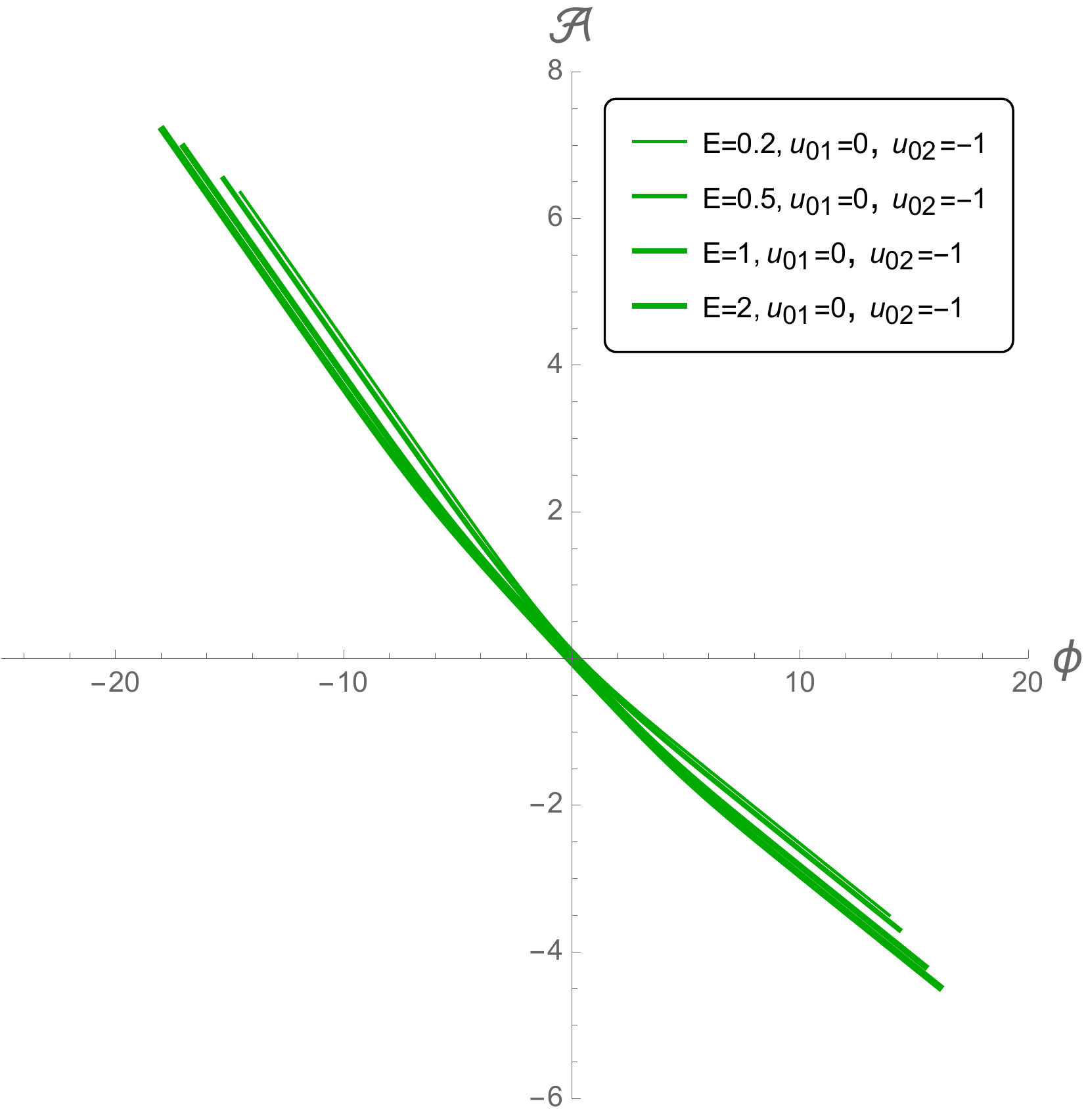}B$\,\,\,$
    \includegraphics[width=4.5cm]{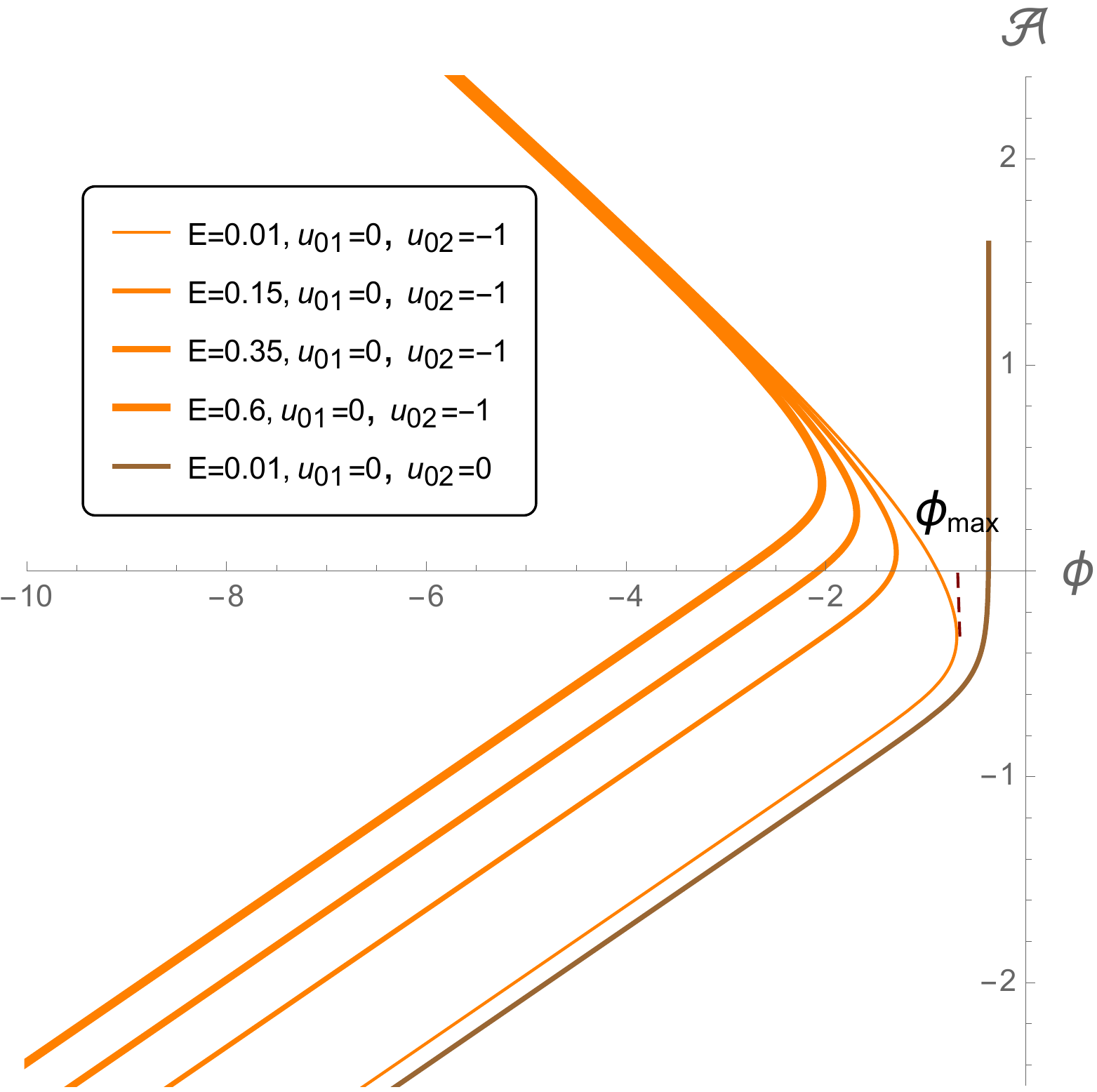}C
  \caption{The behaviour of $\mathcal{A}$ as a function of $\phi$ for the vacuum solutions, we fix values of $u_{01} =0$ and $u_{02} =-1$ and varying $|E_{1}| = |E_{2}|$, denoted as $E$: {\bf A)} the left solutions,  {\bf B)} the middle solutions {\bf C)} the right solutions by orange curves; the solution with $u_{01} =u_{02}$ and $E =0.01$ by the brown curve.}
 \label{fig:X'}
 \end{figure}

 It is also of our interest to see  the function $X(\phi)$ plotted on the solutions with $u_{01} = u_{02} =u_{0}$, particularly for $u_{0} = 0$.
 In Fig.~\ref{fig:Xphi0} we show the function $X(\phi)$ on the dilaton solution (\ref{dil-u012}) with $u_{01} = u_{02}$. 
 We see that the behaviour of the function $X(\phi)$ is the same for all values of  $E_{2}$.
 From Figs.~\ref{fig:X-repr}-\ref{fig:Xphi0} we see that $X$ are regular except those plotted on the left solutions on Fig.~\ref{fig:X-repr} {\bf A)}  at some points  $\phi_{s}$ where $X(\phi_{s})$ takes an infinite value. From eq.~(\ref{XonSOL}) with $\alpha^{1} = 0$ one can see that $X$ is infinite with $\mathcal{A}'=0$, i.e.
\bea\label{singus}
9k^{2}\frac{F'_{2}}{F_{2}} -16\frac{F'_{1}}{F_{1}} = 0.
\eea 
Eq. (\ref{singus}) defines the singular point $u_{s}$, which is related to $\phi_{s}$.
 The singular point $\phi_s$ coincides with the point where $\mathcal{A}'=0$, and its position depends on  
 $E,k,u_{01}$ and $u_{02}$, see Fig.\ref{fig:X'} {\bf A)}.
 From this picture we see that for the left solution the dilaton varies  from $-\infty$ to a special value $\phi _s$, the scale factor is non-monotonic function increases, but after the dilaton passes this special value  $\phi _s$, the scale factor starts to decrease.
 From  Fig.~\ref{fig:X'} {\bf B)} we observe that the scale factor $\mathcal{A}$ has a monotonic behaviour, decreasing with respect to dilaton running from $-\infty$ to $+\infty$.  Fig.~\ref{fig:X'}  {\bf C)} it is demonstrated by orange curves that  the scale factor of the right solution decreases from $+\infty$ to $-\infty$ for all values of the dilaton, which runs from $-\infty$ to some constant value and then goes back to $-\infty$. We also present that the behavior of the scale factor $\mathcal{A}$ on $\phi$ for the solution with $u_{01}=u_{02}=0$ by the brown curve in Fig.~\ref{fig:X'}  {\bf C)}. We see that  the scale factor for this solution starts to decrease  from $+\infty$ with some constant value of the dilaton, then,  passing some value of $\mathcal{A}$, both the scale factor and the dilaton tend to $-\infty$.
 
In the section~\ref{Sect:RG} it was already said  that the system (\ref{dXphi})-(\ref{dYphi}) on the vacuum solutions reduces to eq. (\ref{dXphi0}), which is quite simple and one can treat it in the general form. We remind that the dilaton potential is given by \eqref{1.1b} and in the cases presented at the plots below
\be\label{V2E}
V = -2e^{2k\phi} + 2e^{\frac{32}{9k}\phi},
\ee
with $k=k_1$ and $k_2 = \frac{16}{9k_{1}}$ (a constraint from the solution) and $C_{1} = -2$, $C_{2} = 2$. The values of the dilaton coupling constant have the following restriction $0<k<4/3$.\\

\begin{figure}[h!]
\centering
 \includegraphics[width=6.2cm]{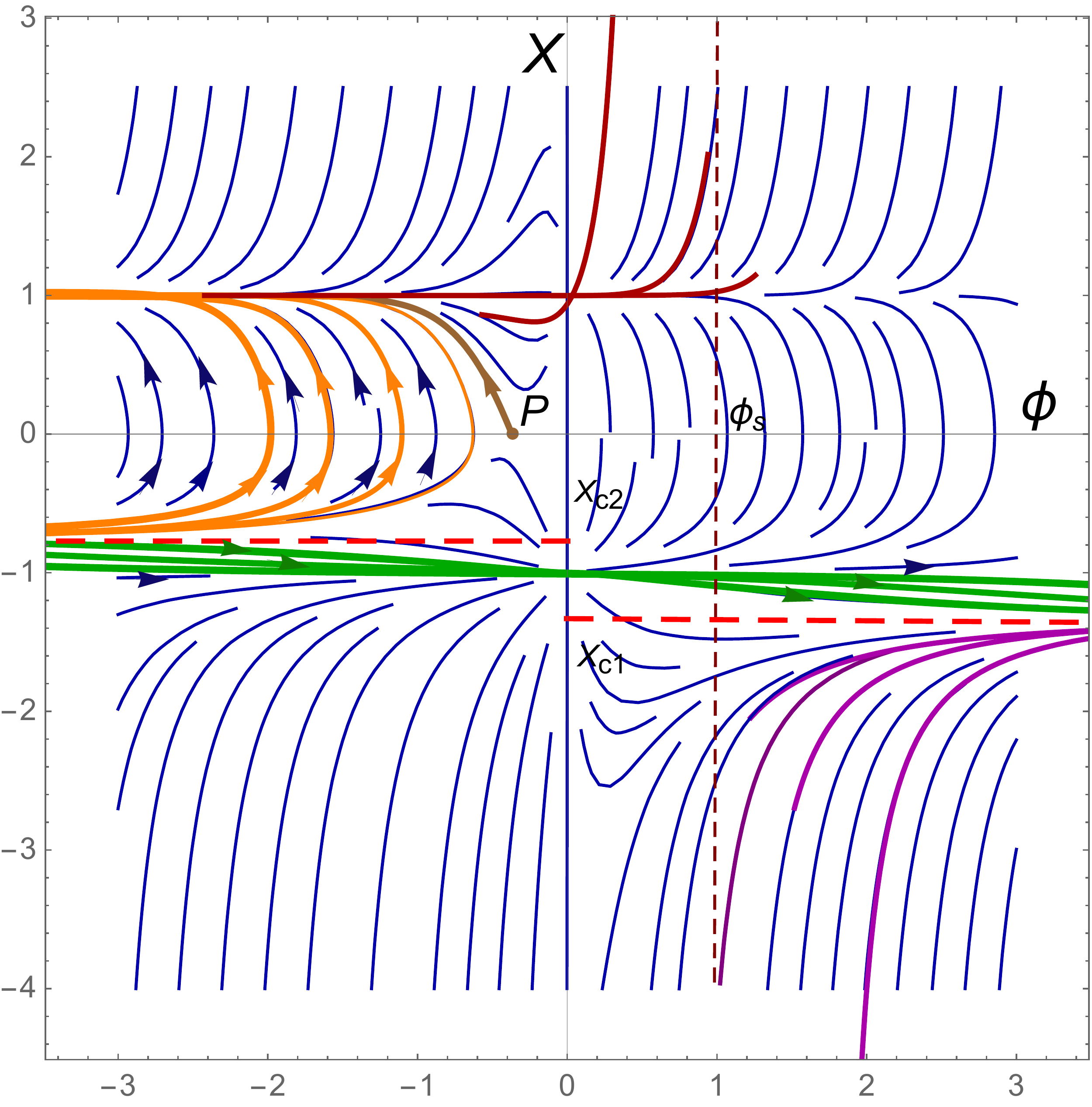}$\,\,\,\,\,$
 \includegraphics[width=6.2cm]{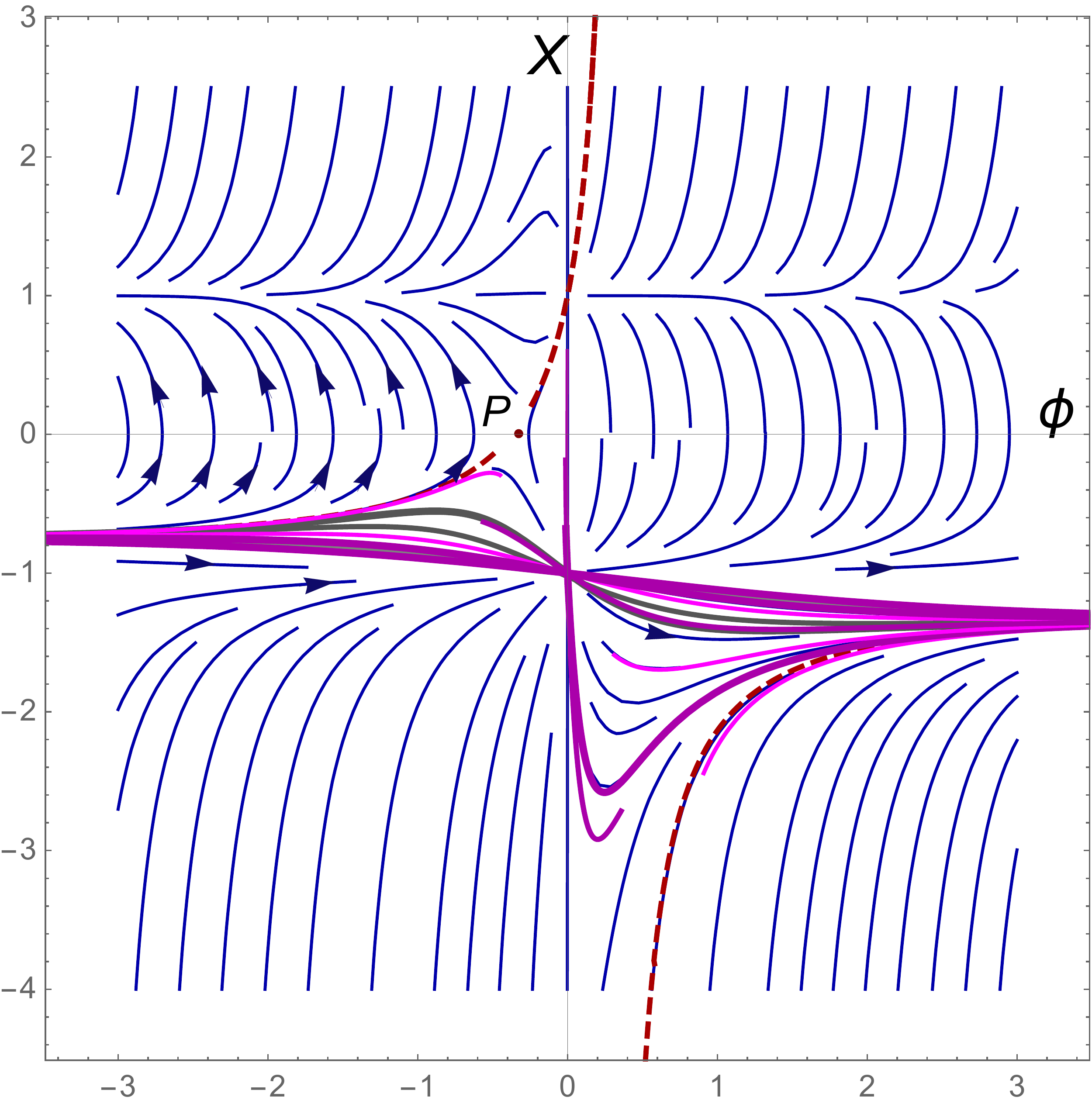}\\
 $\,\,,$ \includegraphics[width=2.5cm]{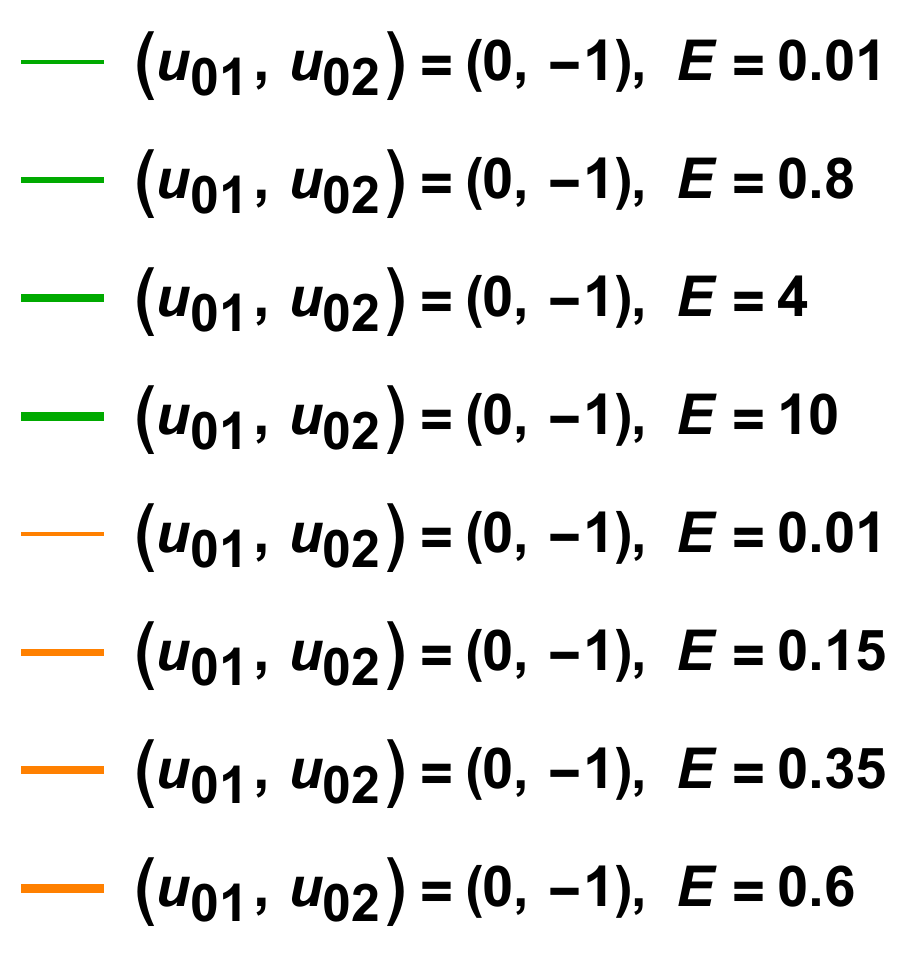}\includegraphics[width=2.5cm]{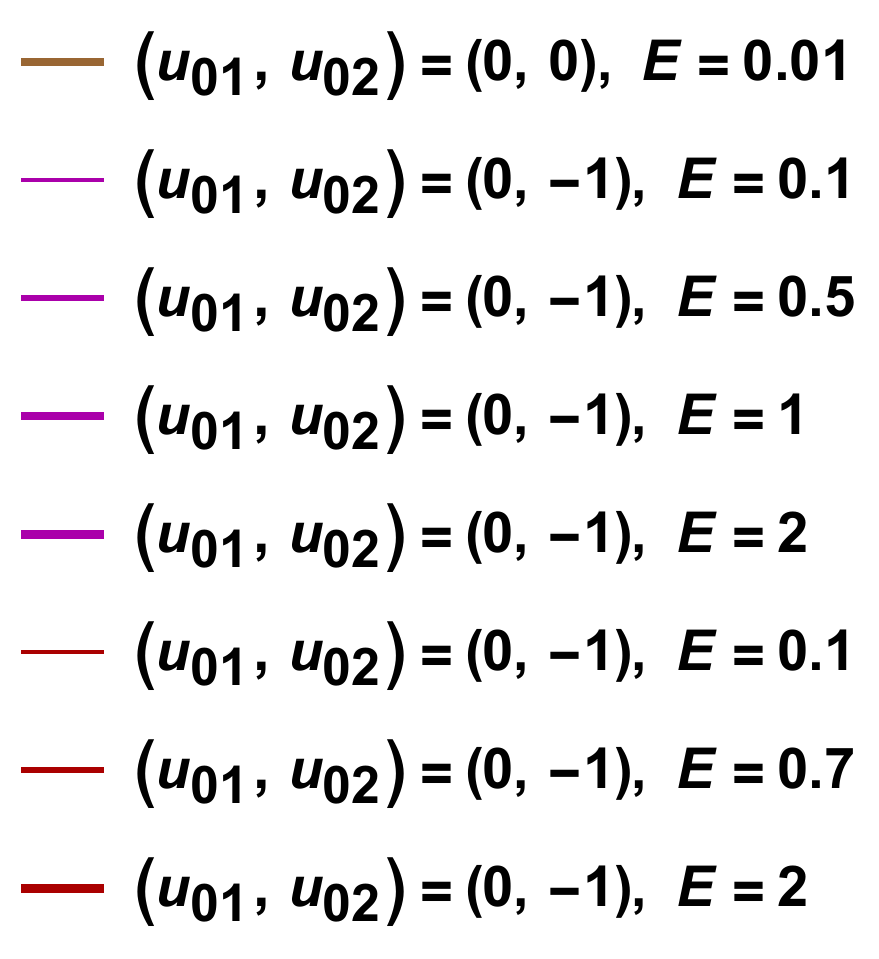}$\,\,\,$A$\,\,\,\,\,\,\,\,\,\,\,\,\,\,\,\,\,$
 \includegraphics[width=2.5cm]{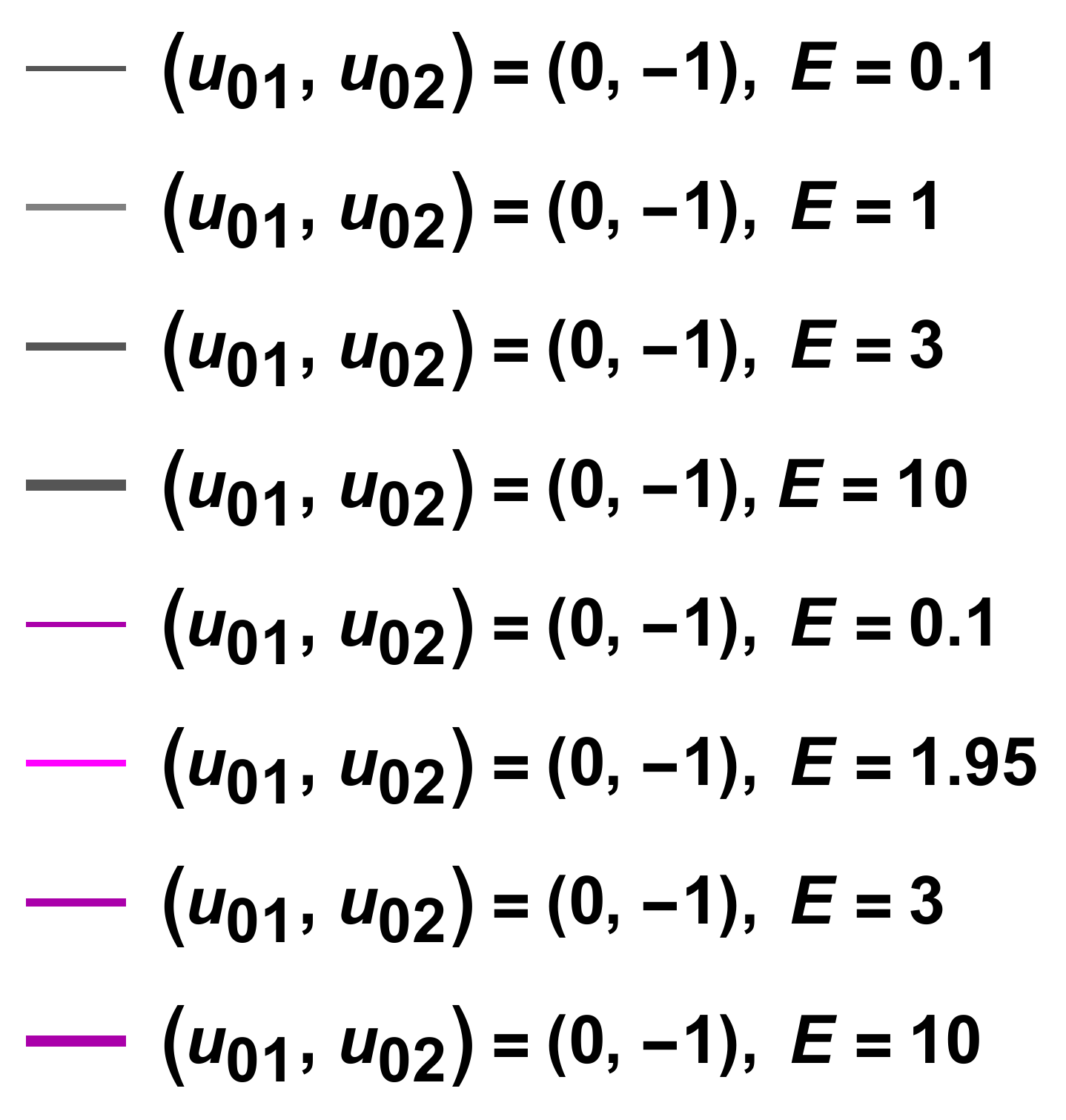}$\,\,\,\,\,\,\,\,\,\,\,\,\,\,\,\,\,\,\,\,\,\,\,\,\,\,\,\,\,\,$ B
    \\$\,$\\
   \includegraphics[width=6.2cm]{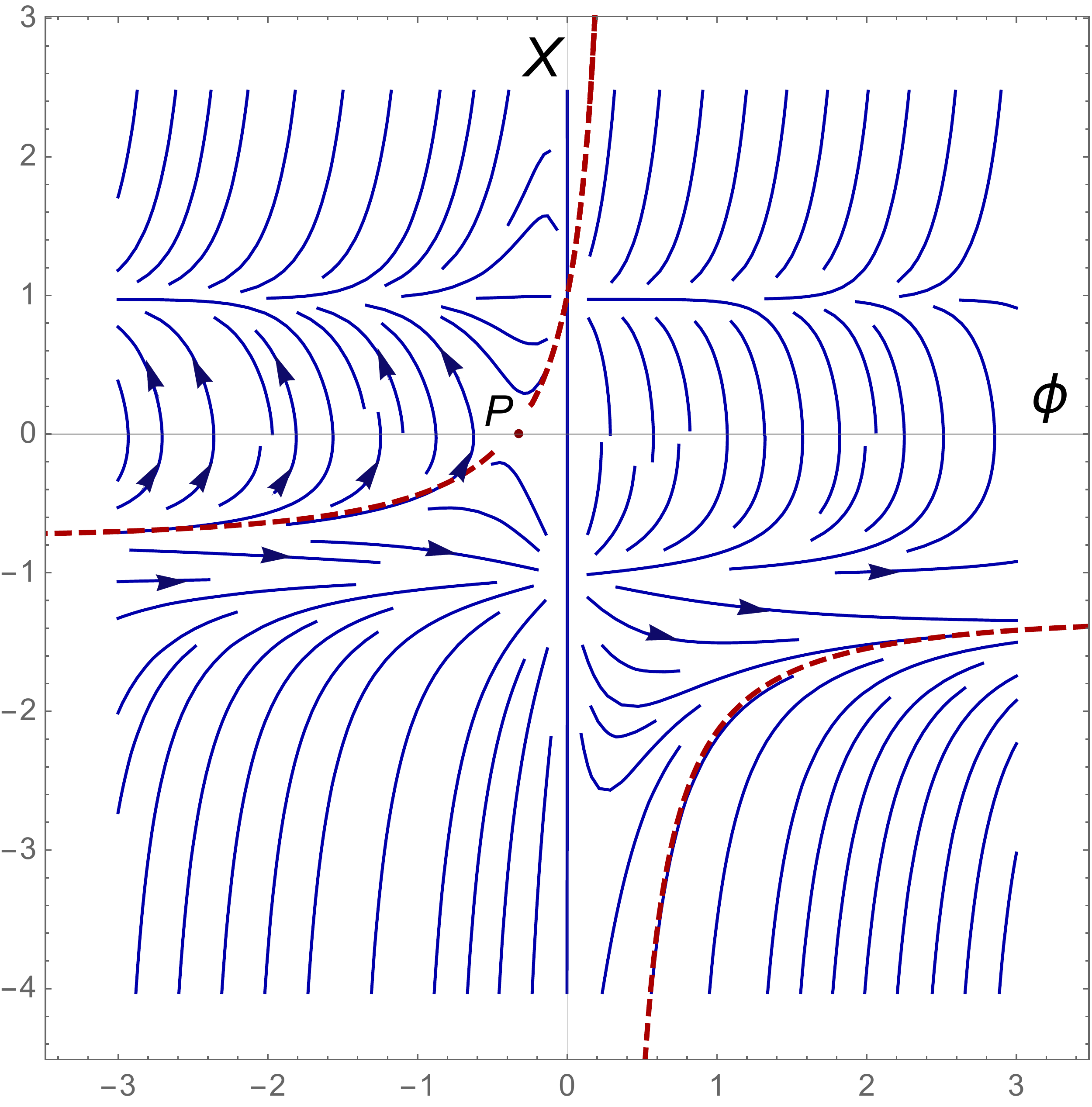}$\,\,\,\,\,$
   \includegraphics[width=6.2cm]{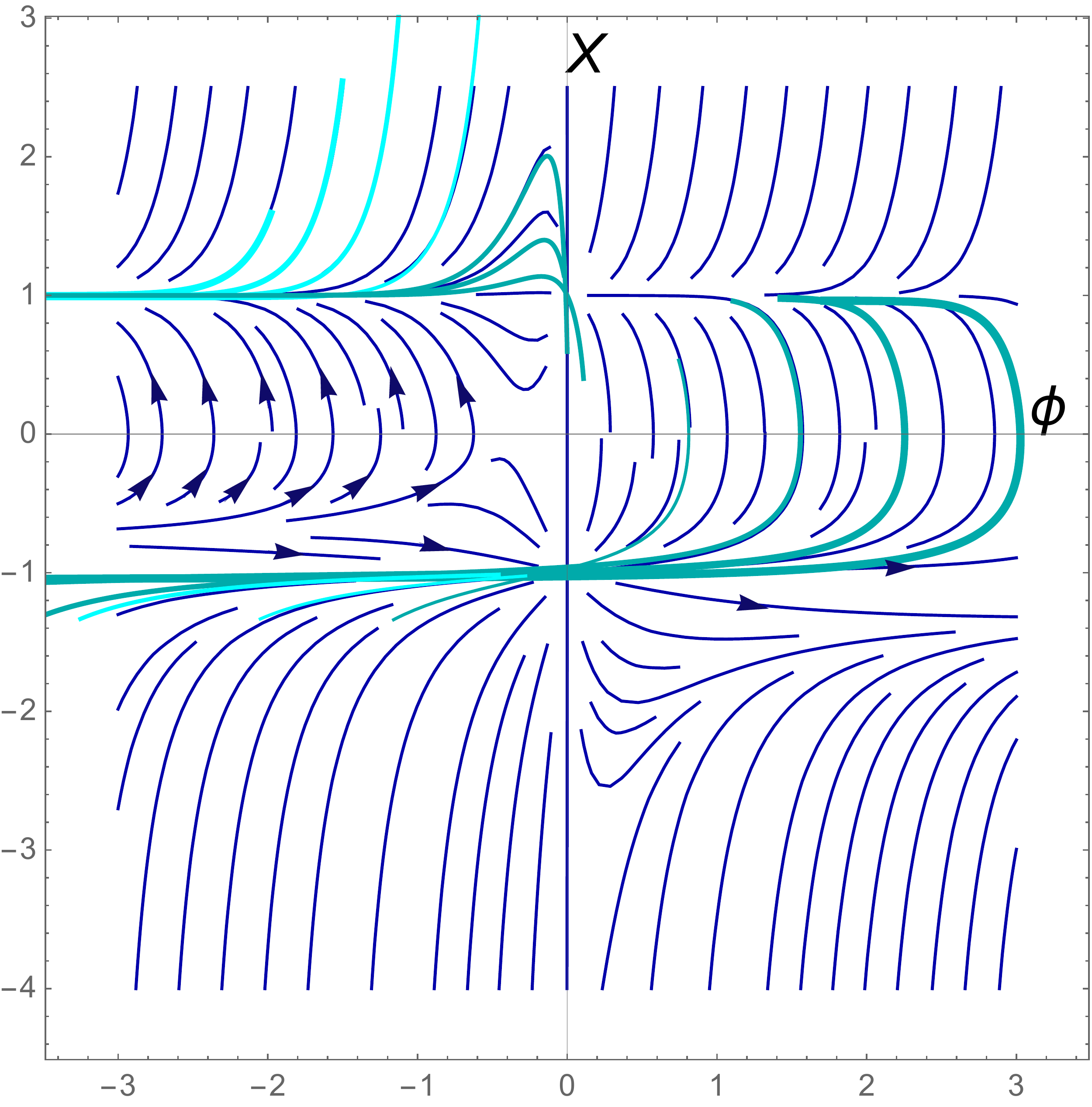}\\  
    $\,\,\,\,\,\,\,\,\,\,\,\,\,\,\,\,\,\,\,\,\,\,\,\,\,\,\,\,\,\,\,\,\,\,\,\,\,\,\,\,\,\,\,\,\,\,\,\,\,\,\,\,\,\,\,\,\,\,\,$ 
   $\,\,\,\,\,\,\,\,\,\,\,\,\,\,\,\,\,\,\,\,\,\,\,\,\,\,\,\,\,$C$\,\,$\includegraphics[width=2cm]{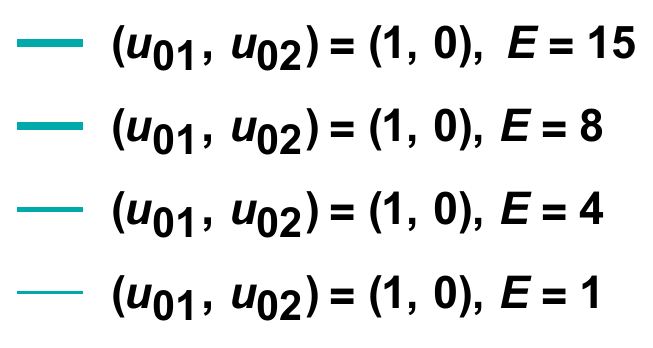} \includegraphics[width=2cm]{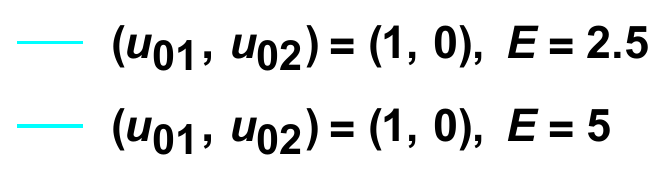} \includegraphics[width=2.5cm]{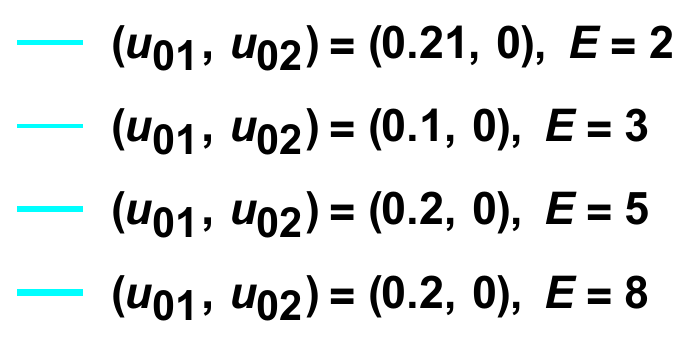}$\,\,$D\\
  \caption{\small All solutions $X$ to \eqref{dXphi0} (blue curves) with potential fixed as $C_1=-C_2=-2$ and $k=1$. {\bf A)} The orange curves show $X=X(\phi)$ plotted on the right solution, the green curves show $X=X(\phi)$ on the middle one, the dark magenta  and red lines show $X=X(\phi)$ one the left solution. {\bf B)} The magenta and grey curves show $X=X(\phi)$ built on the analytic solutions with $\sin$-formula (\ref{F1sin1})-(\ref{F1sin2}), the red dashed lines 
 show $X=X(\phi)$,  given by the "linear"-formula \eqref{1lin}. {\bf C)} Separately, the red lines 
 show $X=X(\phi)$ governed by "linear"-functions \eqref{1lin}. {\bf D)} As for the pictures  {\bf A, B, C} blue lines show the stream, that solves equation \eqref{dXphi0}, 
 while the lines of cyan and darker cyan lines present the behaviour of $X(\phi)$ governed by $cosh$ functions \eqref{cosh1}-\eqref{cosh2}.}
 \label{fig:Xcomparison}
 \end{figure}

Taking into account (\ref{V2E}) we have
\bea\label{VXeq}
\frac{d \log V}{d \phi}  = \frac{-2k e^{2k\phi} + \frac{32}{9k}e^{\frac{32}{9k}\phi}}{-e^{2k\phi} + e^{\frac{32}{9k}\phi}}.
\eea
The solution to (\ref{dXphi0}) with  (\ref{V2E})-(\ref{VXeq}) is represented in Fig.~\ref{fig:Xcomparison} using StreamPlot. 
On this figure we observe all possible solutions for $X$ (blue curves) to our model with the potential (\ref{V2E}).  
We also impose Figs.~\ref{fig:X-repr} and \ref{fig:Xphi0} on Fig.~\ref{fig:Xcomparison}{\bf A)} and see that they can partially cover the plot. In Fig.~\ref{fig:Xcomparison}{\bf A)}
the red lines correspond to the left solution, see Fig.~\ref{fig:X-repr} {\bf A)}, the green lines are those from the middle solution, see Fig.~\ref{fig:X-repr} {\bf B)},
 the orange ones correspond to the right solution, see  Fig.~\ref{fig:X-repr} {\bf C)}. In Fig.~\ref{fig:Xcomparison} {\bf A)} we observe that the function $X$ corresponding to the right solution (the orange curves) interpolates between $X_{c2}$ and $1$. The $X$ for the middle solution (the  green curves ) interpolates between  $X_{c1}$ and  $X_{c2}$ in Fig.~\ref{fig:Xcomparison}.  The dark red curves start at $1$  and go to $+\infty$ as $\phi \to \phi_s-0$, some of these lines have local minimum and are located very close to the brown curve  corresponding to the vacuum flow for the two branch solution $u_{01} = u_{02}$. The flow with local maximum in Fig.~\ref{fig:Xcomparison} {\bf A)} corresponds to our solutions with small difference between $u_{01}$ and  $u_{02}$.  The magenta lines start at $X_{c1}$ and go to $-\infty$
 when $\phi \to \phi_s+0 $.  For the fixed form of the potential the point $\phi_s$ is defined by values of $|E_{1}|=|E_{2}|$ (labeled by $E$ on legend), $u_{01}$, $u_{02}$, i.e. $\phi_s=\phi_s(E,u_{01},u_{02})$.
 
  We see that in Fig.~\ref{fig:Xcomparison}  {\bf A)} some parts of the RG flow (the stream at the left bottom part as well as the stream at the right upper part that interpolates between $-1$ and $1$)  are not covered by our vacuum solutions. 
 However, it was already pointed in Sec.~\ref{Sec:2IntMM} that analytic solutions for $\mathcal{A}$ and $\phi$ can be governed by $F_{1}$ and $F_{2}$ that are $sin$- or linear functions, namely
(\ref{F1sin1})-(\ref{F1sin2}) and (\ref{1lin}).
Since the equation (\ref{dXphi0}) doesn't know about our choice of the solutions and we see on the plot all possible solutions, and the curves of the dependence $X(\phi)$ on $\phi$ built on (\ref{F1sin1})-(\ref{1lin}) should appear on the plot. In Fig.~\ref{fig:Xcomparison} {\bf B)} we present the stream of (\ref{dXphi0}) by blue lines, $X(\phi)$ on $\phi$ plotted using (\ref{F1sin1})-(\ref{F1sin2}) by magenta and grey lines and $X(\phi)$ related to (\ref{1lin}) by the dark red lines, correspondingly.  We see that these lines partially fit
the solutions to equation \eqref{dXphi0}. We show separately the behaviour of $X(\phi)$ with (\ref{1lin}) in Fig.~\ref{fig:Xcomparison} {\bf C)}. From Figs.~\ref{fig:Xcomparison} {\bf A)} and {\bf B)} we see that $X(\phi)$ corresponding to the linear solutions can be considered as a boundary between
$X(\phi)$ corresponding to "sin"-solutions  and $X(\phi)$  goverened by the left $sinh$-solutions (i.e. $sinh$-solutions with $u<u_{02}$).

However, we still observe in Figs.~\ref{fig:Xcomparison} {\bf B)} and {\bf C)} that some regions of the figure are not covered.
If we look at the equation (\ref{dXphi0}) it is evidently that it is invariant under the change of the sign of the potential (\ref{V2E}), i.e. $V \to -V$. 
So the stream of eq.(\ref{dXphi0}) should include also solutions to equations of motions with $-V$. 
In Sec.~\ref{Sec:2IntMM} we see that, indeed, these are solutions with $F_{1}$ and $F_{2}$ governed by $cosh$-functions \eqref{cosh1}-\eqref{cosh2}.

In Fig.~\ref{fig:Xcomparison} {\bf D)} we draw $X(\phi)$ as a solution to  \eqref{dXphi0} (blue lines) and  with the help of the analytical solution related to (\ref{cosh1})-(\ref{cosh2}) (the curves of other colors), that perfectly complete the necessary part of the stream.

\begin{figure}[h!]
\centering
 \includegraphics[width=7cm]{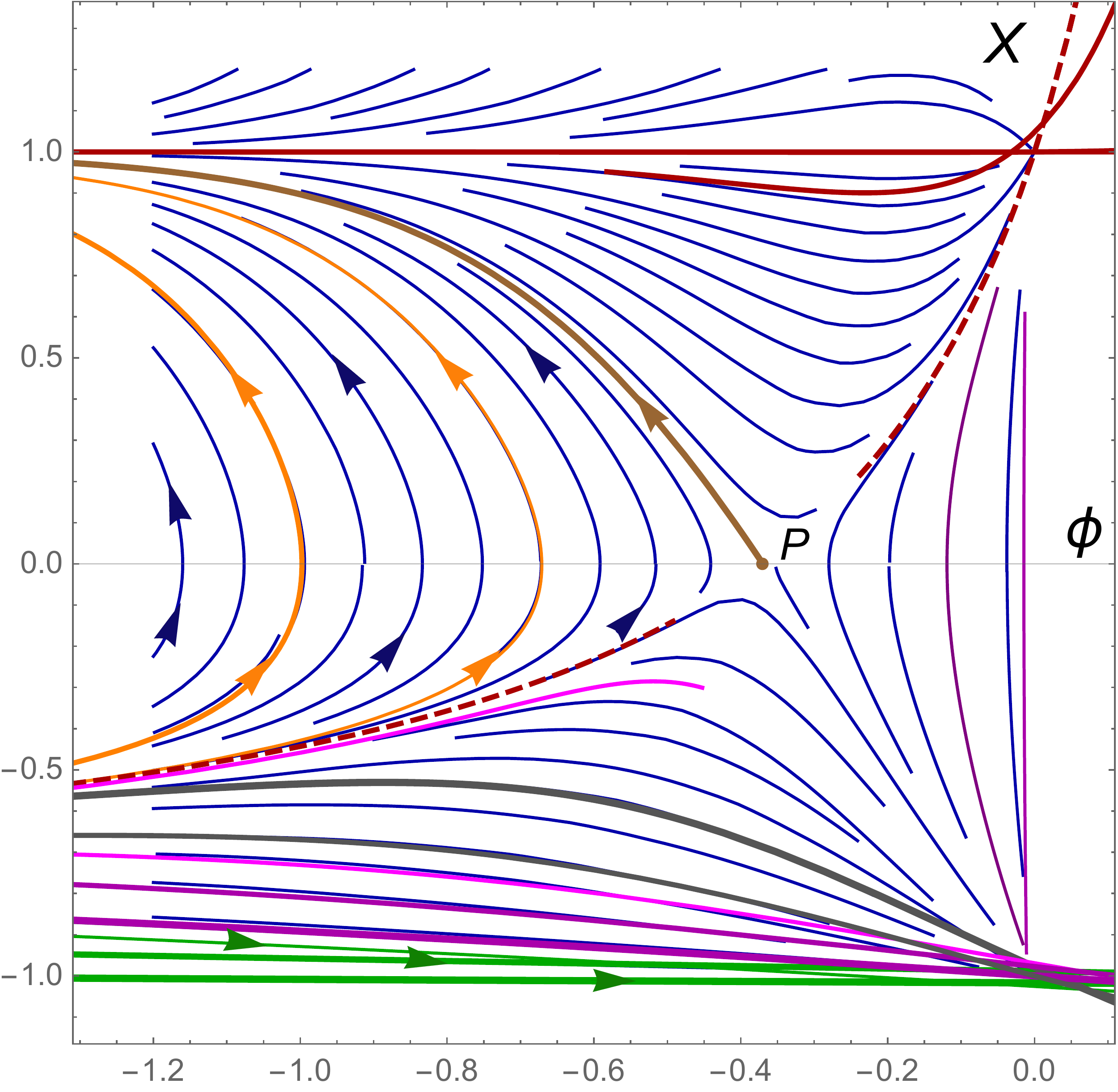}
 \caption{The behaviour of $X(\phi)$ near the point $P$ plotted using exact vacuum solutions and the stream of eq.~\eqref{dXphi0} (blue lines). The legends are the same as in Fig.~\ref{fig:Xcomparison}.}
 \label{fig:pointP}
 \end{figure}

Let's look at Figs.~\ref{fig:Xcomparison} {\bf A)} and {\bf B)} more closely. On these pictures we see a special point  $P$, which is a stationary point  of  our potential $V'(\phi) = 0$.
We zoom the scale near the point $P$  in Fig.~\ref{fig:pointP} to see the behaviour of the vacuum  $X(\phi)$ functions near $P$ in details. As in the previous figures  we show the stream of eq.\eqref{dXphi0} by blue curves and by the other color curves the functions $X(\phi)$ plotted using analytic solutions for $\mathcal{A}$ and $\phi$.
 
It worth to be noted that in Figs.~\ref{fig:Xcomparison} and~\ref{fig:pointP} the arrows always show the direction of  decreasing scale factor $\mathcal{A}$, see Fig.~\ref{fig:X'}, that corresponds to the flow from  UV fixed point to IR.

\subsubsection{The running coupling $\lambda=e^{\phi}$ on the energy scale}\label{Sec:RC0}

It is important to know the behaviour of the running coupling  $\lambda = e^{\phi}$ on the energy scale A$=\exp \mathcal{A}$.
One can trace this using the analytic solutions for $\phi$ and $\mathcal{A}$. For the vacuum case both of them depend on the constant of integration $E_{1}$, $E_{2}$, $u_{01}, u_{02}$ and on the shape of the potential, which is defined by $C_{1}$, $C_{2}$ and $k$. 

\begin{figure}[h!]
\centering
 \includegraphics[width=7cm]{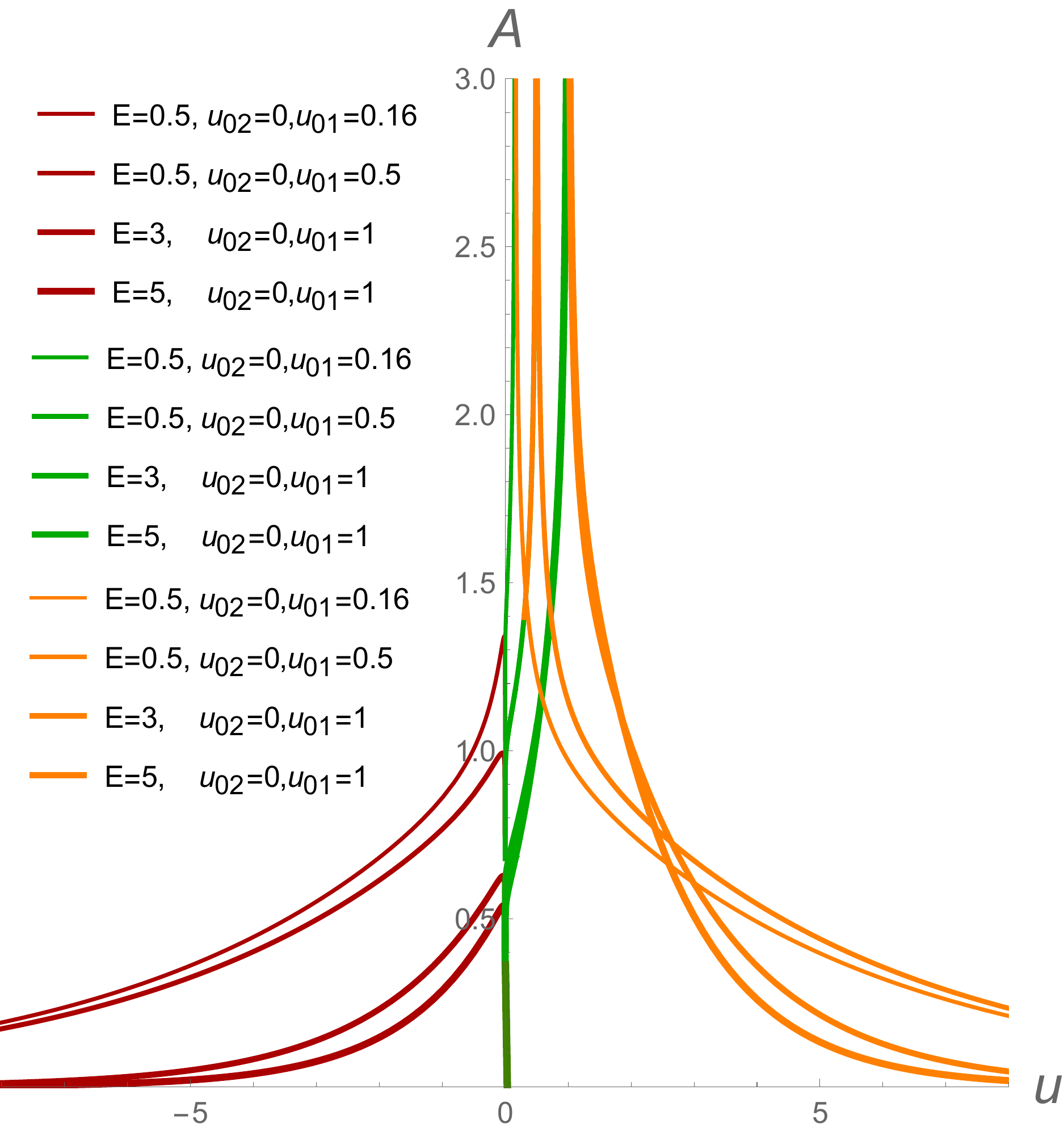}$\,\,\,\,\,\,\,\,\,\,\,\,\,\,\,\,\,\,\,\,$
     \includegraphics[width=5cm]{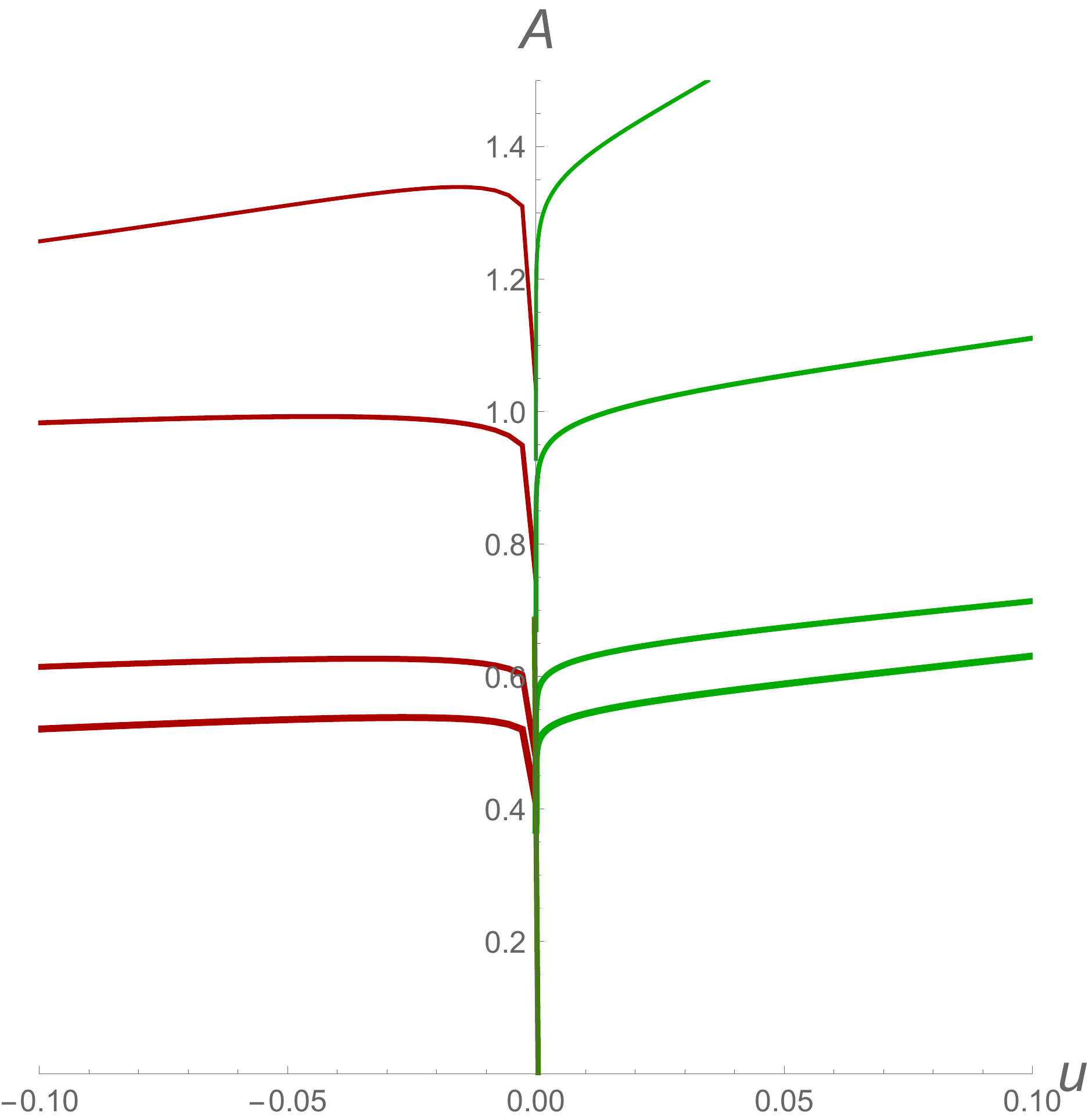}\\
    A$\,\,\,\,\,\,\,\,\,\,\,\,\,\,\,\,\,\,\,\,\,\,\,\,\,\,\,\,\,\,\,\,\,\,\,\,\,\,\,\,\,\,\,\,\,\,\,\,\,\,\,\,\,\,\,\,\,\,\,\,\,\,\,\,\,\,\,\,\,\,\,\,\,\,\,\,\,\,\,\,\,\,\,\,\,\,\,\,\,\,\,\,\,\,\,\,\,\,\,\,\,\,\,\,\,$B    
    \caption{ The behaviour of the energy scale A at all 3 branches (the darker red curves for the left solution, the green ones for the middle and the orange ones for the right solution) and its zoom near $u_{02}=0$. For  all plots  $k=1$, $C_{1} = -2$, $C_{2}=2$, different curves on the same plot correspond to the different values of $|E_{1}| =|E_{2}|$, labeled as $E$,  different $u_{01}$ and $u_{02}$.}
 \label{fig:Au}
 \end{figure}

Let us see how the energy scale A depends on on $u$.  In in Fig.~\ref{fig:Au}{\bf A)} we observe that A is non-monotonic function for the left solutions. First it increases, but near $u_{02}$ it decreases that can be clearly seen in the zoomed picture  Fig.~\ref{fig:Au} {\bf B)}. This non-monotonic behaviour is also read from Fig.~\ref{fig:X'} for the behaviour of the energy A on  the dilaton  for the left branch.

\begin{figure}[t!]
\centering{
      \includegraphics[width=4cm]{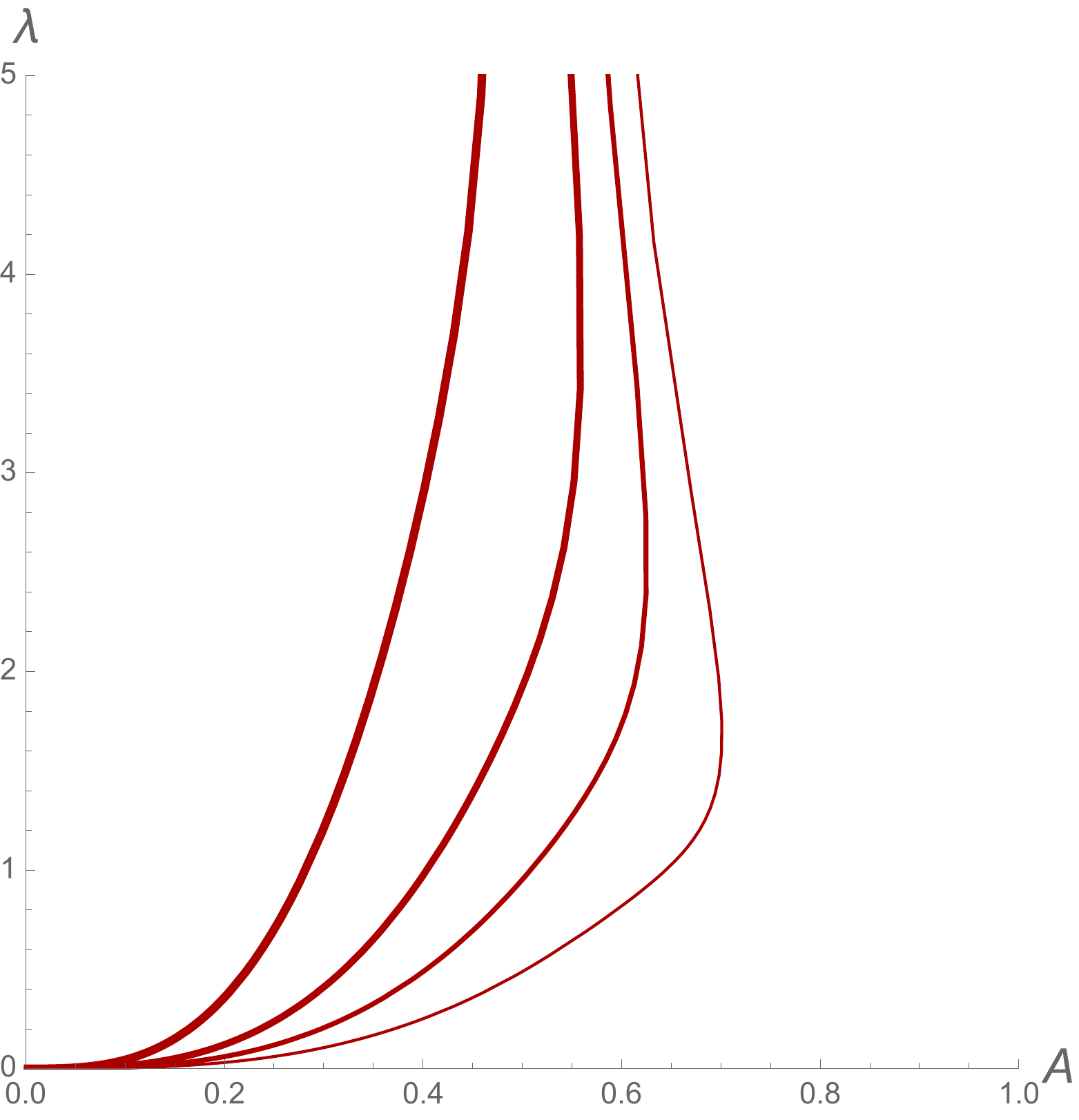}  \includegraphics[width=2cm]{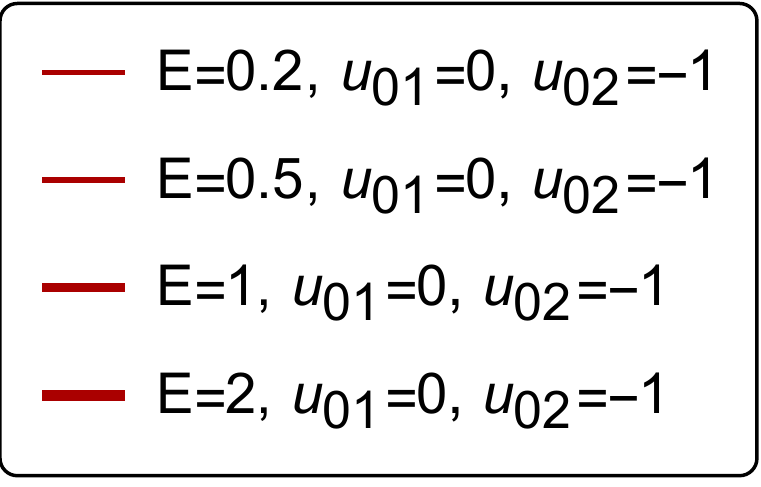}$\,\,\,\,\,$
    \includegraphics[width=4cm]{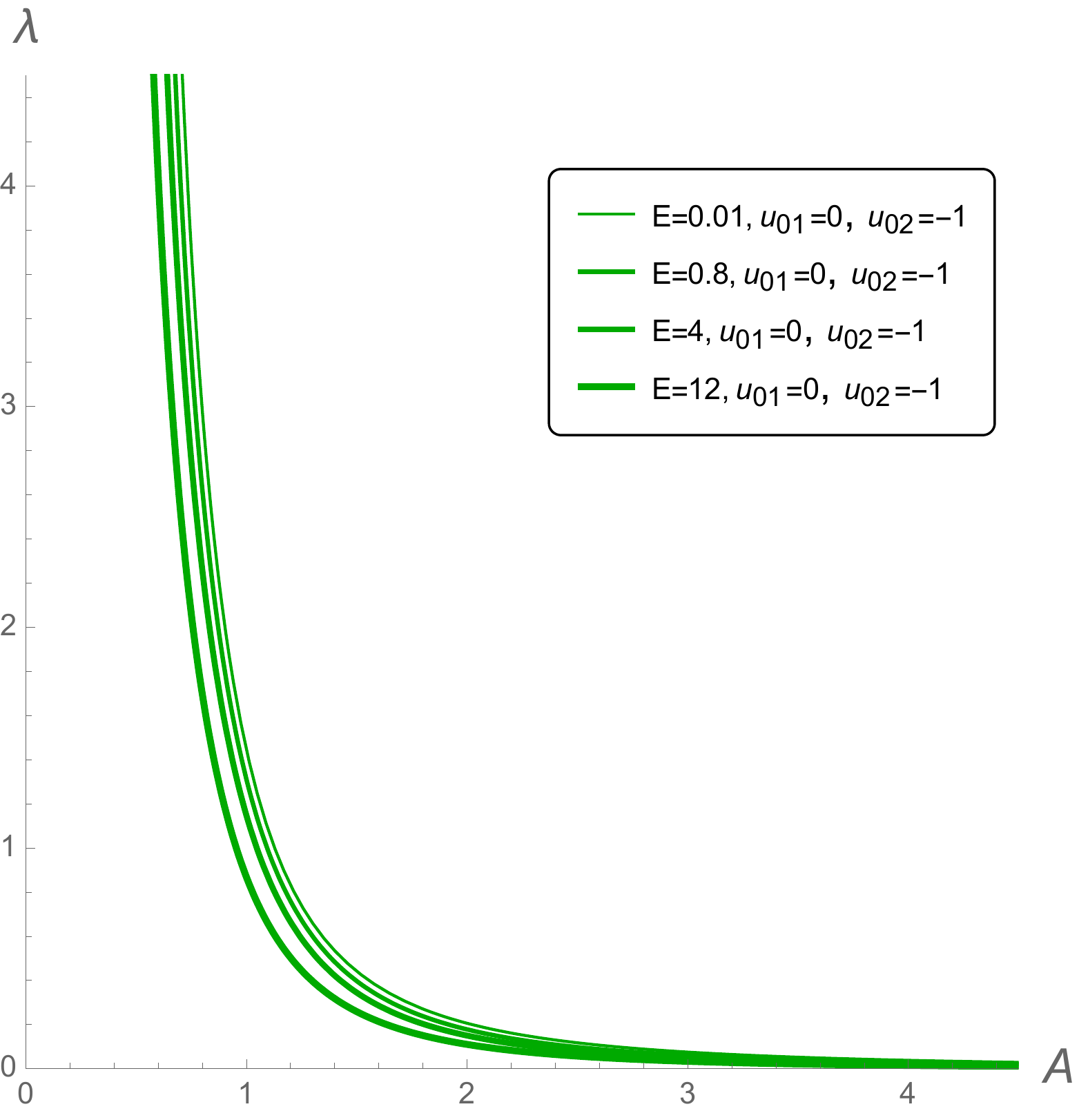}$\,\,\,\,\,$
        \includegraphics[width=4cm]{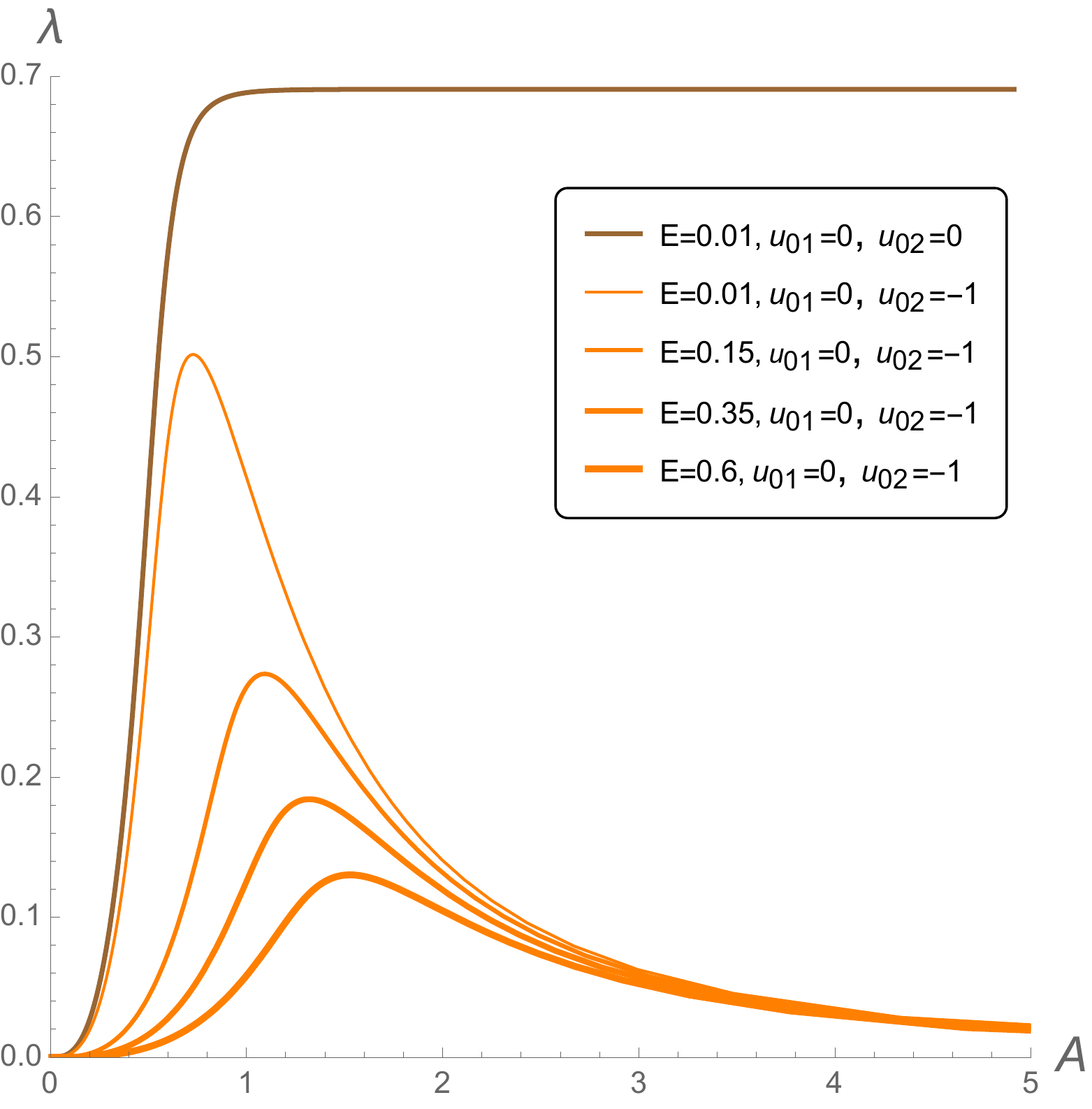}
     \\
    A$\,\,\,\,\,\,\,\,\,\,\,\,\,\,\,\,\,\,\,\,\,\,\,\,\,\,\,\,\,\,\,\,\,\,\,\,\,\,\,\,\,\,\,\,\,\,\,\,\,\,\,\,\,\,\,\,\,\,\,\,\,\,$B$\,\,\,\,\,\,\,\,\,\,\,\,\,\,\,\,\,\,\,\,\,\,\,\,\,\,\,\,\,\,\,\,\,\,\,\,\,\,\,\,\,\,\,\,\,\,\,\,\,\,\,\,\,\,\,\,\,\,\,\,\,\,\,\,\,\,\,\,\,\,\,\,\,\,\,\,\,\,\,\,$C}
    \caption{ The dependence of the coupling constant on the energy A on the dilaton plotted using the solutions for $\mathcal{A}$ and $\phi$: A) the left branch with $u_{02}>u$, B) the middle branch $u_{02}<u<u_{01}$; C) the right branch $u>u_{01}$. For  all plots  $k=1$, $C_{1} = -2$, $C_{2}=2$, different curves on the same plot corresponds to the different values of $|E_{1}| =|E_{2}|$, labeled as $E$ on the legends and different $u_{01}$ and $u_{01}$
    also indicated on the legends.}
 \label{fig:a-lambda}
 \end{figure}

  In Fig.~\ref{fig:a-lambda} the dependence of the coupling constant  $\lambda=\exp \phi$ on the scale A is presented for the zero-temperature solutions with (\ref{chartl})-(\ref{chartr})  and Fig.~\ref{fig:a-lambda} {\bf C)} with (\ref{chart2}).

 We recall that on the left and middle solutions the dilaton interpolates between $\pm \infty$ and for the right solution it starts at $-\infty$ goes to the maximal value 
 $\phi_{max}$ and then goes back to  $-\infty$ (we have observed this behaviour already in the previous sections). In spite of that the dilaton has the similar behaviour on the left and middle solutions, 
 the scale factor has rather different behaviour on these solutions, namely, on the middle solution the scale factor monotonically decreases from large positive values to zero as the dilaton runs from $+\infty$ to $-\infty$. 
The right solution is a bouncing solution with the decreasing scale factor. These behaviours of the scale factors and the dilaton are reflected in the plots for the dependence of the running coupling on the energy scale.

We can summarize the results for the running coupling in the following form

 \begin{itemize}
 \item 
 as expected for solutions with $u<u_{02}$, see Fig.~\ref{fig:a-lambda}{ \bf A)}, where the dilaton tends to $-\infty$ in the IR region and we have IR-free theory, while in the UV region the effective coupling $\lambda\to+\infty$. 
 \item In Fig.~\ref{fig:a-lambda}{ \bf B)} we see that the dependence of $\lambda$ on A plotted on the middle solutions with $u_{02}<u<u_{01}$ mimics the QCD behavior. 
 \item The running coupling plotted as a function of the energy scale at Fig.~\ref{fig:a-lambda}{ \bf C)} for the solutions with (\ref{chartr}) shows that it is  UV free as well as  IR free theory, i.e. the running coupling has the form of the hill.
 \end{itemize}

\section{Non-vacuum solutions}\label{Sect:4NV}
\subsection{The metric and the dilaton}

The metric and the dilaton solutions to the model (\ref{1.1}) in the non-vacuum case are
\bea\label{nvm1}
ds^{2}&=& F^{\frac{8}{9k^{2} -16}}_{1}F^{\frac{9k^{2}}{2(16-9k^{2})}}_{2} \left(- e^{2\alpha^{1}u}dt^{2} + e^{-\frac{2\alpha^{1}}{3}u}d\vec{y}^{~2} \right)+
F^{\frac{32}{9k^{2}-16}}_{1}F^{\frac{18k^{2}}{16-9k^{2}}}_{2}du^{2},\label{metricF1F2alpha} \\\label{dilnvE}
\phi &= &-\frac{9k}{9k^{2}-16}\log{F_{1}} +\frac{9k}{9k^{2}-16}\log{F_{2}},
\eea
where the functions $F_{1}$ and $F_{2}$ are given by  (\ref{nC1pC2.F1E1n})-(\ref{nC1pC2.F2E2p}) as before.

Just as in the vacuum case, we need to separate the solutions in the branches (\ref{chartl})-(\ref{chart2}) with respect to values of $u_{01}$ and $u_{02}$.  
We note that the factors of $\alpha^{1}$ in the metric (\ref{nvm1}) break the Poincar\'e symmetry. 
However, below we will see that the presence of this parameter allows us to define a horizon and to construct black branes.
We also recall that $E_{1}$ and $E_{2}$ must obey the constraint
\bea\label{cond-nv-1}
E_{1} + E_{2} + \frac{2}{3}(\alpha^{1})^{2} = 0.
\eea

The condition (\ref{cond-nv-1}) allows to tune parameters thus we have two additional regimes
\bea\label{mu1mu2e}
1)\mu_{1}  = \mu_{2},\quad 2)\mu_{1}>\mu_{2},
\eea
with
\bea\label{costrE2}
1)E_{2}  = \frac{6k^{2} (\alpha^{1})^{2}}{16- 9k^{2}}, \quad 2)E_{2}< \frac{6k^{2} (\alpha^{1})^{2}}{16- 9k^{2}},
\eea
respectively. 

This leads to new dynamics of the dilaton  (\ref{dilnvE}). Particularly,  the dilaton can be constant, that is inapplicable for the vacuum case where it always holds $\mu_{2} = \frac{4}{3k}\mu_{1}$.

\begin{figure}[h!]
\centering
 \includegraphics[width=4.3cm]{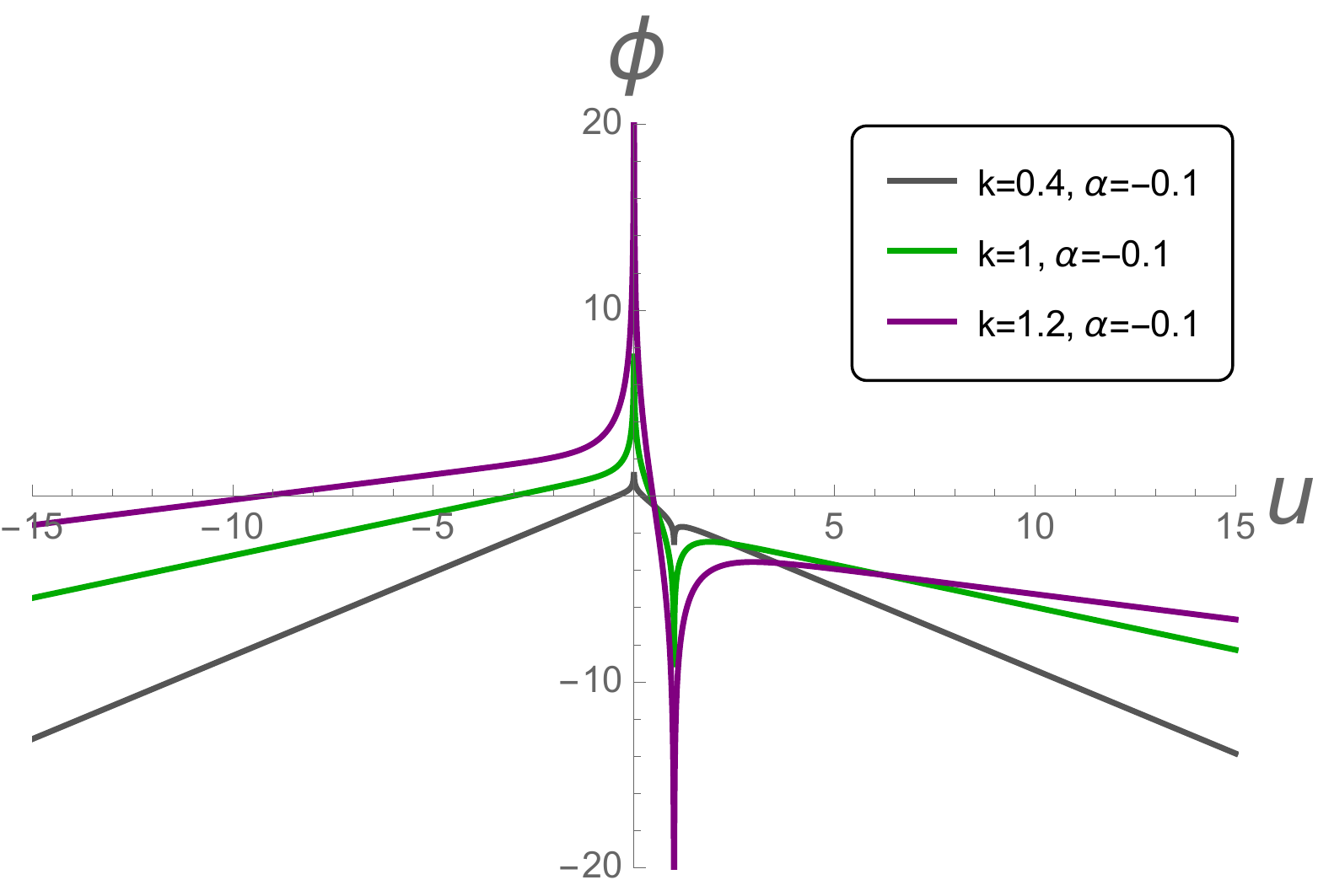}$\,\,$
   \includegraphics[width=4.3cm]{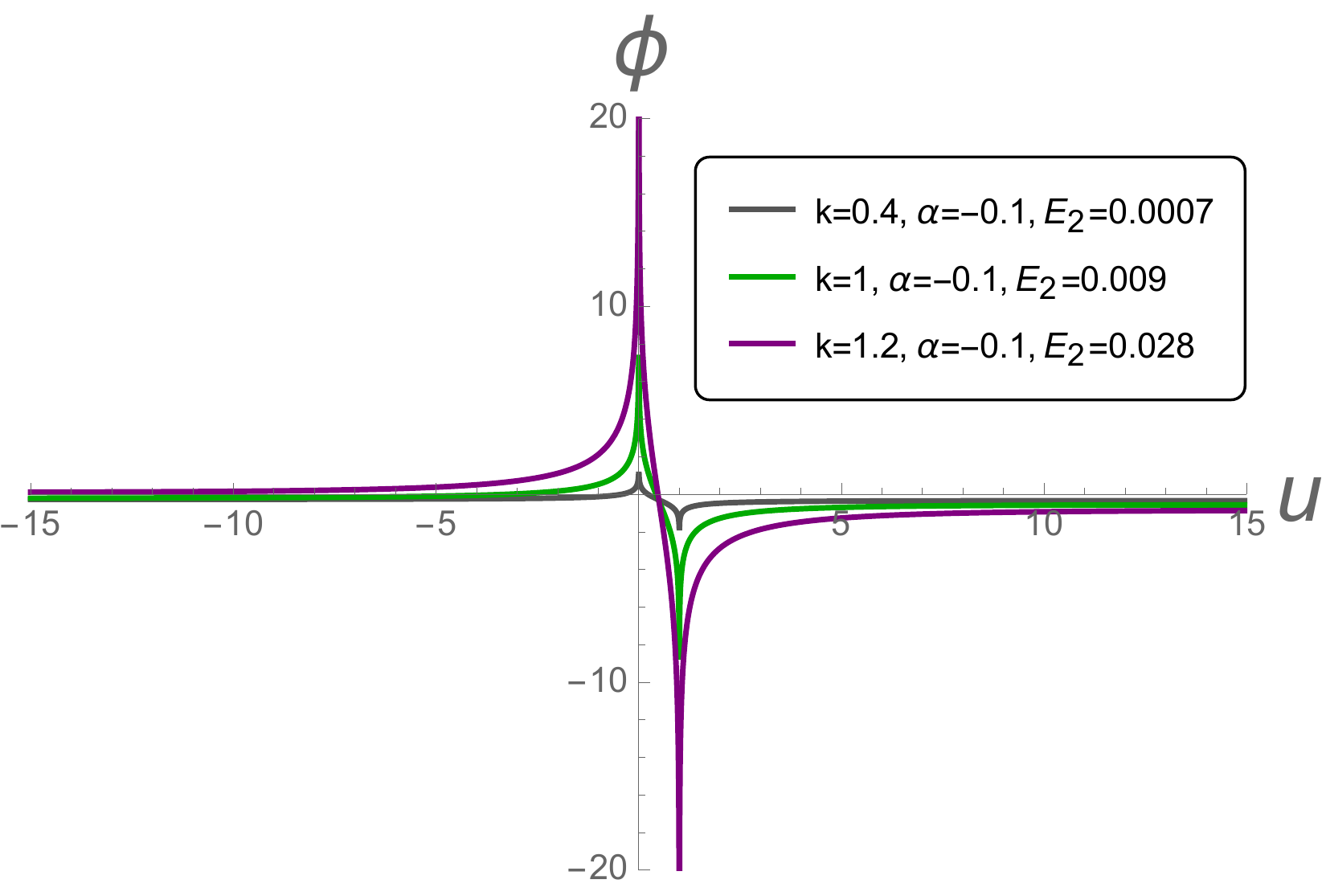}$\,\,$
 \includegraphics[width=4.3cm]{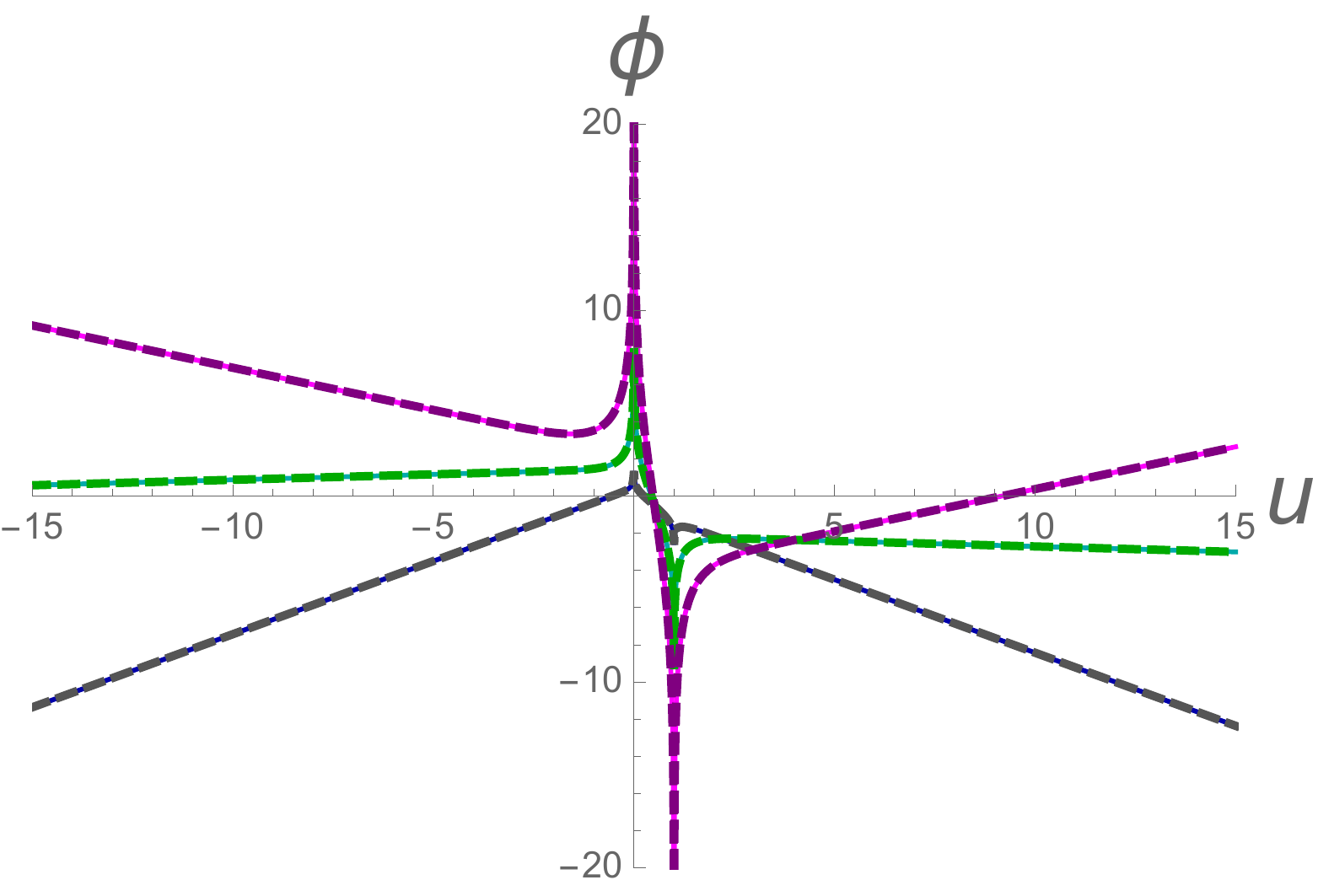}$\,$
  \includegraphics[width=1.5cm]{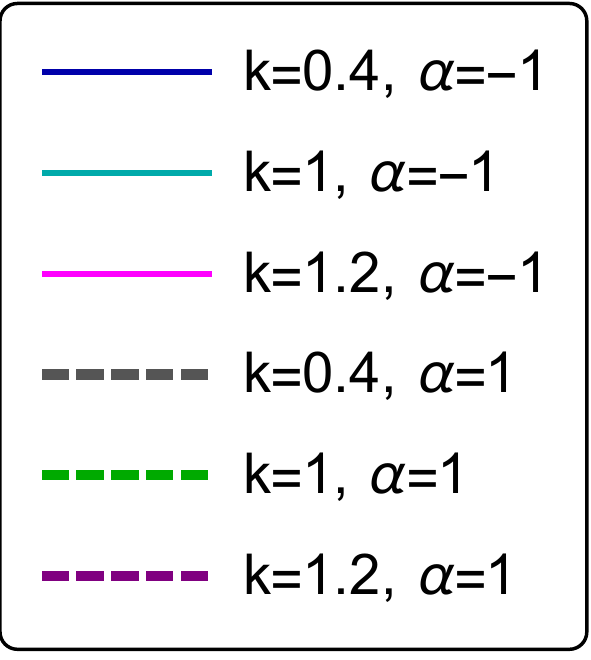}\\
  A$\,\,\,\,\,\,\,\,\,\,\,\,\,\,\,\,\,\,\,\,\,\,\,\,\,\,\,\,\,\,\,\,\,\,\,\,\,\,\,\,\,\,\,\,\,\,\,\,\,\,\,\,\,\,$B$\,\,\,\,\,\,\,\,\,\,\,\,\,\,\,\,\,\,\,\,\,\,\,\,\,\,\,\,\,\,\,\,\,\,\,\,\,\,\,\,\,\,\,\,\,\,\,\,\,\,\,\,\,\,$C
   \caption{The behaviour of the dilaton solution for the 3-branch solutions with $u_{01}=1, u_{02}=0$,  the non-vacuum case: A) 
  $\alpha^{1} = -0.1$, $E_{2} = 1$,  $k =0.4,1,1.2$ ; B)  $E_{2} = 0.0007$ for $k=0.4$, $E_{2} = 0.009$ for $k=1$, $E_{2} = 0.028$ for $k=1.2$; $C_{1}=-1$, $C_{2}=1$, $\alpha^{1} = -0.1$  for all; C) solid lines -- $\alpha^{1} = -1$, dashed lines -- $\alpha^{1} = 1$, $k=0.4,1,1.2$,  $C_{1}=-1$, $C_{2}=1$, $E_{2} = 1$ for all. }
 \label{fig:DIL-NV}
\end{figure}

We illustrate the behaviour of the dilaton solution for the non-vacuum case in Fig.~\ref{fig:DIL-NV} for  branches (\ref{chartl})-(\ref{chartr}) with $u_{01}\neq u_{02}$. As it was expected  the condition (\ref{cond-nv-1}) changes the behaviour of the dilaton. 
In Fig.~\ref{fig:DIL-NV} {\bf B)} we plot the dilaton keeping the same shape of the potential  and the value of $\alpha^{1}$ as for Fig.~\ref{fig:DIL-NV} {\bf A)}, but the values of the parameter $E_{2}$ are changed with respect to (\ref{costrE2}).  In Fig.~\ref{fig:DIL-NV} {\bf C)} we again save the form of the potential, use the same value of $E_{2}$ as in {\bf A)}, take bigger $\alpha^{1}$ and show that the opposite sign of this parameter doesn't change the asymptotics of the dilaton. From Fig.~\ref{fig:DIL-NV} {\bf C)}
one observes that for $u \to \pm\infty$  $\phi$ can tend to $+\infty$ , while for the vacuum case $\phi \to -\infty$ as $u \to \pm\infty$.
% that confirms our discussion of the analytical results for the dilaton asymptotics.

\begin{figure}[h!]
\centering
 \includegraphics[width=4cm]{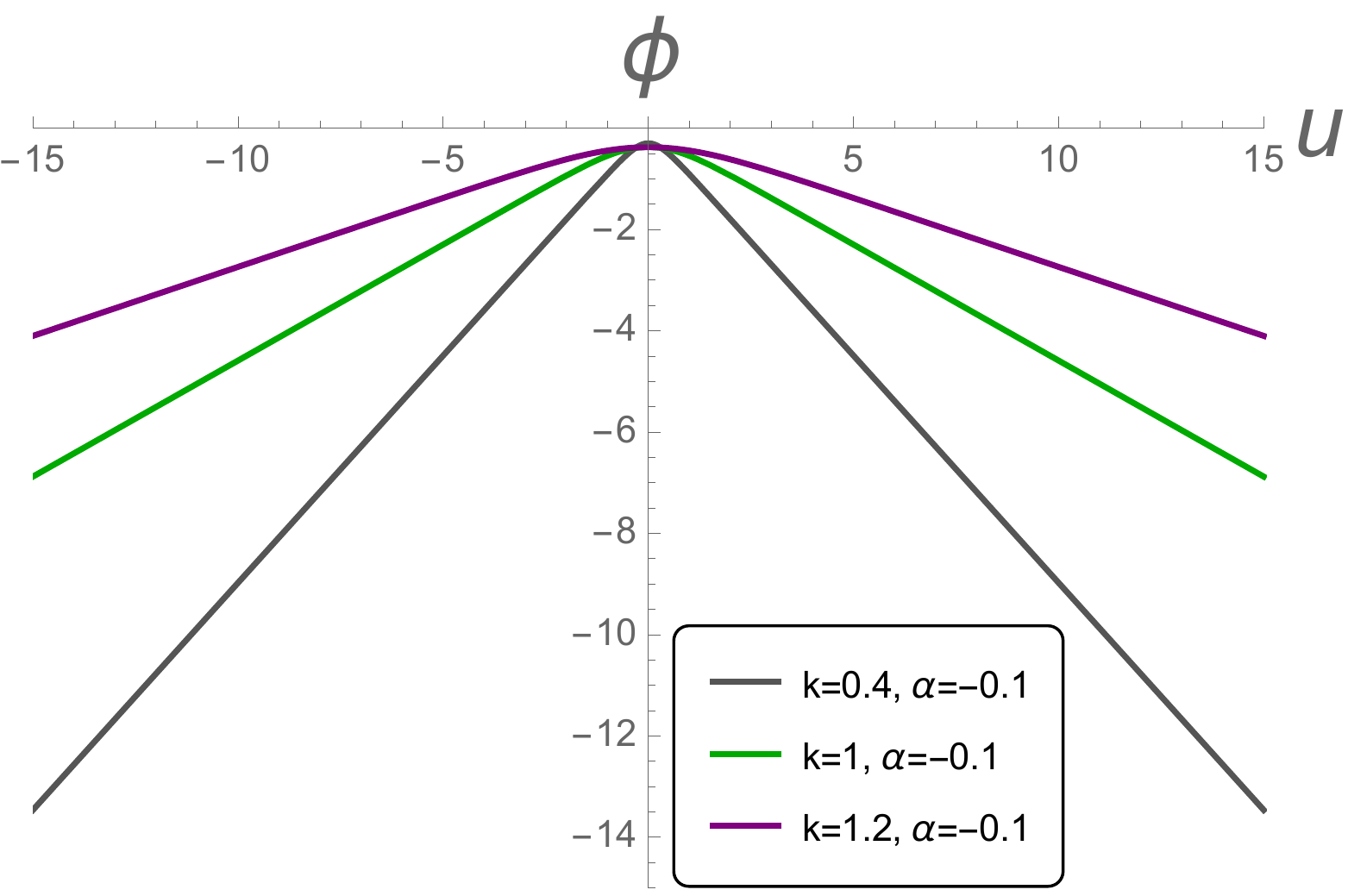}$\,\,\,\,\,$
  \includegraphics[width=4cm]{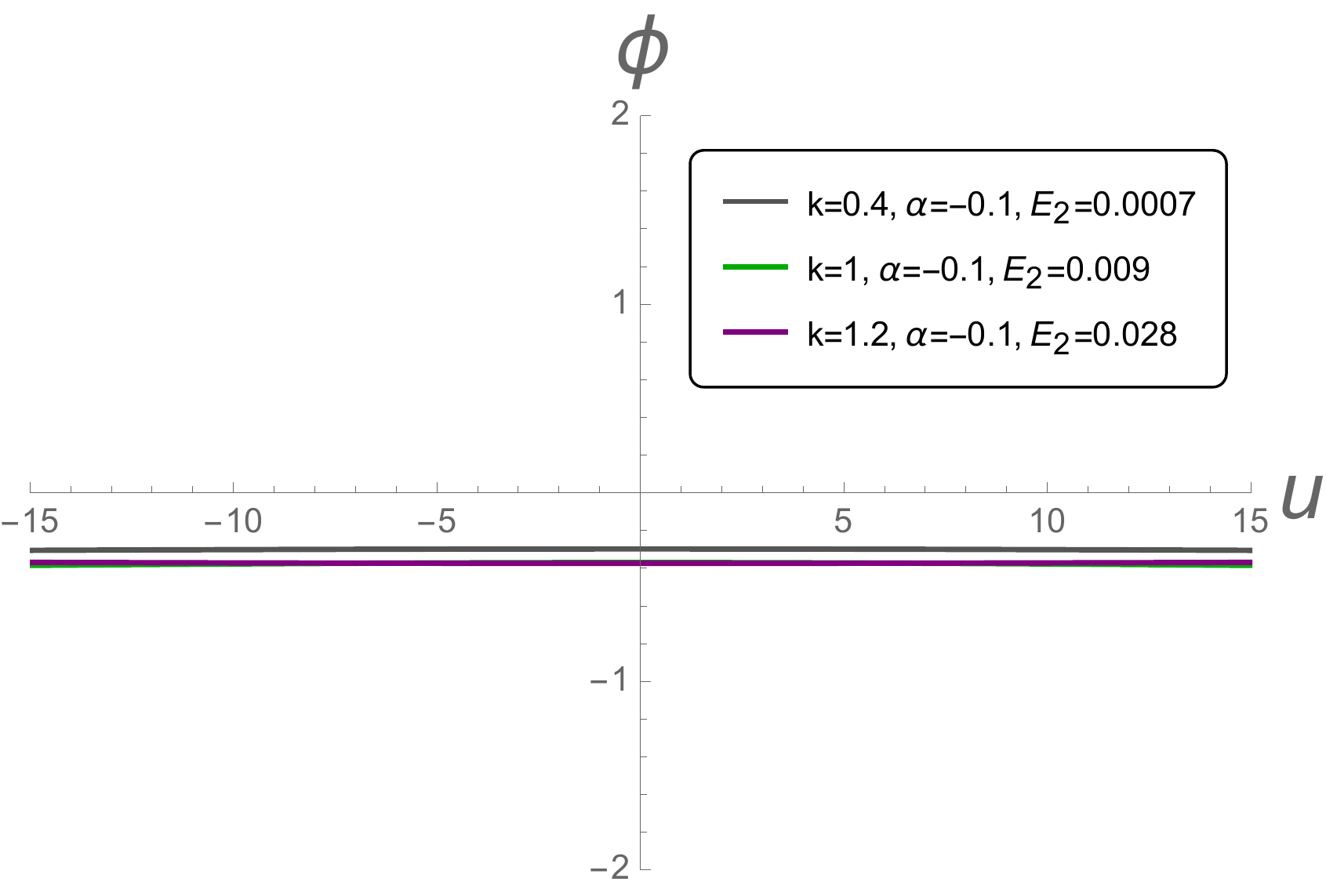}$\,\,\,\,\,$
   \includegraphics[width= 4cm]{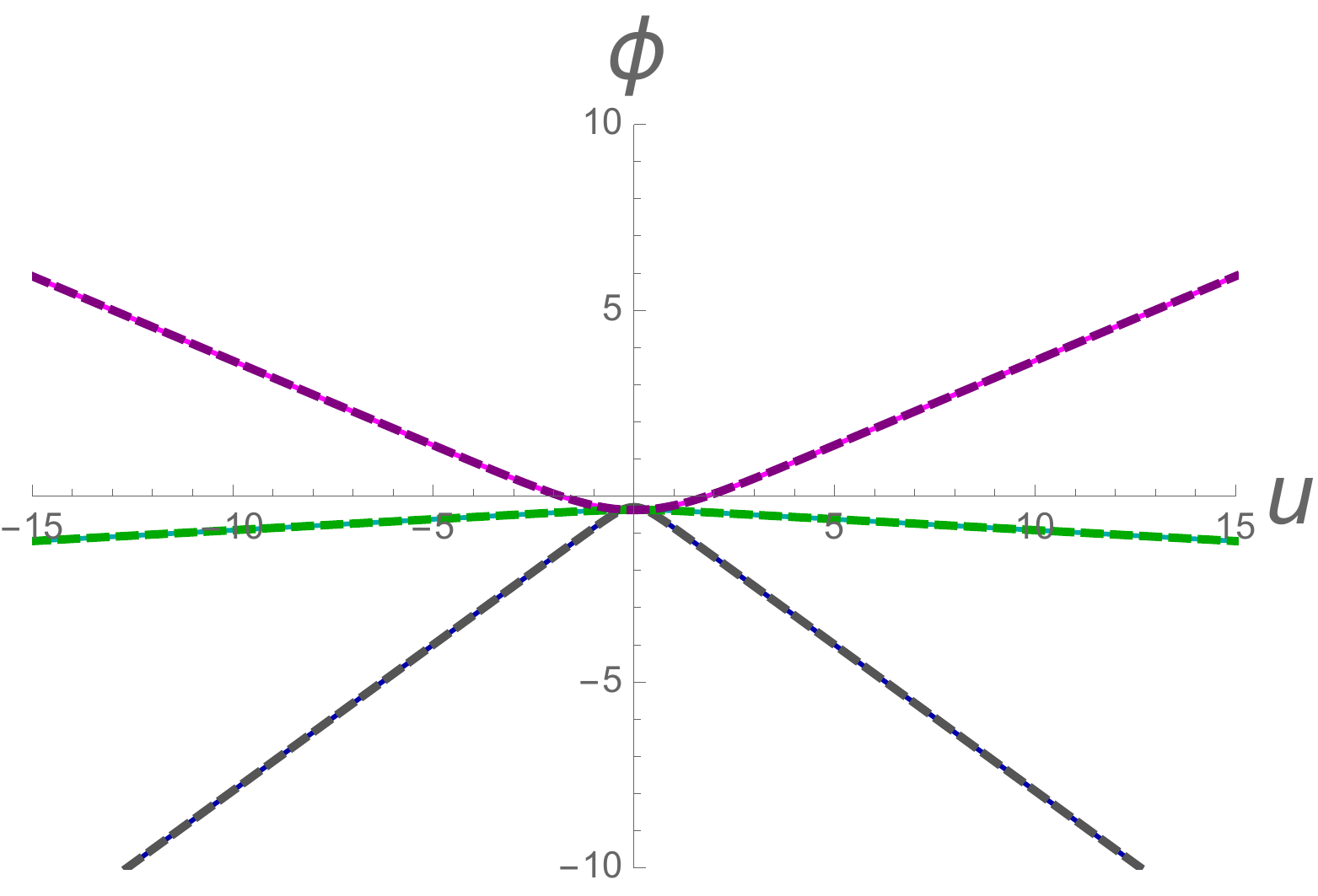}$\,\,\,$ 
  \includegraphics[width=1.5cm]{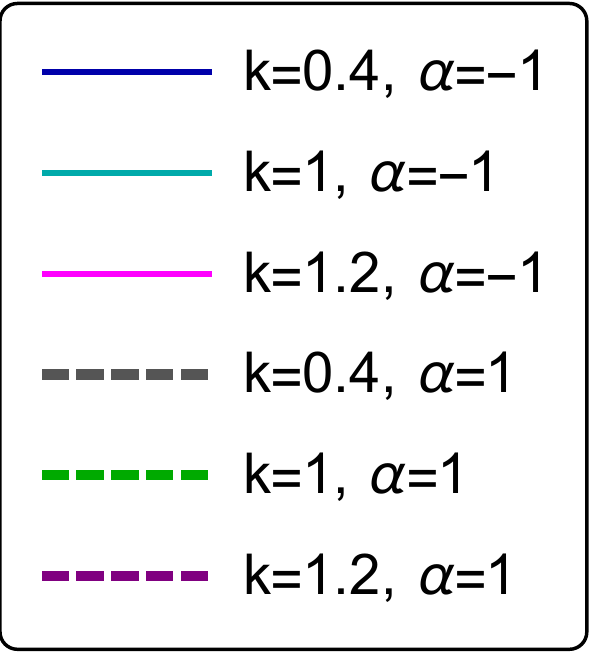}\\
  A$\,\,\,\,\,\,\,\,\,\,\,\,\,\,\,\,\,\,\,\,\,\,\,\,\,\,\,\,\,\,\,\,\,\,\,\,\,\,\,\,\,\,\,\,\,\,\,\,\,\,\,\,\,\,$B$\,\,\,\,\,\,\,\,\,\,\,\,\,\,\,\,\,\,\,\,\,\,\,\,\,\,\,\,\,\,\,\,\,\,\,\,\,\,\,\,\,\,\,\,\,\,\,\,\,\,\,\,\,\,\,\,\,\,\,\,\,\,\,\,\,\,\,\,\,\,\,\,$C
   \caption{The behaviour of the dilaton solution for the 2-branch solutions,  the non-vacuum case: A) 
  $\alpha^{1} = -0.1$, $E_{2} = 1$,  $k =0.4,1,1.2$ , B)  $E_{2} = 0.0007$ for $k=0.4$, $E_{2} = 0.009$ for $k=1$, $E_{2} = 0.028$ for $k=1.2$; $C_{1}=-1$, $C_{2}=1$, $\alpha^{1} = -0.1$ for all; C) solid lines -- $\alpha^{1} = -1$, dashed lines -- $\alpha^{1} = 1$, $k=0.4,1,1.2$,  $C_{1}=-1$, $C_{2}=1$, $E_{2} = 1$ for all.  }
 \label{fig:DIL0-NV}
\end{figure}

In Fig.~\ref{fig:DIL0-NV} we plot the dependences of the dilaton solution with $u_{01} = u_{02}=u_0$ on $u$ (\ref{chart1}). 
As for the 3-branch solutions from Fig.~\ref{fig:DIL0-NV} {\bf B)} we see that the dilaton can be constant with the appropriate choice of parameters agreed with (\ref{cond-nv-1})
and can change its asymptotics from $-\infty$ to $+\infty$ at $u \to \pm \infty$, see Fig.~\ref{fig:DIL0-NV} {\bf C)}.

\begin{figure}[h!]
\centering
 \includegraphics[width=9cm]{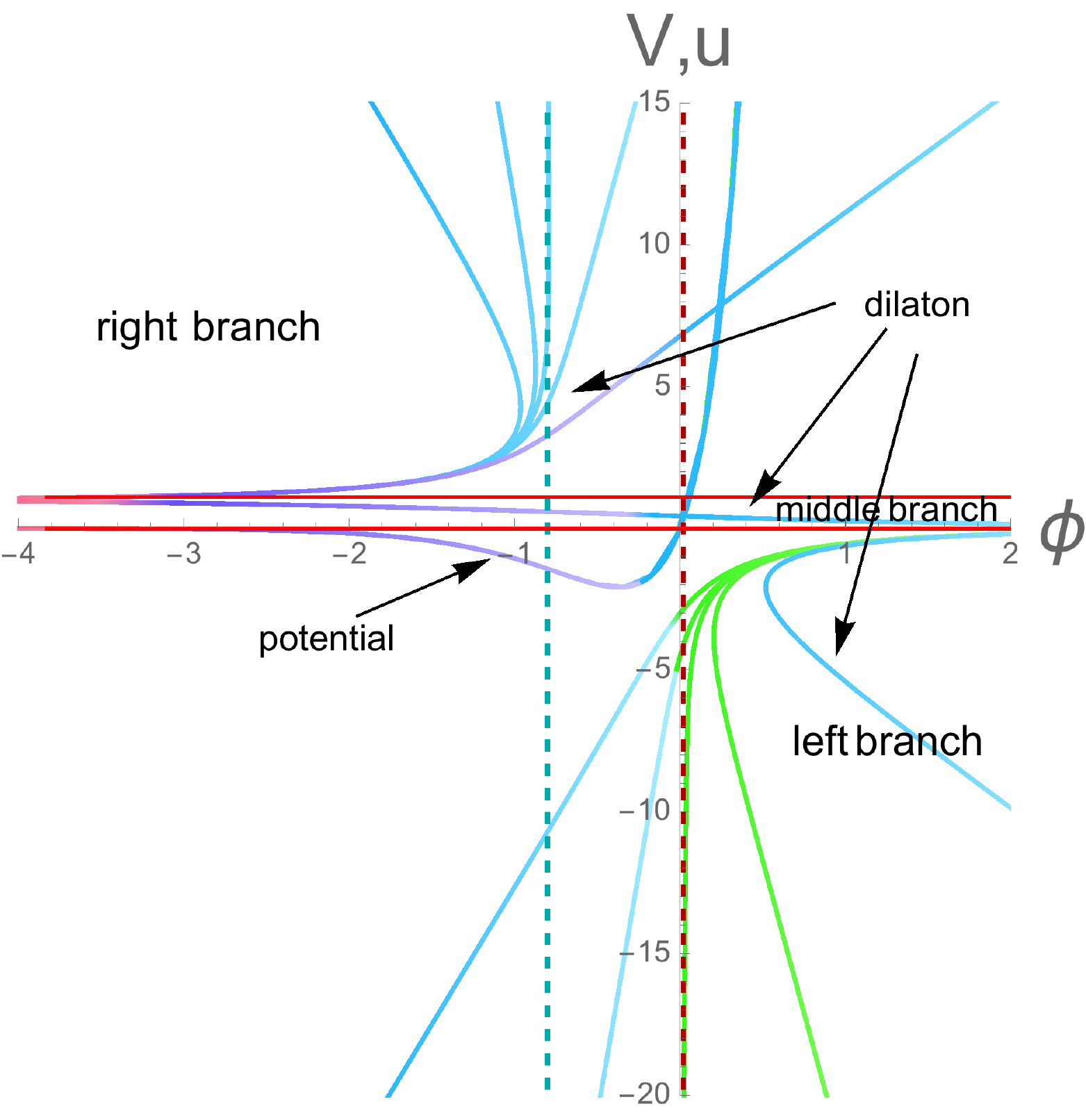}
 \caption{ The dilaton potential  $V=V(\phi)$ on non-vacuum solutions  $\phi=\phi(u)$ with (\ref{chartl})-(\ref{chartr}) and  plots that indicate which values of $u$ 
 correspond to given $\phi$, i.e. $u=u(\phi)$.}
 \label{fig:DIL-POT2}
\end{figure}

In Fig.~\ref{fig:DIL-POT2} we  draw the potential  $V$  as a function of $\phi$ and the dependence $u$ on $\phi$. 
The functions $u(\phi)$ are different on the different branches.  The different values of $u$ corresponding to the same $\phi$ are indicated by points
 at the vertical lines. We see that  the function $u(\phi)$ for $|\alpha^1|<\alpha^{1}_{cr}$ is double-valued on the right branch, and for $|\alpha^1|>
 \alpha^{1}_{cr}$ it is double-valued on the left branch, and for $ |\alpha^1|= \alpha^{1}_{cr}$ the both functions one-to-one functions for
 all branches, and $\phi<\phi_0'$ for the  right branch and  $\phi >\phi_0''$ at the left branch.

The dilaton potential which is plotted in Fig.~\ref{fig:DIL-POT2} on the right solutions for the dilaton is bounded above like for the vacuum right solutions.

 \subsection{The black brane solutions}
Now  we are going to find a black brane representation of the solution (\ref{nvm1})-(\ref{dilnvE}) defined for (\ref{chartr}) with $u_{01}\neq u_{02}$.
The metric \eqref{metricF1F2alpha} can be rewritten in the following form
\bea\label{asm}
ds^{2} =  \mathcal{C}\,\mathcal{X}(u)e^{\kappa u - \frac{2}{3}\alpha^{1}u}\left(  - e^{\frac{8}{3}\alpha^{1}u}dt^{2} + d\vec{y}^{2} +\mathcal{X}(u)^3\mathcal{C}^3e^{(3\kappa+ \frac{2}{3}\alpha^{1})u}du^{2}\right),
\eea
with the constant $\mathcal{C}$ given by
\bea
\mathcal{C}&\equiv&\Big(\frac12\sqrt{\left|\frac{C_{1}}{2E_{1}}\right|}e^{-\mu_1 u_{01}}\Big)^{\frac{8}{9k^{2} - 16}}
\Big(\frac12\sqrt{\left|\frac{C_{2}}{2E_{2}}\right|}e^{-\mu_2 u_{02}}\Big)^{\frac{9k^{2}}{2(16- 9k^{2})}},\label{Cc}
\eea
the function $\mathcal{X}(u)$ written as
\bea
\mathcal{X}(u)&=&(1-e^{-2\mu_1(u-u_{01})})^{-\frac{8}{16-9k^{2} }}(1-e^{-2\mu_2 (u-u_{02})})^{\frac{9k^{2}}{2(16- 9k^{2})}}
%\\&\approx&1+\frac{8}{16-9k^{2} }e^{-2\mu_1(u-u_{01})}-\frac{9k^{2}}{2(16- 9k^{2})}e^{-2\mu_2 (u-u_{02})}+...\label{Xc}
\eea
and the exponent $\kappa$ given by

\bea\label{expkappa}
\kappa&\equiv&\frac{8}{\sqrt{6\left(16-9k^{2}\right)}}\Big(-\sqrt{E_{2}+\frac23 (\alpha^1)^2} +\frac34 k\sqrt{E_{2}} 
\Big),
 \eea
 where we took into account the relations for $\mu_{1}$ and $\mu_{2}$  (\ref{nC1pC2.F1E1n})-(\ref{nC1pC2.F2E2p}). We see that for $0<k<4/3$ one has 
$\kappa <0$.

%To summarize,  at $u \to \infty$ we get
%\bea
%e^{2A}&=&
%\mathcal{C}\,e^{\kappa u}\,\mathcal{X}(u),\,\,\,\,\,\,\kappa<0,\,\,\,\,\,\mathcal{X}(u) {\to} 1\label{2Amm}\eea
%and
 To have the black brane solutions we  need to remove a conical singularity in the metric \eqref{asm} (since with $\alpha^1<0$ we get zero in front of $dt^2$).
Taking 
\be
e^{\frac{4}{3}\alpha^{1}u}=\rho,\,\,\,\,\,\, t=i\tau,
\ee
so the metric reads
\bea\label{conic}
ds^{2}
& \underset{u\to \infty}{\sim} &  \frac{\mathcal{C}^4}{(\frac{4}{3}\alpha^{1})^2} \mathcal{X}e^{\kappa u - \frac{2}{3}\alpha^{1}u}\left(\frac{1}{\mathcal{C}^3} (\frac{4}{3}\alpha^{1})^2 \rho^2d\tau^{2} +\frac{1}{\mathcal{C}^3} (\frac{4}{3}\alpha^{1})^2d\vec{y}^{2} + \mathcal{X}^{3}e^{3(\kappa - \frac23\alpha^{1})u}d\rho^{2}\right),\label{1.3"'}\nonumber\\
\eea
where $\mathcal{X}(u)\to 1$ for $\rho \to 0$ as $u\to \infty$.
%\bea
%\mathcal{X}(u) \approx 1+\frac{8}{16-9k^{2} }e^{-2\mu_1(u-u_{01})}-\frac{9k^{2}}{2(16- 9k^{2})}e^{-2\mu_2 (u-u_{02})}+...\label{Xc}
%\eea

Therefore, there is no conic singularity if the following constraint is satisfied
\bea
&&\kappa - \frac23\alpha^{1}=0\label{NCS}.
\eea
We also fix the periodicity 
\bea
\frac{4}{3\mathcal{C}^{3/2}}\alpha^{1}\beta=2\pi.\eea

Plugging (\ref{expkappa}) in (\ref{NCS}) we come to the condition to the parameters
\bea
\label{E2bb}
E_{1} = - \frac{32(\alpha^{1})^{2}}{3(16- 9k^{2})}, \quad E_{2}  = \frac{6k^{2} (\alpha^{1})^{2}}{16- 9k^{2}},
\eea
that corresponds to $\mu_{1}  = \mu_{2}  = \mu$ with
\bea\label{mualpha1}
\mu = - \frac{4}{3}\alpha^{1}.
\eea

Therefore, we get the black brane, if \eqref{NCS} is satisfied, and the temperature is
\be
\frac{1}{\beta}=T=\frac{2}{3\pi}\frac{|\alpha^{1}|}{\mathcal{C}^{3/2}}.\label{T}\ee
%In {\bf exponentBH.nb} we can check that there are solutions of eq.\eqref{NCS}.
Here $\mathcal{C}$ is taken with the constraint  \eqref{mualpha1}, see \eqref{cbbxbb} below.

Under the condition \eqref{NCS} the black brane metric has the form
   \bea\label{f}
ds^{2}
& = & \mathcal{C}\,\mathcal{X}\Big( - e^{-2\mu u}dt^{2} + d\vec{y}^{2} \Big)+ 
 \mathcal{C}^4\,\mathcal{X}^4
e^{-2\mu u}du^{2},\label{1.3mm}
\eea
where $\mathcal{C}$ and $\mathcal{X}$ are given by
\bea\label{Xbbxbb}
\mathcal{X} &=&(1-e^{-2\mu u})^{-\frac{8}{16-9k^{2} }}(1-e^{-2\mu (u-u_{02})})^{\frac{9k^{2}}{2(16- 9k^{2})}},\\\label{cbbxbb}
\mathcal{C}&\equiv& 2^{\frac{16}{(16-9k^{2})}} (3 \mu)^{\frac{1}{2}}\left|C_{1}\right|^{\frac{8}{2(9k^{2} - 16)}}
\Big(\frac{C_{2}}{k}e^{-2\mu u_{02}}\Big)^{\frac{9k^{2}}{4(16- 9k^{2})}} (16-9k^{2})^{-\frac{1}{4}}
\eea
with the horizon located  at $u=+\infty$ and the near-horizon expansion of $\mathcal{X}(u)$ is
\bea
\mathcal{X} \approx 1 + e^{-2\mu u} \left(\frac{16 -9k^{2}e^{2\mu u_{02}} }{2(16 - 9k^{2})}\right).
\eea
We note that the boundary is at $u_{01}$ and we fixed $u_{01}=0$ to have $f = 1$ at this boundary.
One can check that null geodesics we have the correct behaviour.
Null geodesics imply
 \bea
 ds^{2} = 0,
 \eea
 i.e. for the light moving in the radial direction
 \bea
\frac{dt}{du}  = \pm  e^{3A+\frac{3}{4}\mu},
 \eea
 or
 \bea
 t -t_{0}&\sim&
 \int^{u}_{u_{0}} d\bar{u}e^{(\frac32\kappa+ \frac{3\mu}{4})\bar{u}}\mathcal{C}^{3/2}\,\Big(1+ \frac{3\left(16 - 9k^{2}e^{2\mu u_{02}}\right)}{4(16- 9k^{2})}e^{-2\mu \bar{u}}
 \Big)\nn\\&=&
  \int^{u}_{u_{0}} d\bar{u}\mathcal{C}^{3/2}\,\Big(1+...\Big)\underset{u\to \infty}{\to} \infty.
 \eea 
  This calculation confirms that we have the horizon at $u=+\infty$ .

The scalar curvature and the Kretschmann scalar near horizon $u\to +\infty$ are
\bea\label{R.uinf}
R& =& \left(\frac{C_{1}}{2E_{1}}\right)^{\frac{16}{16 - 9k^{2}}}\left(\frac{C_{2}}{2E_{2}}\right)^{-\frac{9k^{2}}{16- 9k^{2}}}\left(\frac{3(16\mu_{1} - 9k^{2}\mu_{2})^{2}}{4(16-9k^{2})^{2}}- \frac{4}{3}\left(\alpha^{1}\right)^{2}\right)e^{\frac{2(16\mu_{1} - 9k^{2}\mu_{2})}{16- 9k^{2}}u},\\ \label{Krconfinf}
K &=&\frac{\left(4 \alpha^{1} (9 k^2-16)+27 k^2 \mu_{2}-48 \mu_{1}\right)^2 }{864 (16-9 k^2)^4}\left(\frac{C_{1}}{2E_{1}}\right)^{\frac{32}{16-9 k^2}}\left(\frac{C_{2}}{2E_{2}}\right)^{\frac{18 k^2}{9 k^2-16}}e^{\frac{4(16\mu_{1} - 9k^{2}\mu_{2})}{16 - 9k^{2}}u}\nonumber\\
&\cdot&\left(304 (\alpha^{1})^2 (16-9 k^2)^2+168 \alpha^{1} (9k^2-16) (16 \mu_{1}-9 k^2 \mu_{2})+63 (16 \mu_{1}-9 k^2\mu_{2})^2\right).\nonumber\\
\eea
We note that with respect to the constraint to absence of the conic singularity (\ref{conic}) both the scalar curvature (\ref{R.uinf}) and Kretschmann scalar (\ref{Krconfinf}) tend to zero with $u\to +\infty$.
\subsubsection{The Gubser bound}

The dilaton  supporting the geometry (\ref{f}) reads
\bea\label{dilaton-bb}
\phi  = \frac{9k}{9k^{2}-16}\log{\left[\frac{4}{3k}\sqrt{\left|\frac{C_{2}}{C_{1}}\right|}\frac{\sinh(\mu(u-u_{02}))}{\sinh(\mu u)}\right]}.
\eea
and takes the constant value near horizon
\bea\label{dil.horz}
\lim_{u\to +\infty} \phi=  \frac{9k}{9k^{2}- 16}\left(\log{\left(\frac{4}{3k}\sqrt{\left|\frac{ C_{2}}{C_{1}}\right|}\right)} - \mu u_{02}\right) .
\eea
Now one can check if  the Gubser's bound \cite{Gubser} for asymptotically non-AdS solutions  holds 
\bea\label{Gub.nAdS}
V(\phi(u_{h})) < 0,
\eea
where in our case $u_{h} =\infty$.
Plugging the solutions for the dilaton at the horizon (\ref{dil.horz}) in (\ref{1.1b}) the inequality (\ref{Gub.nAdS}) takes the form
\bea\label{GubE2E1}
\frac{E_{2}}{|E_{1}|} -1 < 0, 
\eea
that is valid for our solution due to the constraint (\ref{cond-nv-1}).
The improved Gubser's bound (\ref{Gub.nAdS}) reads
\bea\label{GBimp}
V(\phi(u_{h})) \leq V_{UV}, 
\eea
where $V_{UV}$ is the value of $V(\phi)$ at ultraviolet fixed point. Since $V_{UV} = 0$ with the dilaton $\phi \to - \infty$ at the UV point the constraint (\ref{GBimp}) comes to be (\ref{Gub.nAdS}).\\
In the UV limit, i.e. near $u_{01}=0$, the solutions turns to have the asymptotics as the Chamblin-Reall solution governed by the single exponential potential
\bea
ds^{2} \sim z^{\frac{8}{9k^{2} -4}}\left(- dt^{2}+ d\vec{y}^{2} + dz^{2}\right),
\eea
with the dilaton
\bea\label{dilUV01}
\lim \phi_{u\to u_{01}+ \epsilon} &=&-\frac{9k}{16-9k^{2}}\log\left[\frac{4}{3k}\sqrt{\frac{C_2}{|C_1|}}\frac{\sinh(-\mu\,u_{02})}{\mu\,\epsilon}\right].
\eea

We note that we can construct a black brane background for the left solutions (\ref{chartl}) with $u_{01} \neq u_{02}$ assuming that the horizon is located at $u = - \infty$, and the parameter $\alpha^{1}$ is positive.

\subsubsection{Special case $u_{01} = u_{02}$,  AdS black brane}\label{AdS-black-brane}

Now we turn to the special case of the non-vacuum solutions  with $u_{01}=u_{02} = 0$ with $u > 0$. The construction of a black brane metric is the same as presented before for solutions with $u_{01}\neq u_{02}$ with the horizon located at $u = +\infty$. The metric (\ref{f}) has the form 
\bea
ds^{2} = \mathcal{C}\left(1 - e^{-2\mu u}\right)^{-\frac{1}{2}}\left(-e^{-2\mu u}dt^{2} +d\vec{y}^{2}\right) + \mathcal{C}^{4}\left(1 - e^{-2\mu u}\right)^{-2}e^{-2\mu u}du^{2},\label{AdSbb}
\eea
where we took into account $\mu = - \frac{4}{3}\alpha^{1}$ and the constant $\mathcal{C}$ reads
\bea
\mathcal{C}&\equiv& 2^{\frac{16}{(16-9k^{2})}} (3 \mu)^{\frac{1}{2}}\Big(\left|C_{1}\right|\Big)^{\frac{8}{2(9k^{2} - 16)}}
\Big(\frac{C_{2}}{k}\Big)^{\frac{9k^{2}}{4(16- 9k^{2})}} (16-9k^{2})^{-\frac{1}{4}} 
\eea
Due to the constraint $\mu_{1} = \mu_{2}$ and $u_{01}=u_{02} $ the dilaton (\ref{dilnvE}) becomes constant 
\bea\label{dilAdS}
\phi  = \frac{9k}{9k^{2}-16}\log{\left[\frac{4}{3k}\sqrt{\left|\frac{C_{2}}{C_{1}}\right|}\right]}.
\eea
The curvature of the metric (\ref{AdSbb}) is negative and reads
\bea
R = -\frac{5\mu^{2}}{\mathcal{C}^{4}}.
\eea
Doing the change of coordinates
\bea\label{muhorz0}
z = z_{h}\left(1 - e^{-2\mu u}\right)^{\frac{1}{4}}, \quad \mathcal{C} = z^{-2}_h,
\eea
one gets the usual form for the 5d AdS black brane
\bea
ds^{2} = \frac{1}{z^{2}}\left(-f(z)dt^{2} + d\vec{y}^{2} + \frac{dz^{2}}{f(z)}\right),
\eea
with 
\bea
f = 1 - \left(\frac{z}{z_{h}}\right)^{4}.
\eea
For the dilaton potential we have the saturation of the Gubser's bound (\ref{GBimp})
\bea\label{GubAdS}
V(\phi(u_{h})) = V_{UV}, 
\eea
that is in agreement with the suggestion from \cite{Gubser}, since the solution (\ref{AdSbb})-(\ref{dilAdS}) is anti-de Sitter black brane.

One can summarize our studies on non-vacuum solutions as follows. For $u\to \pm \infty$ the scalar curvature and the dilaton 
for the right and left solutions can be constant for arbitrary value of the temperature which at the same time defines the constant $E_2$ (\ref{costrE2}).
However, the left solutions have a special point $u_{02}$ at which the scalar curvature has a non-removable singularity, 
while the scalar curvature of the right solutions is regular at its special point $u_{01}$.
Finally, from (\ref{GubE2E1}) and (\ref{GubAdS})  we see that the dilaton potential calculated on-shell is bounded only for the right solutions and special solutions with $u_{01}=u_{02}$. 
Therefore, these solutions can satisfy the Gubser's criterion \cite{Gubser}.

\subsection{RG flow  for non-vacuum solutions}

\subsubsection{Details of RG flow  for vacuum solutions}

In Sec.~\ref{Sect:4NV} we showed that non-zero temperature solutions are characterized by the parameter $\alpha^1\neq0$. 
The scale factor of the domain wall for the finite temperature case (\ref{DW}) is
\bea\label{bb-funcA}
\mathcal{A}= \frac{1}{2}\log(\mathcal{C}) + \frac{1}{2}\log(\mathcal{X})
\eea
and
 the energy scale A reads
\bea\label{c}
{\rm A}\equiv e^{\mathcal{A}} = \mathcal{C}^{\frac{1}{2}}\mathcal{X}^{\frac{1}{2}},
\eea
we note that to come to the domain wall form we use  the change of the coordinate
\bea
dw =  \mathcal{C}^2\,\mathcal{X}(u)^2 e^{\frac{8}{3}\alpha^{1}u} du.
\eea
The finite temperature case is described by the additional variable $Y$ defined through the blackening function (\ref{Y}).
 Here we deal with the system (\ref{dXphi})-(\ref{dYphi}), which seems to be rather complicated comparing to the zero-$T$ case and 
 one has to apply analytic solutions for $\phi$ and $\mathcal{A}$  with  the coordinate $u >u_{01}$ to show the RG flow.
  
\begin{figure}[h!]
\centering
\includegraphics[width=5 cm]{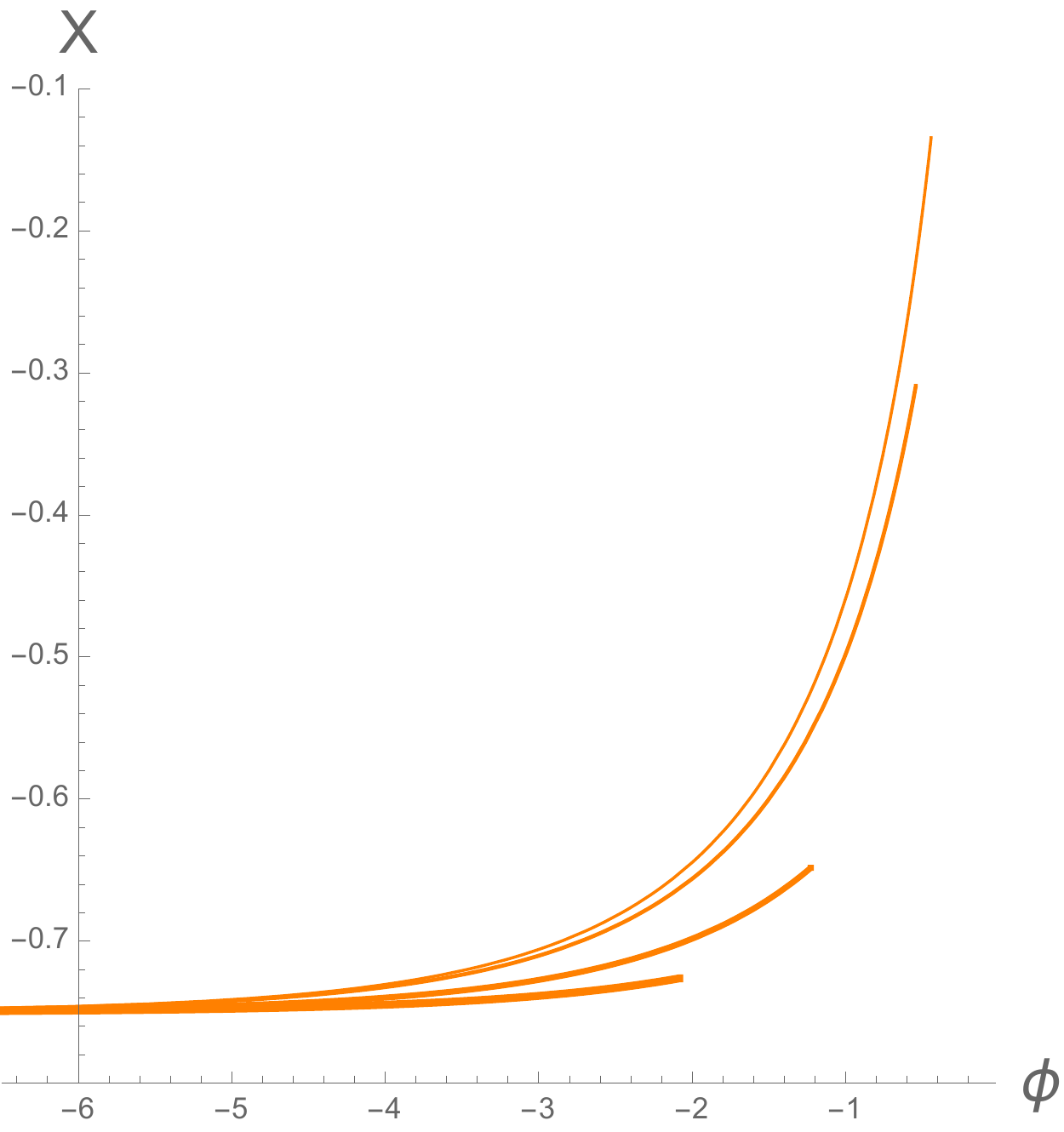}$\,\,\,$
\includegraphics[width=2.5cm]{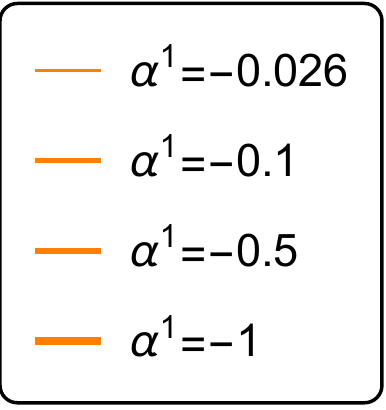}$\,\,\,$
     \includegraphics[width=5 cm]{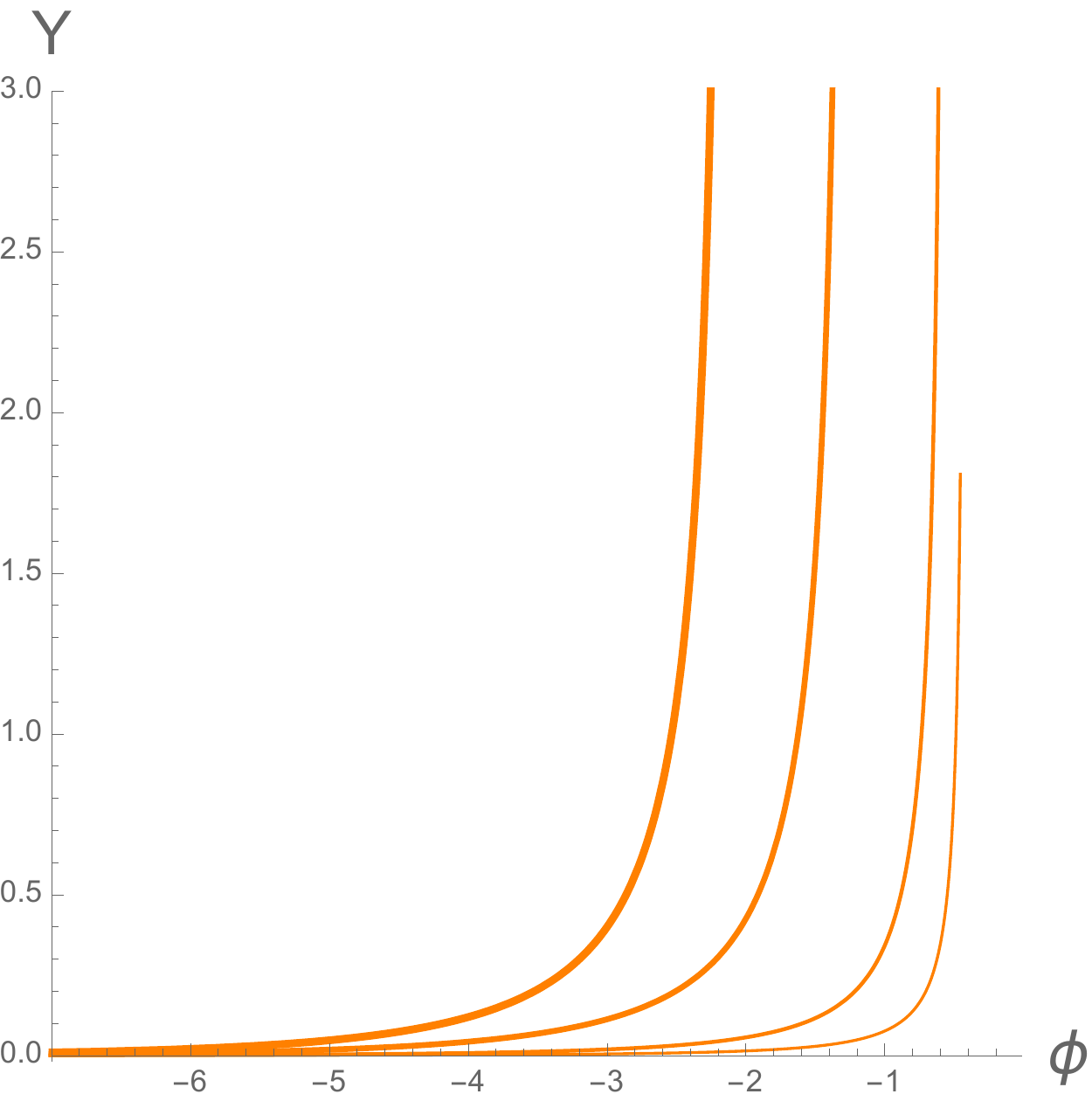}$\,\,\,\,\,\,\,$
\\A$\,\,\,\,\,\,\,$$\,\,\,\,\,\,\,$$\,\,\,\,\,\,\,$$\,\,\,\,\,\,\,$$\,\,\,\,\,\,\,$$\,\,\,\,\,\,\,$$\,\,\,\,\,\,\,$$\,\,\,\,\,\,\,$B
\caption{A) The dependence of the scalar function $X$ on $\phi$. B) The dependence of the scalar function $Y$ on $\phi$.  For both plots $\alpha^1<0$, $u>u_{01}$, $C_{1}=-C_{2}=-2$, $k=1$, $u_{01}=0$, $u_{02}=-1$.}
 \label{fig:XY}
 \end{figure}

To see the behaviour of the RG flow at finite temperature it is useful to plot $X$ and $Y$ as functions of $\phi$ for the analytical solution. 
We remind the horizon of the black brane at $u \to +\infty$ in this case is defined for $\alpha^{1}<0$.
  In Fig.~\ref{fig:XY} we draw $X$ and $Y$ as functions of $\phi$ for different values of the negative $\alpha^{1}$.
 From Fig.~\ref{fig:XY}{\bf A)} we see that the behaviour of $X$ is changed by $\alpha^{1}$, so $X$ becomes to be negative in the finite-$T$ case.
 As for the $Y$ function one can see from Fig.~\ref{fig:XY} {\bf B)} that it is positive.
 In Fig.~\ref{fig:XY}  {\bf A)} we observe that the $X$ function has stop points at some values of $\phi$. These values of $\phi$ correspond to the  asymptotics of the dilaton at the horizon $\phi_{h}$ (\ref{dil.horz}).
One can show that $X$ given by
\bea
X = \frac{1}{3}\frac{\phi'}{\mathcal{A}'}
\eea 
takes a constant value at the horizon with $u\to \infty$ as well and this can be observed as a stop point. For this let us trace the dependence of $\phi$ and $\mathcal{A}$ on $u$ with $u\to +\infty$
\bea
\lim_{u \to + \infty} \phi  &=& \frac{9k}{9k^{2}-16}\left(e^{-2\mu u} - e^{-2\mu(u-u_{02})}\right) ,
 \eea
 then we have
 \bea
\lim_{u \to + \infty} \phi'&= & \frac{-18k\mu}{9k^{2} -16}e^{-2\mu u}\left(1 - e^{2\mu u_{02}}\right).
\eea
The scale factor for the black brane is given by (\ref{bb-funcA}) and have the following asymptotics  at the horizon
\bea\label{mathcalAh}
\lim_{u \to + \infty} \mathcal{A}&=& \frac{1}{2}e^{-2\mu u}\left(\frac{16 - 9k^{2}e^{2\mu u_{02}}}{2(16 - 9k^{2})}\right)+ \frac{1}{2}\log(\mathcal{C}),
\eea
so we have
\bea
\lim_{u \to + \infty} \mathcal{A}^{\,'} = - \mu\left(\frac{16 - 9k^{2}e^{2\mu u_{02}}}{2(16 - 9k^{2})} \right)e^{-2\mu u}.
\eea

   Then the value of $X$ at the horizon is 
\bea
\lim_{u \to + \infty} X= - 12k \frac{e^{-2\mu u_{02}}-1}{16e^{-2\mu u_{02}} - 9k^{2}}.
\eea
We see that
\bea
X \to 0,
\eea
as
 $ \mu \to 0$.% \quad u_{01}-u_{02}\to 0.

As for the $Y$-function (\ref{Y}), it takes  infinite values at the horizon that can be read from Fig.~\ref{fig:XY} {\bf B)}.

\begin{figure}[h!]
\centering
     \includegraphics[width=5cm]{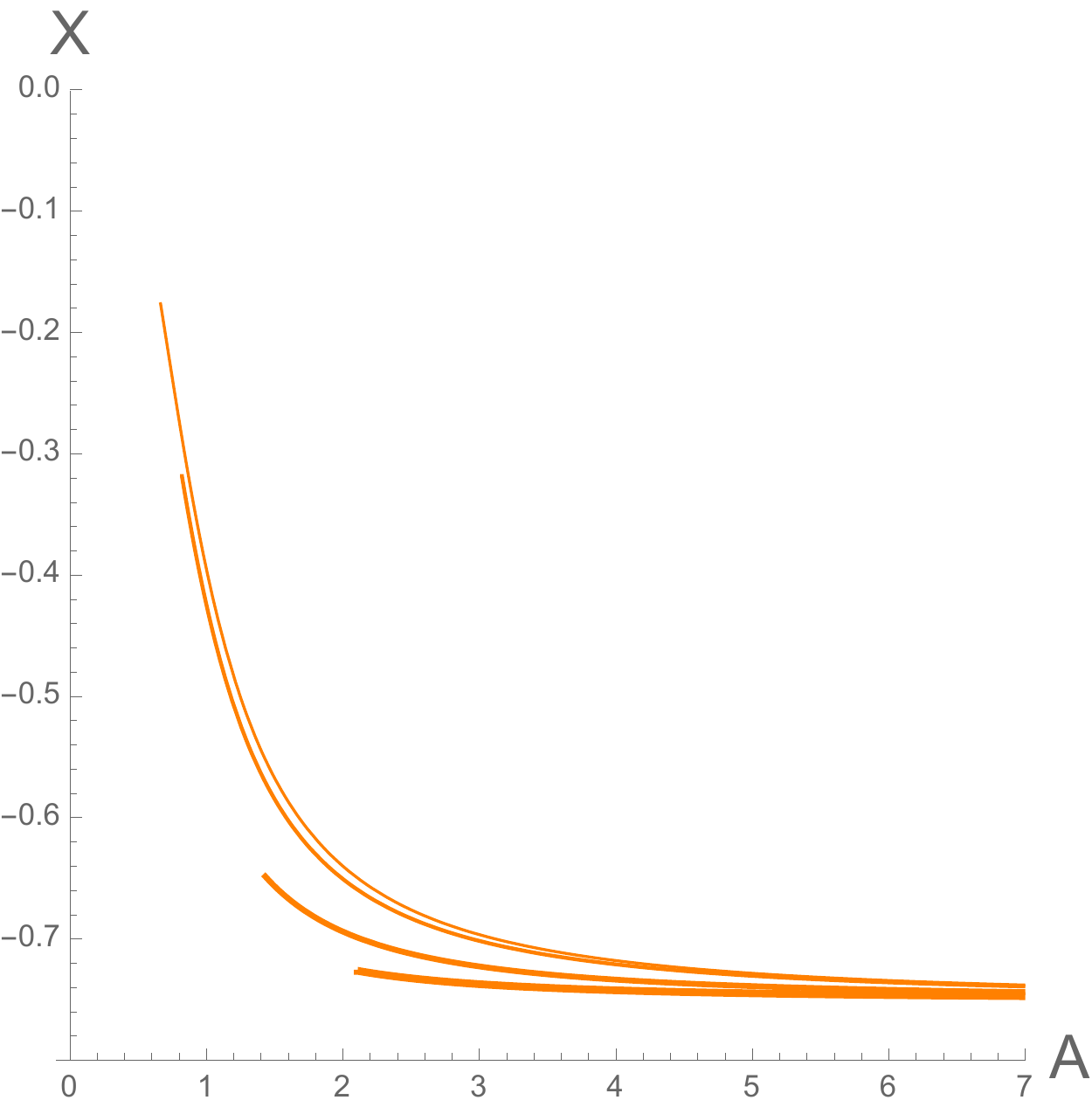} $\,$
     \includegraphics[width=2cm]{XRG/PhiXRM-in2-leg}$\,\,$
     \includegraphics[width=5cm]{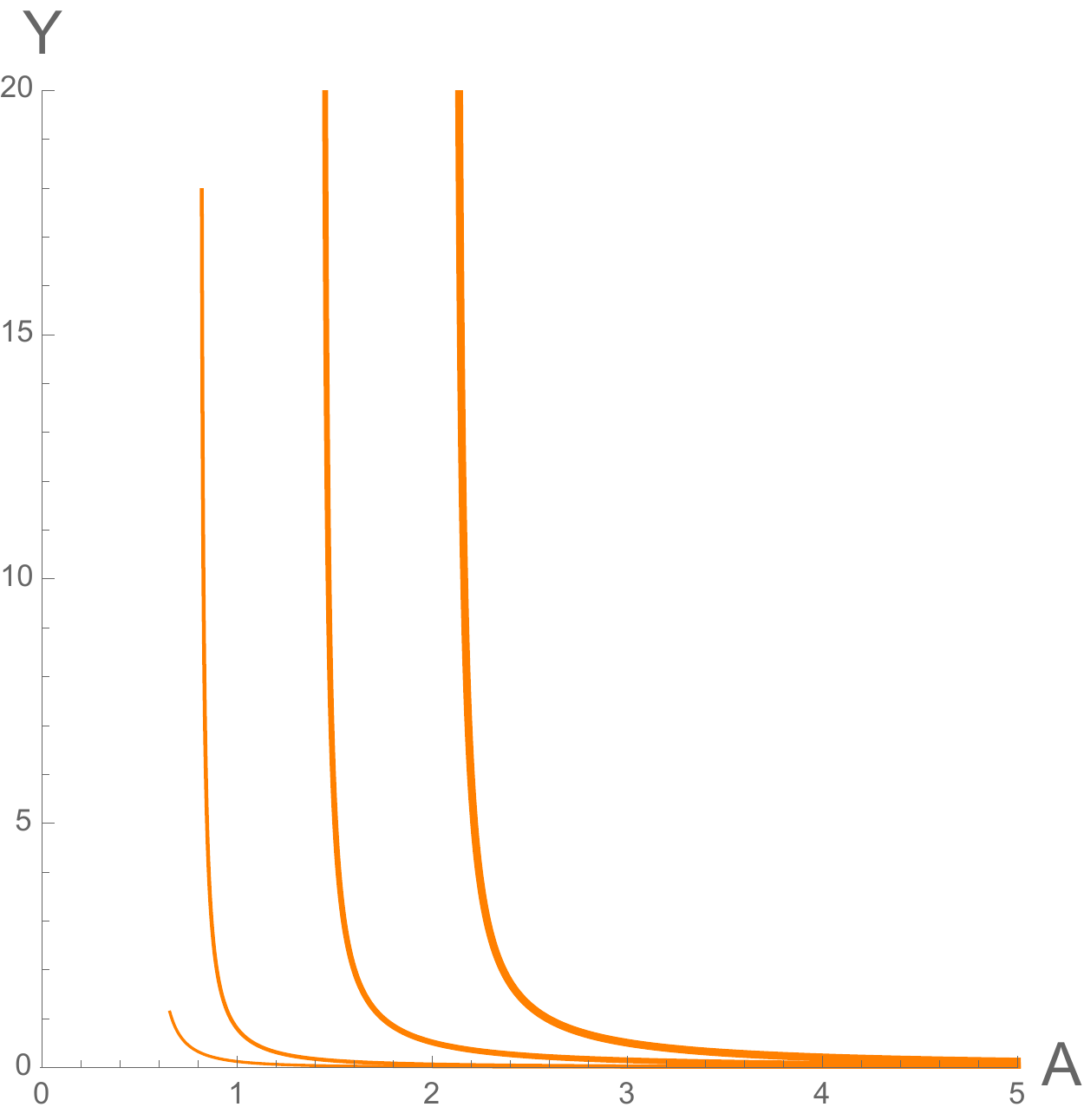}$\,$
      \\A$\,\,\,\,\,\,\,\,\,\,\,\,\,\,\,\,\,\,\,\,\,\,\,\,\,\,\,\,\,\,\,\,\,\,\,\,\,\,\,\,\,\,\,\,\,\,\,\,\,\,\,\,\,\,\,\,\,\,\,\,\,\,\,\,\,\,\,\,\,\,\,\,\,\,\,\,\,\,\,\,\,\,\,\,\,\,\,\,\,\,\,\,\,\,\,\,$B
\caption{ The  RG flows for the right solution with $\alpha^1<0$. {\bf A)} The flow in the (A,X) plane.
    {\bf B)} The flow in the (A,Y) plane. $C_{1}=-C_{2}=-2$, $k=1$, $u_{01}=0$, $u_{02}=-1$. }
 \label{fig:XARM}
 \end{figure}

 In Fig.~\ref{fig:XARM} we present the dependences of $X$ and $Y$ on the energy scale A. In Fig.~\ref{fig:XARM} {\bf A)} we again observe stop points of $X$ at some $\mathcal{A}$ (\ref{mathcalAh}) with $u\to +\infty$. 

\subsubsection{The running coupling  $\lambda = e^{\phi}$ on the energy scale for $T\neq 0$ flow} \label{Sec:RCN0}

Now let us look what happens with the behaviour of the running coupling on the energy scale at finite temperature.  As for the vacuum case we have the parametric dependence on $E_{1}$, $E_{2}$, $\alpha^{1}$, the position of  poles $u_{01}$ and $u_{02}$. We note that $E_{1}$, $E_{2}$ and $k$ are not independent. We have also seen before the temperature is related to the parameter $\alpha^{1}$ (\ref{NCS}). As in the previous section, Sect.\ref{Sec:RC0},  to have an insight to possible behavior of the running coupling on the constructed solutions, we start with presenting the  behavior of the energy scale A as a function of the  coordinate $u$ with (\ref{chartl})-(\ref{chartr}) and then incorporated found in Sect. \ref{Sect:4NV}, see Fig.
\ref{fig:DIL-NV}, the behavior of the dilaton as a function of $u$. 
  \begin{figure}[h!]
\centering
  \includegraphics[width=6cm]{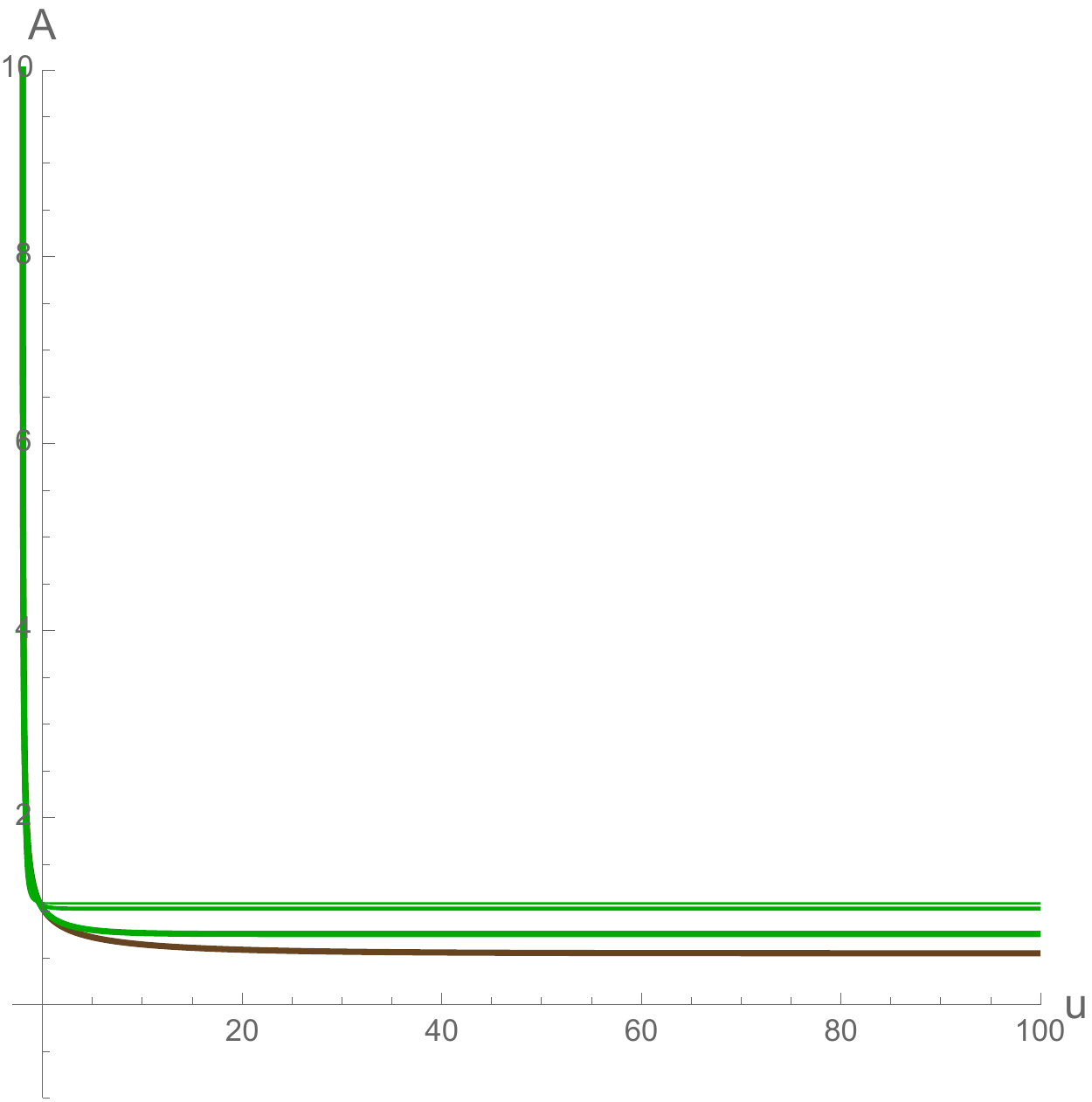}$\,\,\,\,\,\,\,\,\,\,\,\,\,\,$
  \includegraphics[width=5cm]{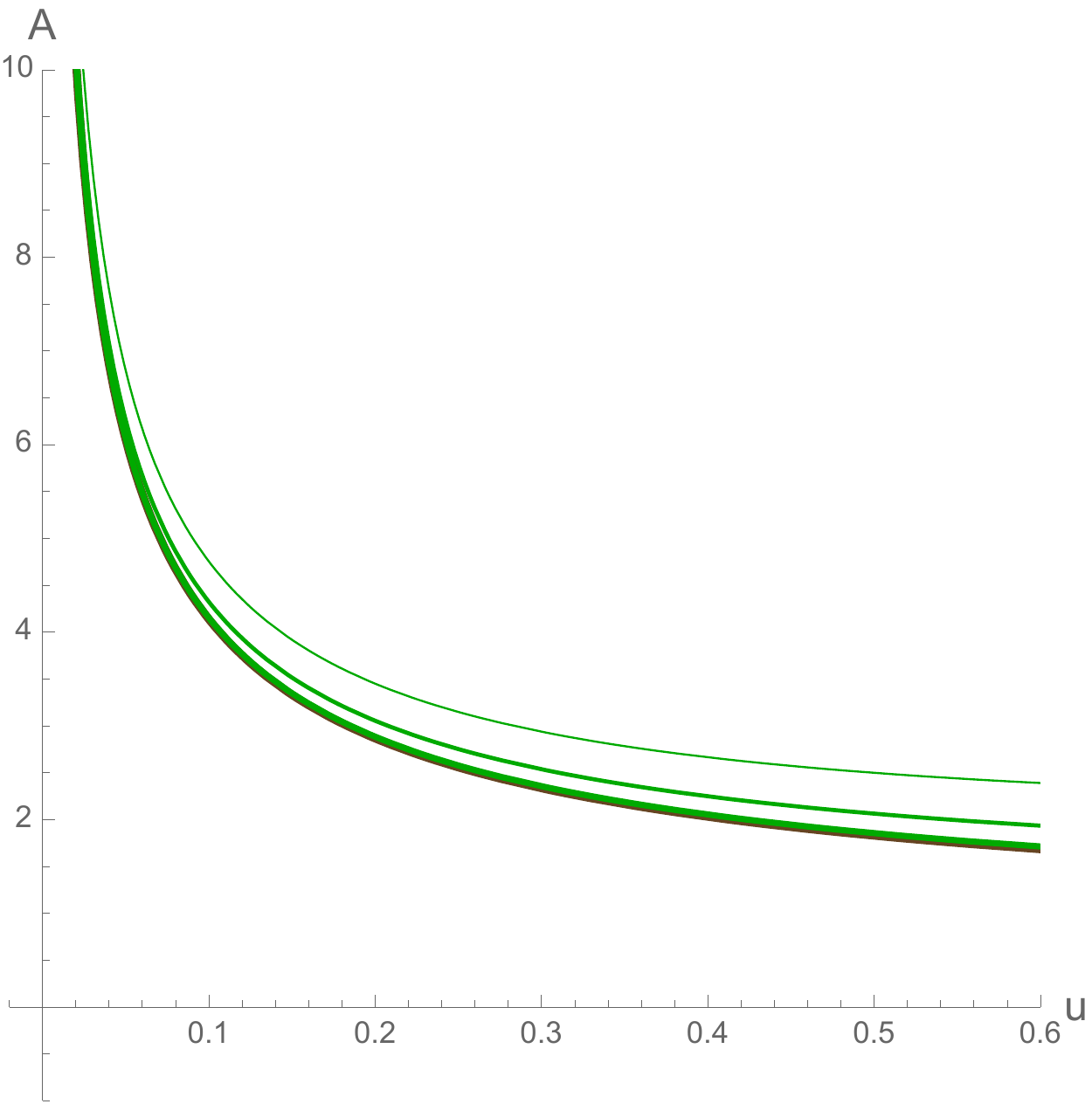}$\,\,\,\,\,\,\,$
  \includegraphics[width=2cm]{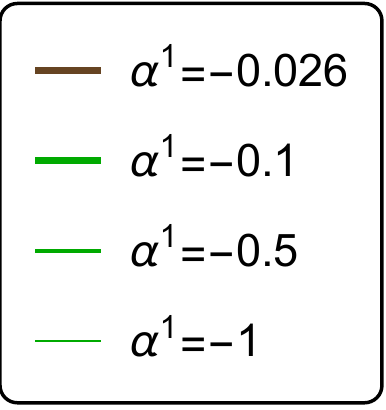}
  \\$\,\,\,\,\,\,\,\,\,\,\,\,\,\,\,\,\,\,\,\,\,\,\,\,\,\,\,\,\,\,\,\,\,\,\,\,\,\,\,\,\,\,\,\,\,\,\,\,\,\,\,\,\,\,\,\,\,\,\,\,\,\,\,\,\,\,\,\,\,$A$\,\,\,\,\,\,\,\,\,\,\,\,\,\,\,\,\,\,\,\,\,\,\,\,\,\,\,\,\,\,\,\,\,\,\,\,\,\,\,\,\,\,\,\,\,\,\,\,\,\,\,\,\,\,\,\,\,\,\,\,\,\,\,\,\,\,\,\,\,\,\,\,\,\,\,\,\,\,\,\,\,\,\,\,\,\,\,\,\,\,\,\,\,\,\,\,\,$B
  \caption{The plots show A for  $u_{01}=0$ and $u_{02}=-1$, $C_{1} = -2$, $C_{2}=2$, $k=1$ and negative $\alpha^1$. The plot in the right panel zooms the area closed to zero of  the left panel plot. }
 \label{fig:A0uinfty}
 \end{figure}

 In Fig.\ref{fig:A0uinfty} we see  that A tends to some constant value at $u\to\infty$ that  can be supported  by  (\ref{mathcalAh}).

\begin{figure}[h!]
 \centering
\includegraphics[width=5.5cm]{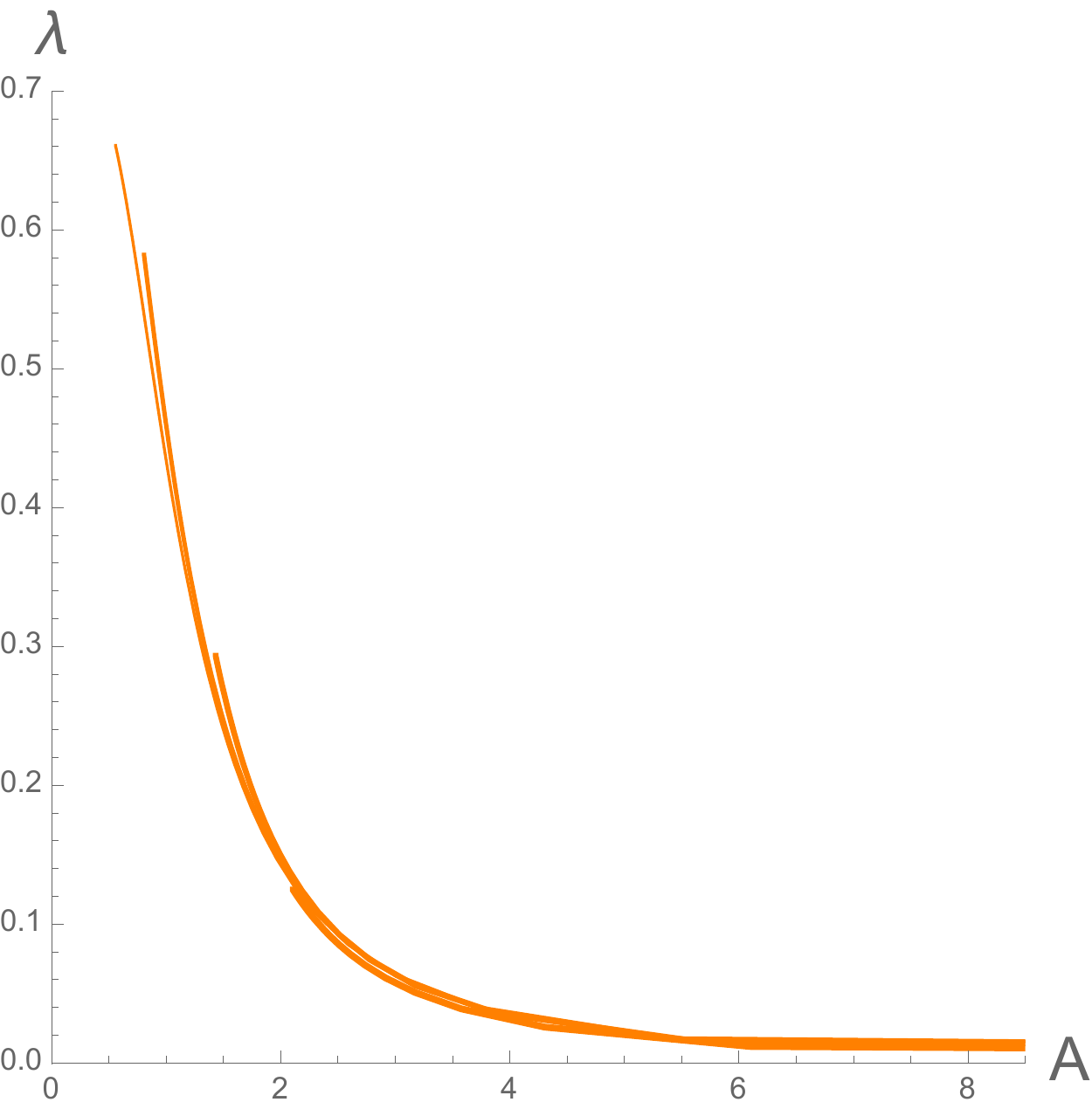} \includegraphics[width=2cm]{XRG/PhiXRM-in2-leg}$\,\,\,$\includegraphics[width=5.5cm]{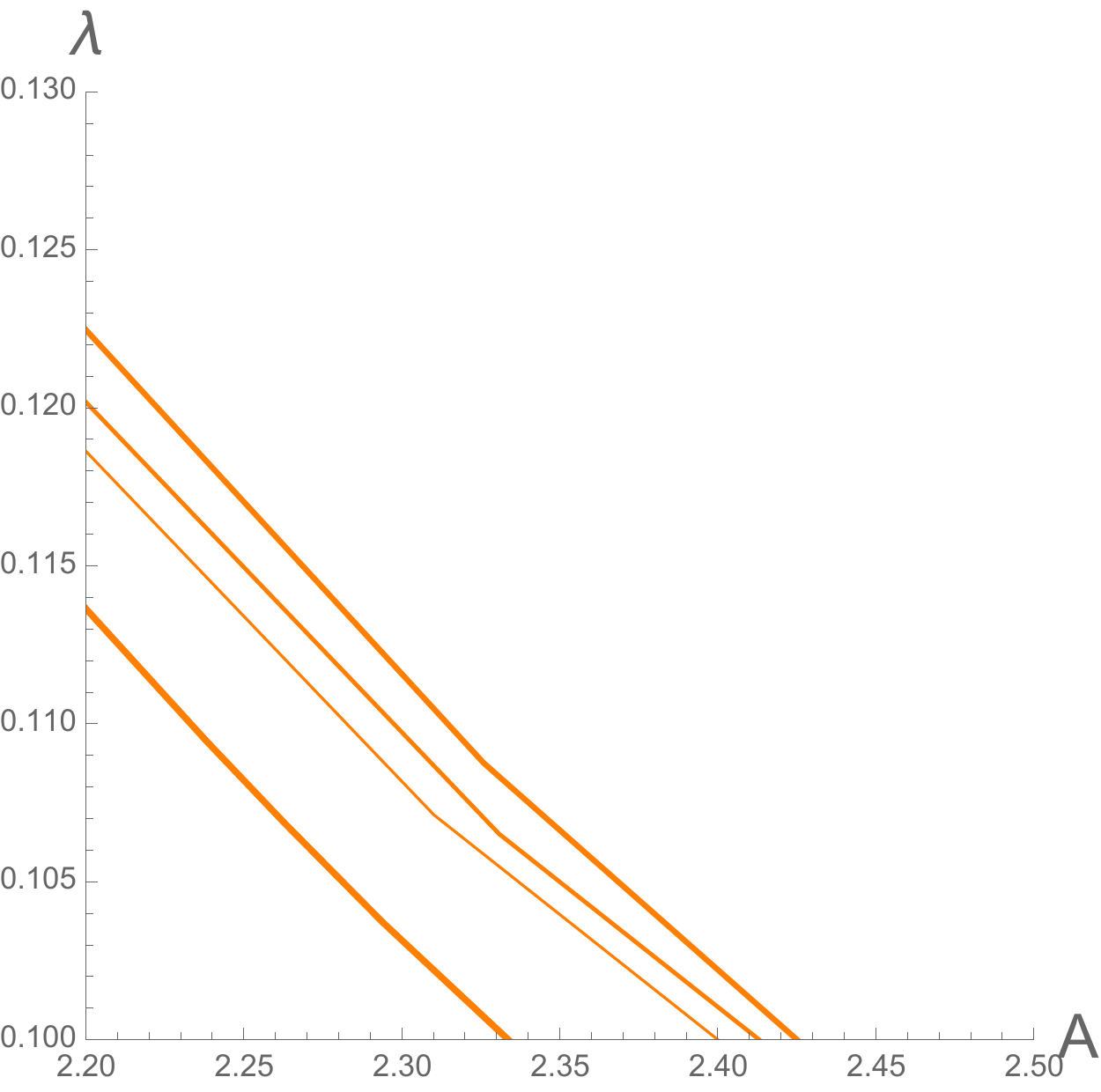}\\
 $\,\,\,\,\,\,\,\,\,\,\,\,\,\,\,\,\,\,\,\,\,\,\,\,\,\,\,\,\,\,\,\,\,\,\,\,\,\,\,\,\,\,\,\,\,\,\,\,\,\,\,\,\,\,\,\,\,\,\,\,\,\,\,\,\,\,\,\,\,$A$\,\,\,\,\,\,\,\,\,\,\,\,\,\,\,\,\,\,\,\,\,\,\,\,\,\,\,\,\,\,\,\,\,\,\,\,\,\,\,\,\,\,\,\,\,\,\,\,\,\,\,\,\,\,\,\,\,\,\,\,\,\,\,\,\,\,\,\,\,\,\,\,\,\,\,\,\,\,\,\,\,\,\,\,\,\,\,\,\,\,\,\,\,\,\,\,\,$B
   \caption{ A) The dependence of $\lambda$ on the energy scale A$=e^{\mathcal{A}}$.  In all cases constants that the potential is fixed with $k=1$, $C_{1} = -2$, $C_{2} =2$ and we vary 
  $\alpha^{1}$,  $u_{01}=0$ and $u_{02}=-1$. B) A zoomed region of {\bf A)}.}
 \label{fig:RC2}
 \end{figure}

 In Fig.~\ref{fig:RC2} we present the  parametric dependence of the running coupling $\lambda$ as a function of the energy scale A for the black brane solutions with $u_{01} =0$, $u_{02}= -1$ 
 and  different values of the parameter $\alpha^{1}$.  
 We see that the IR dynamics is changed by  $\alpha^{1}$. This comes from the fact the dilaton can change the asymptotics from $-\infty$ to some constant value
with (\ref{mualpha1}) as has been explained in Sect.~\ref{Sect:4NV}. So we observe  the increasing coupling constant in the IR regions.
For A$\to +\infty$ the running coupling  $\lambda$ goes to 0 for all chosen parameters providing the UV freedom.  Or in others words, 
 we can mimic the QCD RG flow for negative $\alpha^1$ .

\subsection{Free energy} 

The free energy corresponding to the black brane solution  is given by  the renormalized on-shell action. This can be computed directly, but the computation is simplified by the following observations. We use the domain wall coordinates here. 
First, the trace of the Einstein equations gives 
\begin{equation}
R = {5 \over 3} V + {4 \over 3} (\partial \phi)^2,
\end{equation} 
so that the bulk Lagrangian on-shell is 
\begin{equation}
\sqrt{g} (R -  {4 \over 3} (\partial \phi)^2 - V) = {2 \over 3} e^{4 A} V \,.
\end{equation} 
The finite temperature potential equations imply that 
\begin{equation}
 V = -(12 \mathcal{A}'^2 + 3 \mathcal{A}'')f - 3 \mathcal{A}' f' \,, 
 \end{equation} 
and from this relation one can see that the bulk term is a total derivative 
\begin{equation}
{\cal L}_{bulk} = - 2 {d \over dw} (e^{4 \mathcal{A}} \mathcal{A}' f) \,. 
\end{equation} 
The normal vector to the cutoff surface $ w = \epsilon$ is $n^w = \sqrt{f}, n^i=0$. The extrinsic curvature reads 
\begin{equation}
 K = {1 \over 2} h^{ab} n^w \partial_w h_{ab} = {\sqrt{f} \over 2} (8 \mathcal{A}' + {f' \over f} ) \,. 
 \end{equation} 

The on-shell Einstein action with regularization
  \bea
 I^{\epsilon}_{E} &=&-2 V_{3}\int^{\beta}_{0}dt\int^{w_{h}}_{\epsilon}dw {d \over dw} (e^{4 \mathcal{A}} \mathcal{A}' f)  \nonumber\\ 
 &=&-2V_{3}\beta(e^{4 \mathcal{A}(w_{h})} \mathcal{A}'(w_{h}) f(w_{h}) - e^{4\mathcal{ A}(\epsilon)} \mathcal{A}'(\epsilon) f(\epsilon)\nonumber\\
 &=&
 2V_{3}\beta e^{4\mathcal{ A}(\epsilon)} \mathcal{A}'(\epsilon) f(\epsilon),
 \eea
 while the Gibbons-Hawking term with regularization reads
 \bea
 I^{\epsilon}_{GH}  = V_{3}\int^{\beta}_{0}dt \frac{f e^{4\mathcal{A}}}{2}\left[8 \mathcal{A}' + {f' \over f}\right]_{\epsilon} =V_{3}\beta e^{4\mathcal{A}(\epsilon)}(8\mathcal{A}'(\epsilon)f(\epsilon) + f'(\epsilon)),
 \eea
 so we get
\begin{equation}
 \frac{I_{reg}}{\beta V_{3}} = - e^{4 \mathcal{A}} (6 \mathcal{A}' f + f') |_{w=\epsilon} \,.
\end{equation}  

We need to evaluate it on the regular black brane solutions given by eqs. (\ref{f})-(\ref{cbbxbb}), and (\ref{dilaton-bb}). 
Taking into account the change of coordinates $dw = e^{4 \mathcal{A}} f du$, the action becomes 
\begin{equation}\label{OnShell-1}
 \frac{I_{reg}}{\beta V_{3}}  = -\left( 6 \mathcal{A}'(u) +{f'(u)\over f(u)} \right) |_{u=\epsilon} \,.
\end{equation} 
The expansion of the scale factor $\mathcal{A}$ near $u\sim 0 $ reads 
\bea\label{UVasymp1}
\mathcal{A} & \sim - {4 \over 16-9k^2} \log u + \mathcal{A}_0 + \mathcal{A}_1 u + \ldots \,,
\eea
with 
\bea\label{UVasymp2}
\mathcal{A}_0 & =&  {1 \over 2} \log {\mathcal C} - {4 \over 16-9k^2} \log(2\mu) + {9 k^2 \over 4(16-9k^2)} \log(1-e^{2 \mu u_{02}}),\\ \label{UVasymp3}
\mathcal{A}_1 & =& {4 \mu \over 16-9k^2}   +  {9 k^2 \over 2(16-9k^2)}{\mu \over e^{-2\mu u_{02}}-1}.
\eea
Plugging (\ref{UVasymp1}) in (\ref{OnShell-1}) we obtain the regularised on-shell action
\bea
 \frac{I_{reg}}{\beta V_{3}} &= & {24 \over 16-9k^2} {1 \over \epsilon} - 6 \mathcal{A}_1 + 2 \mu \nonumber \\
 &=& {1 \over 16-9k^2} \left( {24 \over \epsilon } + \mu \Bigl(8 - 18k^2 - {27 k^2 \over e^{-2\mu u_{02}}-1}\Bigr)
 \right).
\eea 
The regularized on-shell is divergent and one needs to add the counterterms before removing the cut-off.

 The counterterms for the general dilaton-gravity system have been derived in 
\cite{Papadimitriou:2011qb}. For our homogeneous solutions the only relevant term is 
\begin{equation}
I_{ct} = -{8 \over 3} \int d^4 x \sqrt{h} \, U(\phi),
\end{equation} 
where $U$ is any function that satisfies the equation of the zero-temperature superpotential. We use then the superpotential appropriate for the regular solution, i.e. (see App. \ref{App:SuperW})
\begin{equation}
I_{ct} = -{8 \gamma \over 3} \int d^4 x \sqrt{h} e^{k \phi} \,.  
\end{equation} 
The asymptotics of the dilaton is given by
\bea \label{dil-asympt-1}
\phi & \sim {9k \over 16-9k^2} \log u + \phi_0 + \phi_1 u + \ldots 
\eea
with
 \bea\label{dil-asympt-2}
\phi_0 & =& -{9k \over 16-9k^2} \log  \left( {4 \over 3k} \sqrt{{C_2 \over C_1}} {\sinh(-\mu u_{02}) \over \mu}    \right) \,,
\\ \label{dil-asympt-3}
\phi_1 & =& - {9k \over 16-9k^2} \mu \coth (-\mu u_{02}) \,.
\eea 
Using the asymptotics (\ref{UVasymp1})-(\ref{UVasymp3}) and (\ref{dil-asympt-1}) -(\ref{dil-asympt-3}) we find 
\begin{equation}
{\cal L}_{ct} = - {24 \over 16-9k^2}( {1\over \epsilon}+ 4 \mathcal{A}_1 + k \phi_1)(1 - \mu \epsilon) =  - {24 \over 16-9k^2} {1 \over \epsilon} + {\it o}(\epsilon) \,.
\end{equation}
The renormalized action is then 
\bea\label{Iren}
  \frac{I_{ren}}{\beta V_{3}} &=&  \frac{I_{reg}+I_{ct}}{\beta V_3} =  {\mu \over 16-9k^2} \left(8 - 18k^2 - {27 k^2 \over e^{-2\mu u_{02}}-1} \right)   = \nonumber \\
& = & {1 \over 2} \left( \mu - {27 k^2 \over 16-9k^2} \sqrt{\Lambda^2 +\mu^2} \right), \,
\eea
where we defined the UV scale in terms of the scale factor at the boundary, i.e. by setting 
\bea \label{scale-fixing}
\frac{\mu}{\Lambda} = \sinh(- \mu u_{02}) \,. 
\eea
The free energy can be computed through the renormalized on-shell action (\ref{Iren}); the difference between the free energy of the black brane solution and the free energy of the vacuum, obtained at $\mu = 0$, is 
\bea
\mathcal{F} \sim -  {1 \over 2} \left( \mu - {27 k^2 \over 16-9k^2} (\sqrt{\Lambda^2+ \mu^2} - \Lambda) \right).
\eea
On the other hand one can calculate the free energy using black brane thermodynamics, that involves the following relation for the free energy,  the entropy density and the temperature of the black brane
\bea\label{bbtherm}
d\mathcal{F} =-sdT.
\eea
At the same time the black brane entropy density reads
\bea
s=\frac14 \int \sqrt{h_{ind}}\,dy_1dy_{2} dy_{3} \Big|_{u\to \infty},
\eea
since $\mathcal{X} \to 1$ at the horizon we have
\bea
s =\frac{V_3}{4}\, \mathcal{C}^{\frac{3}{2}}.\label{entropy}
\eea
Comparing \eqref{entropy} with \eqref{T} and taking into account the relation
\be
\mu= -\frac{4 \alpha^1 }{3},\label{mu-alpha}
\ee 
we get
\be
s\,T=\frac{V_3}{2\pi}\,\mu.\ee

Integrating (\ref{bbtherm}) we get the expression for the free energy 
\bea\label{fe-int}
\mathcal{F}& = - \displaystyle{\int s\, d T} = - \frac{V_3}{2\pi}\displaystyle{\int_{0}^{\mu} \cfrac{\mu'}{T}\, \cfrac{dT}{d\mu'} \,d\mu' }.
\eea
The temperature as function of $\mu$ is 
 \bea
T = \frac{2}{3\pi Q^{3/2}}|\frac{3  }{4}\mu|^{1/4}e^{\frac{27k^{2}}{4(16 - 9k^{2})}\, u_{02}\,\mu},\eea
but we have to express $u_{02}$ in terms of the scale $\Lambda$, using \eqref{scale-fixing}; 
 we get
\bea\label{Temp-sc}
T=\cfrac{\sqrt{2}}{3^{3/4}\pi Q^{3/2}} \mu^{1/4} e^{-\frac{27 k^{2}}{4(16 - 9k^{2})}\textrm{arcsinh}(\frac{\mu}{\Lambda})}.
\eea

Performing integration in \eqref{fe-int} with $T$ given by \eqref{Temp-sc} we obtain for the free energy
\bea
\mathcal{F}& = 
 &  -\frac{V_3}{8\pi} \left( \mu - \frac{27k^2}{16-9k^2}( \sqrt{\Lambda^2 + \mu^2}-\Lambda )\right) \,, 
\eea
in agreement with the result from the renormalized action.

The temperature and the free energy exhibit a qualitatively different behavior depending on the value of $k$. 
In the case $0<k<{2\over 3}$ , both are monotonic functions, so the black brane solution exists for any temperature, and there is no phase transition, so the black brane is thermodynamically favored over the thermal gas solution. 

In the other case, for ${2\over 3}<k<{4 \over 3}$, the temperature increases as function of $\mu$ up to a maximum value and then decreases back to zero. The maximum is attained at 
\bea
\frac{ \mu_{max}}{\Lambda} = {16-9k^2 \over \sqrt{8(9k^2-4)(9k^2+8)}} \,. 
\eea
The free energy is initally negative, but changes sign at 
\bea
\frac{\mu_{cr}}{\Lambda} = {54 k^2 (16-9k^2) \over 8(9k^2-4)(9k^2+8)}  > \frac{\mu_{max}}{\Lambda} \,.
\eea

We notice that considering  $u_{02} \to 0$, we get $\Lambda \to 0$ and the free energy comes to
\bea
\mathcal{F} = - \frac{V_3}{8 \pi} \mu \, 
\eea
the dependence on the scale disappears, consistently with the fact that the solution becomes the AdS black brane, with the constant dilaton, and there is no phase transition. \\

\begin{figure}[h!]
 \centering
\includegraphics[width=6 cm]{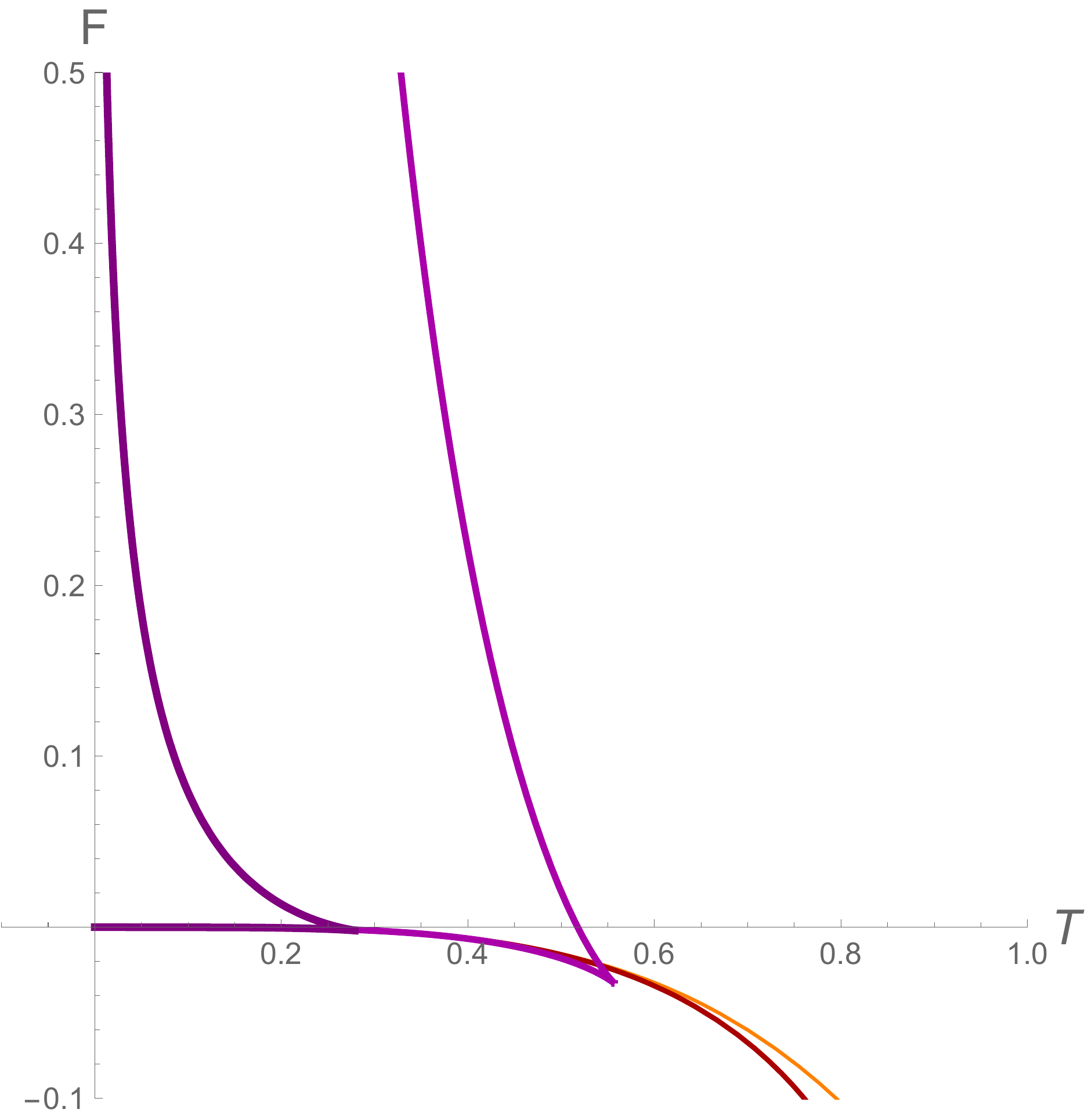}$\,\,\,\,$ \includegraphics[width=2cm]{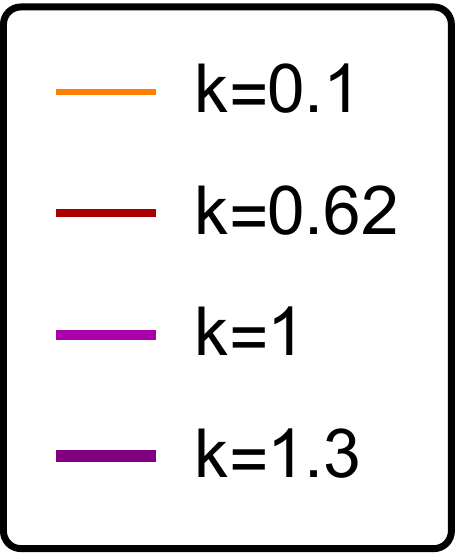}
   \caption{The dependence of the free energy $F$ on the temperature $T$  for the different shapes of the potential (different  $k$, $C_{1} = -2$, $C_{2} =2$). }
 \label{fig:FT}
 \end{figure}

In Fig.~\ref{fig:FT} we present the behaviour of the free energy on the temperature for different values of $k$. As we discussed above for $0<k<{2\over 3}$ the free energy is monotonic, while  for ${2\over 3}<k<{4 \over 3}$ the free energy decreases up to $T_{max}$ and then starts to increase, while the temperature is decreasing. This phase diagram cannot be complete, as the free energy would be discontinuous at $T_{max}$. We do not know whether this means that the dual theory does not make sense for this range of $k$, or there is some other solution that connects to the ones that we know and restores the continuity.  It is worth to notice that a similar behaviour of the free energy was observed in \cite{Gursoy:2018umf} for black hole solutions which are the finite temperature generalizations of the bouncing vacuum solutions.

\section{Conclusion and Discussion}

We have presented some analytic solutions of Einstein equations coupled to a dilaton field, at zero and non-zero temperatures, that correspond to holographic RG flows between different fixed points.
 The non-trivial form of the potential allows for a rich variety of different behaviours for the dilaton and the scale factor, corresponding to a coupling that can run to zero or to infinity in the UV and in the IR. We mainly considered solutions which are governed by sinh-functions. However we have two more classes of solutions for our choice of the potential, namely the sin-class and linear class. All solutions depend on two parameters,  $u_{01}$ and $u_{02}$, splitting the solutions into three branches: "left", "middle" and "right". Moreover, a special solution appears when $u_{01} = u_{02}$. In the zero-temperature case the dilaton  flows from $+\infty$ to $-\infty$ both for the left and middle solutions. However, the scale factor of the left solution is non-monotonic function while the scale factor of the middle solution  monotonically decreases. Correspondingly, the left vacuum solution does not have a holographic interpretation and the middle one is supposed to describe a holographic RG flow that mimics QCD behaviour. The right solution at $T=0$ has a bouncing dilaton and a scale factor, that monotonically decreases, i.e. it has both IR and UV free limits.  We note that the left, middle and right solutions interpolate between the hyperscaling violating boundaries. A special family of solutions with $u_{01} = u_{02}$ interpolates between an AdS boundary in the UV limit given by (3.28) and the hyperscaling violating boundary  in the
IR limit. Another solution with an AdS boundary but in the IR limit (and the hyperscaling violating boundary  in the UV limit) is the one from the linear class. For the right vacuum solution we could find its black brane analogue, which again interpolates between the hyperscaling violating boundaries  with a constant dilaton at the horizon. The special solution with coinciding point turns to be AdS-Schwarzschild black hole with the corresponding constant dilaton. We also showed that the finite temperature solutions can behave very differently from the corresponding zero-temperature ones.

The next step in our program is to understand better these solutions from the point of view of the putative dual field theory, and to investigate possible applications of these solutions to the physics of QCD-like theories. 

It would be interesting to determine the spectrum of fluctuations around these solutions, in particular the quasi-normal modes of the black brane solutions, and also to extend our ansatz in order to look for analytic time dependent solutions that describe out-of-equilibrium dynamics. 

It would also be worthwile to understand if the method we used, that reduces the Einstein equations to an integrable Toda chain, could be used for other case, for instance whether it is possible to relax the constraints (\ref{kconstr1})-(\ref{kconstr2}) on the parameters of the potential, and in general achieve a classification of all cases that can be solved in this way. One way to approach this problem could be to reverse-engineer the dilaton potential and understand if it can come from some dimensional reduction. 

It would also be interesting to consider the coupling of a Maxwell field in the bulk and find charged black hole solutions. \\

\vskip 10pt

{\bf Acknowledgments}

We thank  E. Kiritsis for valuable discussions and suggestions.
GP would like to thank M. Jarvinen and F. Nitti for giving useful suggestions, and
the Steklov Mathematical Institute for hospitality during the initial stages of this work. 
AG  thanks  Gleb Arutyunov, Eoin \'O Colg\'ain, Bum-Hoon Lee, Chanyong Park and Sunyoung Shin for useful conversations. 
AG is also very grateful for the warm hospitality DESY Theory Group,  APCTP (Pohang) and Sogang University (Seoul).

\newpage
\appendix 
{\Large{\bf Appendix}}

 \section{The curvature invariants of the background}\label{Sec:Cur}
 The ansatz for the metric is given by
 
 \begin{eqnarray}\label{app1}
ds^{2} = -  e^{2A(u)}dt^{2} + e^{2B(u)}\sum^{3}_{i =1}dy^{2}_{i} + e^{2C(u)}du^{2}.
\end{eqnarray}

The non-zero components of the Ricci tensor are
\bea\label{app2}
R_{00} &=& e^{2(A-C)} \left((\dot{A})^{2} + 3\dot{A}\dot{B} - \dot{A}\dot{C} + \ddot{A}\right), \\
R_{11} &=& R_{22} =R_{33} = -e^{2(B-C)}\left(\dot{A}\dot{B} + 3\dot{B}^{2}-  \dot{B}\dot{C} + \dot{B} \right),\\
R_{44}&=& - \dot{A}^{2} - 3 \dot{B}^{2}+ \left(\dot{ A} + 3\dot{ B}\right)\dot{C}-  \ddot{A} - 3\ddot{B}.
\eea

The scalar curvature reads
\begin{eqnarray}\label{app3}
R = -2e^{-2C}\left[\dot{ A}^{2} + 3 \dot{A} \dot{ B} + 6(\dot{B})^{2} - \dot{ A}\dot{ C} - 3 \dot{ B}\dot{C} +  \ddot{A} + 3\ddot{B} \right].\nonumber\\
\end{eqnarray}

Thus, we have
\begin{eqnarray}\label{app4}
\sqrt{|g|}R = -e^{A+3B - C}\Bigl[2(\dot{A})^{2} + 6\dot{A}\dot{B} + 12(\dot{B})^{2} - 2 \dot{A} \dot{C} - 6  \dot{ B}\dot{C} + 2 \ddot{ A}  + 6\ddot{B} \Bigr].
\end{eqnarray}

To find the solutions to the model we use the gauge $C= A + 3B$ for (\ref{app1}).
The generic form of the obtained solutions is
\bea
e^{A} =e^{A_{1}}e^{\alpha^{1}u}, \quad e^{B} = e^{B_{1}}e^{-\frac{1}{3}\alpha^{1}u},
\eea
where the functions $A_{1}=B_{1}$ and $\alpha^{1}$ is a constant, which is equal to zero for the vacuum case.
Now the scalar curvature (\ref{app3}) reads
\bea\label{GSCapp}
R = -\frac{4}{3}\left((\alpha^{1})^{2} - 9\dot{A}^{2}_{1} + 6\ddot{A}_{1}\right)e^{-8A_{1}},
\eea
that covers the vacuum case with $\alpha^{1} = 0$
\bea\label{GSCapp2}
R =\left(12\dot{A}^{2}_{1} - 8\ddot{A}_{1}\right)e^{-8A_{1}}.
\eea

The Kretschmann scalar is defined as follows
\bea
K= R_{abcd}R^{abcd}.
\eea
Pluggin the metric (\ref{app1}) we have
\bea
K& =& 4(\dot{A}^{4} + 3\dot{A}^{2}\dot{B}^{2} + 6 \dot{B}^{4} + 2\dot{A}^{2}\ddot{A} + 6 \dot{B}^{2} \ddot{B} + (\dot{A}^{2} + 3 \dot{B}^{2})\dot{C}^{2} +  \ddot{A}^{2} + 3\ddot{B}^{2}\nonumber\\
&- &2(\dot{A}^{3} + 3 \dot{B}^{3} + \dot{A}\ddot{A} + 3\dot{B}\ddot{B})\dot{C})e^{-4C}.
\eea

 One can write down the Kretschmann scalar of the non-vacuum solution ($\alpha^{1}\neq 0$)
 \bea
 K = \frac{8}{27}\left((\alpha^{1} - 3\dot{A}_{1})^{2}(19\alpha^{1} + 42\alpha^{1}\dot{A}_{1} + 63\dot{A}^{2}_{1}) + 36(\alpha^{1} -9\dot{A}^{2}_{1})\ddot{A}_{1} + 54\ddot{A}^{2}_{1}\right)e^{-16A_{1}}.\nonumber\\
 \eea

Taking into account that   $C = 4A_{1}$  and $\alpha^{1} = 0$ for the vacuum solution one obtains
\bea\label{KS1}
K = 8(21\dot{A}^{4}_{1} -12\dot{A}^{2}_{1}\ddot{A}_{1} + 2\ddot{A}^{2}_{1})e^{-16A_{1}}.
\eea

\subsection{The equations of motion in the harmonic gauge}\label{App:EOM}
The Einstein equations of motion which follow from the action with the harmonic gauge $A + 3B =C$ are
\bea
&\,& e^{-6B}\left[3 B^{\prime 2} + 3 A' B' -  3B''\right]= \frac{2}{3}\phi^{\prime 2}e^{-6B} + \frac{1}{2}e^{2A}V, \\
&\,&e^{-2A - 4B} \left[-3B^{\prime2} - 3A'B'+ \frac{\partial^{2}A}{\partial u^{2}}\ +2 B''\right]= -\frac{2}{3}{\phi}^{\prime 2}e^{-2A-4B} - \frac{1}{2}e^{2B}V, \\
&\,&3 B^{\prime2} + 3 B'A'= \frac{2}{3}\phi^{\prime 2}- \frac{1}{2}e^{2A + 6B}V
\eea
and the dilaton equation of motion reads
  \be
 \frac{8}{3} e^{-2A - 6B}\phi '' -V'_{\phi}=0.
 \ee
\subsection{The scalar curvature of the vacuum solutions.}\label{App:VacR}

Using the expression for the scalar curvature (\ref{GSCapp2}) and taking into account (\ref{FABC.1nv}) and (\ref{nC1pC2.F1E1n})-(\ref{nC1pC2.F2E2p}) one can write
\bea \label{gen-R2}
R &=& \frac{(C_{1}/2E_{1})^{\frac{16}{16-9k^{2}}}(C_{2}/2E_{2})^{\frac{-9k^{2}}{16 - 9k^{2}}}}{4(16 -9k^{2})^2}\Bigl(8(16 - 9k^{2})(16\mu^{2}_{1} - 9k^{2} \mu^{2}_{2}) \nonumber\\
&+& 128(9k^{2} - 10)\mu^{2}_{1}\coth(\mu_{1}(u - u_{01}))^{2} \nonumber\\
&-& 864k^{2}\mu_{1}\mu_{2}\coth(\mu_{1}(u - u_{01}))\coth(\mu_{2}(u - u_{02})) \nonumber\\
&+& 9k^{2}(128- 45k^{2})\mu^{2}_{2}\coth(\mu_{2}(u- u_{02}))^{2}\Bigr)\sinh(\mu_{1}(u- u_{01}))^{\frac{32}{16-9k^{2}}}\sinh(\mu_{2}(u- u_{02}))^{\frac{-18k^{2}}{16-9k^{2}}}. \nonumber\\
\eea
   
It is exemplarily to look how the scalar curvature behaves for each branch of the solution.

\begin{itemize}
\item  For the  left solution which is defined for $u< u_{02}$ we have the following limits
\begin{itemize}
\item $u \to -\infty$, so (\ref{gen-R2}) can be rewritten as
\bea\label{leftsc1}
R = \left(\frac{C_{1}}{2E_{1}}\right)^{\frac{16}{16 - 9k^{2}}}\left(\frac{C_{2}}{2E_{2}}\right)^{-\frac{9k^{2}}{16- 9k^{2}}}\frac{3(16\mu_{1} - 9k^{2}\mu_{2})^{2}}{4(16 - 9k^{2})^{2}}e^{-\frac{2(16\mu_{1} -9 k^{2}\mu_{2})}{16- 9k^{2}}u},
\eea
\end{itemize}
The quantity $(16\mu_{1} - 9k^{2}\mu_{2})$, with $\mu_{1}=\sqrt{\left|\frac{3E_{1}}{2}(k^{2}- \frac{16}{9})\right|}$, $\mu_{2} = \sqrt{\left|\frac{3E_{2}}{2}((\frac{16}{9})^{2}\frac{1}{k^{2}} - \frac{16}{9})\right|}$, $|E_{1}| = |E_{2}|$, $0<k<4/3$ is always positive and the scalar curvature grows as  $u\to - \infty$.\\

The scalar curvature in the conformal coordinates with $z\to 0$
\bea\label{leftsconf}
R = \left(\frac{C_{1}}{2E_{1}}\right)^{\frac{16}{16 - 9k^{2}}}\left(\frac{C_{2}}{2E_{2}}\right)^{-\frac{9k^{2}}{16- 9k^{2}}}\frac{3(16\mu_{1} - 9k^{2}\mu_{2})^{2}}{4(16 - 9k^{2})^{2}}\left(\frac{3\mu_{1}}{4+3k}z\right)^{-\frac{8}{3}}.
\eea

\item $u \to u_{02} - \epsilon$, then the scalar curvature (\ref{gen-R2}) takes the form
\bea\label{scu02}
R &=& \frac{(C_{1}/2E_{1})^{\frac{16}{16-9k^{2}}}(C_{2}/2E_{2})^{\frac{-9k^{2}}{16 - 9k^{2}}}}{4(16 -9k^{2})^2}\Bigl(8(16 - 9k^{2})(16\mu^{2}_{1} - 9k^{2} \mu^{2}_{2}) \nonumber\\
&+& 128(9k^{2} - 10)\mu^{2}_{1}\coth(\mu_{1}(u_{02} - u_{01}))^{2} 
- 864k^{2}\mu_{1}\coth(\mu_{1}(u_{02} - u_{01}))(u - u_{02})^{-1} \nonumber\\
&+& 9k^{2}(128- 45k^{2})(u- u_{02})^{-2}\Bigr)\sinh(\mu_{1}(u_{02}- u_{01}))^{\frac{32}{16-9k^{2}}}(\mu_{2}(u- u_{02})^{\frac{-18k^{2}}{16-9k^{2}}}, \nonumber\\
\eea
where one can see that the scalar curvature  (\ref{scu02}) has divergencies. \\
In the conformal coordinates  the scalar curvature at $u \to u_{02} - \epsilon$ is 
\bea\label{Rconfl02}
R&=& \frac{(C_{1}/2E_{1})^{\frac{16}{16-9k^{2}}}(C_{2}/2E_{2})^{\frac{-9k^{2}}{16 - 9k^{2}}}}{4(16 -9k^{2})^2}\Bigl(8(16 - 9k^{2})(16\mu^{2}_{1} - 9k^{2} \mu^{2}_{2}) \nonumber\\
&+& 128(9k^{2} - 10)\mu^{2}_{1}\coth(\mu_{1}(u_{02} - u_{01}))^{2} \nonumber\\
&-& 864k^{2}\mu_{1}\coth(\mu_{1}(u_{02} - u_{01}))\left(\frac{4(16-9k^{2})}{64-9k^{2}}z\right)^{\frac{4(16-9k^{2})}{9k^{2} -64}} \nonumber\\
&+& 9k^{2}(128- 45k^{2})\left(\frac{4(16-9k^{2})}{64-9k^{2}}z\right)^{\frac{8(16-9k^{2})}{9k^{2} -64}}\Bigr)\nonumber\\
&\cdot&\sinh(\mu_{1}(u_{02}- u_{01}))^{\frac{32}{16-9k^{2}}}\left(\mu_{2}\frac{4(16 -9k^2)}{64 - 9k^2}z\right)^{\frac{72k^{2}}{9k^{2}-64}}. \nonumber\\
\eea

\item The middle solution with $u\in (u_{01};u_{02})$ can be characterized by the following limits

\begin{itemize}
\item For  $u \to u_{02} + \epsilon$ the scalar curvature  is the same as for the left solution  (\ref{scu02})-(\ref{Rconfl02}), so divergencies as  (\ref{scu02}) for the left solution.

\item  $u \to u_{01} - \epsilon$, so (\ref{gen-R2}) reads as
\bea\label{middleR01}
R &=& \frac{(C_{1}/2E_{1})^{\frac{16}{16-9k^{2}}}(\sqrt{C_{2}/2E_{2}}\sinh(\mu_{2}(u_{01}- u_{02})))^{\frac{-18k^{2}}{16 - 9k^{2}}}}{4(16 -9k^{2})^2}\Bigl(8(16 - 9k^{2})(16\mu^{2}_{1} - 9k^{2} \mu^{2}_{2}) \nonumber\\
&+& 128(9k^{2} - 10)|u- u_{01}|)^{-2} - 864k^{2}\mu_{2}(|u -u_{01}|)^{-1}\coth(\mu_{2}(u_{01} - u_{02})) \nonumber\\
&+& 9k^{2}(128- 45k^{2})\mu^{2}_{2}\coth(\mu_{2}(u_{01}- u_{02}))^{2}\Bigr)(\mu_{1}(u_{01} - u))^{\frac{32}{16-9k^{2}}},\nonumber\\
\eea
\end{itemize}
that is regular.
As for the scalar curvature written in the conformal coordinates
\bea\label{Rscconfu01}
R&=& \frac{(C_{1}/2E_{1})^{\frac{16}{16-9k^{2}}}(\sqrt{C_{2}/2E_{2}}\sinh(\mu_{2}(u_{01}- u_{02})))^{\frac{-18k^{2}}{16 - 9k^{2}}}}{4(16 -9k^{2})^2}\Bigl(8(16 - 9k^{2})(16\mu^{2}_{1} - 9k^{2} \mu^{2}_{2}) \nonumber\\
&+& 128(9k^{2} - 10)(\frac{9k^{2} -4}{16-9k^{2}}z)^{\frac{2(16 - 9k^{2})}{9k^{2} - 4}} - 864k^{2}\mu_{2} \left(\frac{9k^{2}-4}{16-9k^{2}}z\right)^{\frac{16-9k^{2}}{9k^{2} -4}}\coth(\mu_{2}(u_{01} - u_{02})) \nonumber\\
&+& 9k^{2}(128- 45k^{2})\mu^{2}_{2}\coth(\mu_{2}(u_{01}- u_{02}))^{2}\Bigr)\left(\mu_{1}\frac{9k^{2}-4}{16 - 9k^{2}}z\right)^{\frac{32}{4-9k^{2}}},
\eea
where $z$ is given by (\ref{zmsr}).
\item For the right solution one has
\begin{itemize}
\item $u \to u_{01} + \epsilon$
The  scalar curvature matches with that one  (\ref{middleR01})- (\ref{Rscconfu01})  for the middle solution near  the point $u_{01}$. 

\item $u \to +\infty$
\bea\label{sccrs2}
R = \left(\frac{C_{1}}{2E_{1}}\right)^{\frac{16}{16 - 9k^{2}}}\left(\frac{C_{2}}{2E_{2}}\right)^{-\frac{9k^{2}}{16- 9k^{2}}}\frac{3(16\mu_{1} - 9k^{2}\mu_{2})^{2}}{4(16 - 9k^{2})^{2}}e^{\frac{2(16\mu_{1} -9 k^{2}\mu_{2})}{16- 9k^{2}}u}.
\eea
As for the left solution the scalar curvature (\ref{sccrs2}) grows due to $(16\mu_{1} - 9k^{2}\mu_{2})>0$ as for (\ref{leftsc1}) with $u \to  - \infty$.
In the conformal coordinates (\ref{sccrs2}) is
\bea\label{Rrconfinf}
R = \left(\frac{C_{1}}{2E_{1}}\right)^{\frac{16}{16 - 9k^{2}}}\left(\frac{C_{2}}{2E_{2}}\right)^{-\frac{9k^{2}}{16- 9k^{2}}}\frac{3(16\mu_{1} - 9k^{2}\mu_{2})^{2}}{4(16 - 9k^{2})^{2}}\left(\frac{3\mu_{1}}{4+3k}z\right)^{-\frac{8}{3}}.
\eea
with  $z \to 0$.

\end{itemize}
\end{itemize}

$\bullet$ Let's find the scalar curvature for the special case of the solution with $u_{01}= u_{02} = u_{0}$.\\

The general formula for the scalar curvature with $u_{01}= u_{02} = u_0$ reads
\bea
R &=& \frac{(\sqrt{\frac{C_{1}}{2E_{1}}}\sinh(\mu_{1}(u-u_{01})))^{\frac{32}{16-9k^{2}}}(\sqrt{\frac{C_{2}}{2E_{2}}}\sinh(\mu_{2}(u-u_{01})))^{\frac{18k^{2}}{-16 + 9k^{2}}}}{4(16- 9k^{2})^{2}}\cdot\Bigl(8(9k^{2} -16)(9k^{2}\mu^{2}_{2} - 16\mu^{2}_{1}) \nonumber\\
&+& 128(-10+9k^{2})\mu^{2}_{1}\coth^{2}(\mu_{1}u) - 864k^{2}\mu_{1}\mu_{2}\coth(\mu_{1}u)\coth(\mu_{2}u)\nonumber\\\
& +& 9k^{2}(128 - 45k^{2})\mu^{2}_{2}\coth^{2}(\mu_{2}u)\Bigr).
\eea
Here one has to study the behaviour of the scalar curvature in the limits of small $u$ and $u \to +\infty$.\\
For $u \to  0$  one has
\bea\label{sclimit0}
R = -\frac{5}{4}\left(\sqrt{\frac{C_{1}}{2E_{1}}} \mu_{1}\right)^{\frac{32}{16-9k^{2}}}\left(\sqrt{\frac{C_{2}}{2E_{2}}}\mu_{2}\right)^{\frac{18k^{2}}{9k^{2} -16}}.
% + \frac{2(16\mu^{2}_{1} - 9k^{2}\mu^{2}_{2})}{16-9k^{2}}u^{2}.
\eea
So,  one comes to a background with the constant negative curvature in the case of small values of $u$.\\
For $u\to +\infty$ we onbtain
\bea\label{sclimitinf}
R =  \frac{3}{4}\frac{(16\mu_{1} - 9k^{2}\mu_{2})^{2}}{(16- 9k^{2})^{2}}\left(\frac{C_{1}}{2E_{1}}\right)^{\frac{32}{16-9k^{2}}}\left(\frac{C_{2}}{2E_{2}}\right)^{\frac{18k^{2}}{-16 + 9k^{2}}}e^{\frac{2(16\mu_{1} - 9k^{2}\mu_{2})}{16-9k^{2}}u},
\eea
that is an agreement with (\ref{sccrs2}). The corresponding expression in the scalar coordinates matches with (\ref{Rrconfinf}).

\subsection{The scalar curvature of the non-vacuum solutions.} 

Now we turn to the non-vacuum background (\ref{nvm1}) with (\ref{dilnvE})-(\ref{cond-nv-1}), which is  characterized by a non-zero parameter $\alpha^{1}$. 
The scalar curvature for the non-vacuum case reads
\bea\label{curv-main}
R &=& - \Bigl(8(9k^{2} - 16)(\frac{2}{3}(\alpha^{1})^{2}(9k^{2} -16) + 16\mu^{2}_{1}- 9k^{2}\mu^{2}_{2})  + 128(10 -9k^{2})\mu^{2}_{1}\coth(\mu_{1}(u - u_{01}))^{2}\nonumber\\
&-&864 k^2 \mu_1 \mu_2 \coth (\mu_1 (u- u_{01})) \coth (\mu_2 (u- u_{02}))+ 9k^2 (-128+45 k^2)  \mu_2^2 \coth ^2(\mu_2 (u-u_{02}))\Bigr)\nonumber\\
&\cdot&\frac{\left(\sqrt{\frac{C_{1}}{2E_{1}}} \sinh (\mu_1 (u- u_{01}))\right)^{\frac{32}{16-9 k^2}} \left(\sqrt{\frac{C_{2}}{2E_{2}}} \sinh (\mu_2 (u- u_{02}))\right)^{\frac{18 k^2}{9 k^2-16}}}{4(16- 9k^{2})^{2}}.
\eea
As for the vacuum solution it is instructive to see the scalar curvature for each of the branches with certain limits.

\begin{itemize}
\item  The left solution with $u< u_{02}$ has the following scalar curvature
\begin{itemize}
\item $u \to -\infty$ one has
\bea\label{Rnvl02}
R = \left(\frac{C_{1}}{2E_{1}}\right)^{\frac{16}{16 - 9k^{2}}}\left(\frac{C_{2}}{2E_{2}}\right)^{-\frac{9k^{2}}{16- 9k^{2}}}\left(\frac{3(16\mu_{1} - 9k^{2}\mu_{2})^{2}}{4(16-9k^{2})^{2}}- \frac{4}{3}\left(\alpha^{1}\right)^{2}\right)e^{-\frac{2(16\mu_{1} - 9k^{2}\mu_{2})}{16- 9k^{2}}u}. \nonumber\\
\eea

\item for the limit $u \to u_{02} - \epsilon$
\bea\label{Rnvl02u}
R &=& - \Bigl(8(9k^{2} - 16)\left(\frac{2}{3}(\alpha^{1})^{2}(9k^{2} -16) + 16\mu^{2}_{1}- 9k^{2}\mu^{2}_{2}\right)  \nonumber\\ 
&+& 128(10 -9k^{2})\mu^{2}_{1}\coth(\mu_{1}(u_{01}-u_{02}))^{2}\nonumber\\
&-&864 k^2 \mu_1 \coth (\mu_1 (u_{01}-u_{02}))  (u- u_{02})^{-1}+ 9k^2 (-128+45 k^2)  (u-u_{02})^{-2}\Bigr)\nonumber\\
&\cdot&\frac{\left(\sqrt{\frac{C_{1}}{2E_{1}}} \sinh (\mu_1 (u_{01} - u_{02}))\right)^{\frac{32}{16-9 k^2}} \left(\sqrt{\frac{C_{2}}{2E_{2}}}\mu_2 (u- u_{02})\right)^{\frac{18 k^2}{9 k^2-16}}}{4(16- 9k^{2})^{2}}.
\eea
as  for the vacuum case (\ref{scu02})  the scalar curvature is divergent at $u_{02}$.

In the conformal coordinates (\ref{Rnvl02u}) reads
\bea\label{Rnvl02conf}
R &=& - \Bigl(8(9k^{2} - 16)\left(\frac{2}{3}(\alpha^{1})^{2}(9k^{2} -16) + 16\mu^{2}_{1}- 9k^{2}\mu^{2}_{2}\right)  \nonumber\\ 
&+& 128(10 -9k^{2})\mu^{2}_{1}\coth(\mu_{1}(u_{01}-u_{02}))^{2}\nonumber\\
&-&864 k^2 \mu_1 \coth (\mu_1 (u_{01}-u_{02}))  \left(\frac{4(16-9k^{2})}{64-9k^{2}}z\right)^{\frac{4(16-9k^{2})}{9k^{2} -64}}\nonumber\\
&+& 9k^2 (-128+45 k^2)  \left(\frac{4(16-9k^{2})}{64-9k^{2}}z\right)^{\frac{8(16-9k^{2})}{9k^{2} -64}}\Bigr)\nonumber\\
&\cdot&\frac{\left(\sqrt{\frac{C_{1}}{2E_{1}}} \sinh (\mu_1 (u_{01} - u_{02}))\right)^{\frac{32}{16-9 k^2}} \left(\sqrt{\frac{C_{2}}{2E_{2}}}\mu_{2}\frac{4(16 -9k^2)}{64 - 9k^2}z\right)^{\frac{72k^{2}}{9k^{2}-64}}}{4(16- 9k^{2})^{2}}.
\eea
\end{itemize}

\item  The middle solution defined for $u\in (u_{01};u_{02})$.
\begin{itemize}
\item  For $u \to u_{02} + \epsilon$ the scalar curvature matches with the curvature given by (\ref{Rnvl02u}). As well as in the conformal coordinates it is the same as  (\ref{Rnvl02conf}).

One can seem that scalar curvature of the middle solution in the non-vacuum case has also a singularity at $u_{01}$.
\item  $u \to u_{01} - \epsilon$
\bea\label{Rnvm01u}
R &=& - \Bigl(8(9k^{2} - 16)(\frac{2}{3}(\alpha^{1})^{2}(9k^{2} -16) + 16\mu^{2}_{1}- 9k^{2}\mu^{2}_{2})  + 128(10 -9k^{2})(u - u_{01})^{-2}\nonumber\\
&-&864 k^2  \mu_2  (u- u_{01})^{-1} \coth (\mu_2 (u_{01}- u_{02})) \nonumber\\
&+& 9k^2 (-128+45 k^2)  \mu_2^2 \coth ^2(\mu_2 (u_{01}-u_{02}))\Bigr)\nonumber\\
&\cdot&\frac{\left(\sqrt{\frac{C_{1}}{2E_{1}}} \mu_1 (u- u_{01})\right)^{\frac{32}{16-9 k^2}} \left(\sqrt{\frac{C_{2}}{2E_{2}}} \sinh (\mu_2 (u_{01}- u_{02}))\right)^{\frac{18 k^2}{9 k^2-16}}}{4(16- 9k^{2})^{2}},
\eea
or in the conformal coordinates with $z$ given by (\ref{zmsr})
\bea\label{Rnvm01}
R &=& - \Bigl(8(9k^{2} - 16)(\frac{2}{3}(\alpha^{1})^{2}(9k^{2} -16) + 16\mu^{2}_{1}- 9k^{2}\mu^{2}_{2})  \nonumber\\
&+& 128(10 -9k^{2})(\frac{9k^{2} -4}{16-9k^{2}}z)^{\frac{2(16 - 9k^{2})}{9k^{2} - 4}}- 864 k^2  \mu_2  (\frac{9k^{2}-4}{16-9k^{2}}z)^{\frac{16-9k^{2}}{9k^{2} -4}} \coth (\mu_2 (u_{01}- u_{02}))\nonumber\\
&+& 9k^2 (-128+45 k^2)  \mu_2^2 \coth ^2(\mu_2 (u_{01}-u_{02}))\Bigr)\nonumber\\
&\cdot&\frac{\left(\sqrt{\frac{C_{1}}{2E_{1}}} \mu_1\right)^{\frac{32}{16-9 k^2}}  (\frac{9k^{2}-4}{16-9k^{2}}z)^{\frac{32}{4-9k^2}} \left(\sqrt{\frac{C_{2}}{2E_{2}}} \sinh (\mu_2 (u_{01}- u_{02}))\right)^{\frac{18 k^2}{9 k^2-16}}}{4(16- 9k^{2})^{2}}.
\eea
\end{itemize}
\item The right solution with $u >u_{01}$
\begin{itemize}
\item $u\to u_{01} + \epsilon$  the scalar curvature  coincides with (\ref{Rnvm01u})-(\ref{Rnvm01}).
\item $u \to +\infty$
\bea\label{Rrconfinf2}
R = \left(\frac{C_{1}}{2E_{1}}\right)^{\frac{16}{16 - 9k^{2}}}\left(\frac{C_{2}}{2E_{2}}\right)^{-\frac{9k^{2}}{16- 9k^{2}}}\left(\frac{3(16\mu_{1} - 9k^{2}\mu_{2})^{2}}{4(16-9k^{2})^{2}}- \frac{4}{3}\left(\alpha^{1}\right)^{2}\right)e^{\frac{2(16\mu_{1} - 9k^{2}\mu_{2})}{16- 9k^{2}}u}. \nonumber\\
\eea
\end{itemize}
It should be noted that the scalar curvatures given by (\ref{Rnvl02}) and (\ref{Rrconfinf}) can be equal to zero if the parameter $\alpha^{1}$ is taken as
\bea\label{alphaR0}
\alpha^{1} = \pm \sqrt{\frac{(16 - 9k^{2})E_{2}}{6k^{2}}}.
\eea

\end{itemize}

\subsection{The Kretschmann scalar for the solution with $u>u_{01}$}

For the vacuum solution defined for $u>u_{01}$ (the right solution) with $u\to u_{01} + \epsilon$ one has 
\bea\label{KS2a}
K \sim k_{0} (u - u_{01})^{\frac{64}{16 - 9k^{2}}}\sum^{4}_{i = 0}c_{i}(u - u_{01})^{-i},
\eea
where $k_{0}$ and $c_{i}$ $i = 0,..,4$ are some constants.

As for $u\to +\infty$ the Kretschmann scalar for the right solution with (\ref{nC1pC2.F1E1n})-(\ref{nC1pC2.F2E2p})  takes the form
\bea\label{KS3}
K =   \frac{21 (16 \mu_{1}-9 k^2 \mu_{2})^4}{32 (16-9k^2)^4}\left(\frac{C_{1}}{2E_{1}}\right)^{\frac{32}{16-9 k^2}}\left(\frac{C_{2}}{2E_{2}}\right)^{\frac{18 k^2}{9 k^2-16}}e^{\frac{4(16\mu_{1} - 9k^{2}\mu_{2})}{16 - 9k^{2}}u}.
\eea

 %\bea
%K = 8(21(\alpha^{1} + \dot{A})^{4} -12(\alpha^{1} + \dot{A})^{2}\ddot{A} + 2\ddot{A}^{2})e^{-16A},
 %\eea
%that seems  not destroy the  behaviours (\ref{KS2a})-(\ref{KS3}).
For the non-vacuum solutions (\ref{nvm1}) with (\ref{dilnvE})-(\ref{cond-nv-1}) with (\ref{chartr}) the Kretschmann scalar with reads  
$u \to  u_{01} +\epsilon$
\bea\label{KS2aNV}
K \sim k_{0} (u - u_{01})^{\frac{64}{16 - 9k^{2}}}\sum^{4}_{i = 0}c_{i}( \#(\alpha^{1})^{i})(u - u_{01})^{-i},
%K \sim \#(u- u_{01})^{\frac{64}{16-9k^{2}}} \left(c_{0} + \# (\alpha^{1})^{2} + \#(\alpha^{1})^{3} + \#(\alpha^{4})^{2}\right).
%\frac{(b\sinh(\mu_{2}(u-u_{02})))^{\frac{64}{9k^{2} - 16}}}{32(16-9k^{2})^{4}}(a\mu_{1}(u- u_{01}))^{\frac{36k^{2}}{16-9k^{2}}} \left(\#(u - u_{01})^{2}+\#(u- u_{01})^{4}(c_{0} + \alpha^{1}c_{1}+ (\alpha^{1})^{2}c_{2} + (\alpha^{1})^{3}c_{3}+(\alpha^{4})^{2}c_{4})\right)
\eea
while with $u\to + \infty$ is
\bea
K &=&\frac{\left(4 \alpha^{1} (9 k^2-16)+27 k^2 \mu_{2}-48 \mu_{1}\right)^2 }{864 (16-9 k^2)^4}\left(\frac{C_{1}}{2E_{1}}\right)^{\frac{32}{16-9 k^2}}\left(\frac{C_{2}}{2E_{2}}\right)^{\frac{18 k^2}{9 k^2-16}}e^{\frac{4(16\mu_{1} - 9k^{2}\mu_{2})}{16 - 9k^{2}}u}\nonumber\\
&\cdot&\left(304 (\alpha^{1})^2 (16-9 k^2)^2+168 \alpha^{1} (9k^2-16) (16 \mu_{1}-9 k^2 \mu_{2})+63 (16 \mu_{1}-9 k^2\mu_{2})^2\right).\nonumber\\
\eea

\section{The scalar field}\label{Sec:phi}

We have the following relation for the scalar field
\begin{eqnarray}\label{app5}
\left( \partial \varphi \right)^{2} = e^{-2C} \left(\frac{\partial \phi}{\partial u}\right)^{2}.
\end{eqnarray}
The dilaton solution defined for $u > u_{01}$ reads
\bea
\phi = \frac{9k}{9k^{2} - 16}\log\left(\sqrt{\frac{C_{2}}{C_{1}}}\frac{\sinh(\mu_{2}(u - u_{02}))}{\sinh(\mu_{1}(u - u_{01}))}\right).
\eea
The derivative of the dilaton with respect to $u$-variable is
\bea
\frac{\partial \phi}{\partial u}= \frac{9k}{9k^{2} - 16} \left(-\mu_{1} \frac{\cosh(\mu_{1} (u - u_{01}))}{\sinh(\mu_{1} (u - u_{01}))} + \mu_{2}\frac{\cosh(\mu_{2} (u - u_{02}))}{\sinh(\mu_{2} (u - u_{02}))}\right),
\eea

\bea
\left(\frac{\partial \phi}{\partial u}\right)^{2} = \frac{81k^{2}}{(16 - 9 k^2)^2}(\mu_{1} \coth(\mu_1 (u - u_{01})) - \mu_2 \coth(\mu_2(u - u_{02})))^2.
\eea
For the vacuum case one has
\bea
e^{-2C}  = \left(\sqrt{\frac{C_{1}}{2E_{1}}}\sinh(\mu_{1}(u - u_{01}))\right)^{\frac{32}{16-9k^{2}}}\left(\sqrt{\frac{C_{2}}{2E_{2}}}\sinh(\mu_{2}(u- u_{02}))\right)^{\frac{18k^{2}}{9k^{2} -16}}.
\eea
The dilaton for $u \to  u_{01} +\epsilon$ reads
\bea
\left( \partial \varphi \right)^{2}& =&  \frac{81k^{2}}{(16 - 9 k^2)^2}( (u - u_{01})^{-1} - \mu_2 \coth(\mu_2(u_{01} - u_{02})))^2\nonumber\\
&\cdot&\left(\sqrt{\frac{C_{1}}{2E_{1}}}\mu_{1}(u - u_{01})\right)^{\frac{32}{16-9k^{2}}}\left(\sqrt{\frac{C_{2}}{2E_{2}}}\sinh(\mu_{2}(u_{01}- u_{02}))\right)^{\frac{18k^{2}}{9k^{2} -16}}.
\eea
Since $0<k<4/3$ the quantity $\left( \partial \varphi \right)^{2}$ has a good behaviour.\\

In the limit $u \to + \infty$ the dilaton behaves as
\bea
\left( \partial \varphi \right)^{2} = \frac{81k^{2}}{(16 - 9k^{2})^{2}}\left(\mu_{1} - \mu_{2}\right)^{2}\left(\frac{C_{1}}{2E_{1}}\right)^{\frac{16}{16 -9k^{2}}}\left(\frac{C_{2}}{2E_{2}}\right)^{\frac{9k^{2}}{9k^{2}-16}}e^{\frac{2(16\mu_{1} - 9k^{2}\mu_{2})}{16-9k^{2}}u}.
\eea

The divergence disappears for $\mu_{1} = \mu_{2}$ in the non-vacuum case.

\section{The superpotential in the UV}\label{App:SuperW}

In this section, following the analysis of \cite{GKMN}, we look at the superpotential in the neighborhood of the maximum of the potential, which is at $\phi \to - \infty$. Their analysis was done assuming that $V<0$ so that the vacuum is asymptotically AdS, but we will show that their conclusions remain valid also in our case when $V \to 0$. 

The equation that determines the superpotential in the vacuum is, in the domain wall coordinates, 
\bea
V = {4 \over 3} \left( {d W \over d \phi} \right)^2 -{64 \over 27} W^2 \,, \\ \label{eq:superpot-1}
A'(w) = - {4 \over 9} W \,, \quad \phi'(w) = {d W \over d \phi}\,.  \label{eq:superpot-2}
\eea
These equations imply that ${\phi' \over A'} = -{9 \over 4} {W' \over W}$. 
We want to solve it asymptotically in the UV, where $V = - |C_1| e^{2 k \phi}$. There are two solutions: if one makes the ansatz that $W = \gamma e^{k \phi}$, substituting in the equation gives 
\begin{equation} \label{gamma}
\gamma^2 = {27 |C_1| \over 4 (16-9k^2)} \,,
\end{equation} 
so the solution is uniquely determined. This is the ``regular" solution in the terminology of \cite{GKMN}, and on this solution 
$X = {\phi'\over 3 A'} \to - {3 k \over 4}$.  

Another solution is obtained by assuming that the superpotential terms are dominant in \eqref{eq:superpot-1}, so one can set $V=0$ and finds $W = c e^{- 4 \phi/3}$. Since $k<4/3$, $W^2 \gg V$ so the assumption is self-consistent.  In this case $c$ is arbitrary, so there is a 1-parameter family of solutions for which $X \to 1$. These are singular 
solutions. 

The plots in Fig. \ref{fig:Xcomparison} show that these two cases correspond to the asymptotics of the right solutions, in the UV and IR respectively. We can check this explicitly by computing the superpotential from the asymptotics of the solution given in \ref{MABC.pC}- \ref{dil-u012}. In these formulas the coordinate $u$ is not the domain wall coordinate, which is obtained by $dw = e^{4 A} du$. The superpotential is then $W = -{9 \over 4} e^{-4 A} A'(u)$. 
For convenience we set $u_{01}=0$. The UV asymptotics, for $u\sim 0$, are 
\bea
A & \sim - {4 \over 16-9k^2} \log u + A_0 \,, \\
\phi & \sim {9k \over 16-9k^2} \log u + \phi_0 \,, 
\eea
where 
\bea
A _0 & = - {4 \over 16-9k^2} \log \left(  \sqrt{{|C_1| \over 2 |E|}}   \mu_1  \right)   + 
{9k^2 \over 4(16-9k^2)} \log \left(  \sqrt{{|C_2| \over 2 |E|}}  \sinh( -\mu_2 u_{02} ) \right)  \,, \\
\phi_0 & = {9k \over 16-9k^2} \log  \left(  \sqrt{{|C_1| \over  |C_2|}} {\mu_1 \over  \sinh( -\mu_2 u_{02}) } \right) \,. 
\eea
Then 
\begin{equation}
W = -{9 \over 4} e^{-4 A} A' = {9 \over 16-9k^2} e^{-4 A_0} e^{{9k^2 \over 16-9k^2} \log u} = {9 \over 16-9k^2} e^{-4 A_0-k \phi_0} e^{k \phi} 
\end{equation} 
and a rather tedious computation shows that the coefficient is equal to $\gamma$ in \eqref{gamma}, so the dependence on all the parameters of the solution cancels out. 

At $u \to \infty$ we have the asymptotics 
\bea
A & \sim - {\mu_1 \over 4+3k}  u + A_\infty \,, \\
\phi & \sim -{3 \mu_1 \over 4+3k}  u + \phi_\infty \,, 
\eea
where 
\bea
A _\infty & = - {4 \over 16-9k^2} \log   \sqrt{{|C_1| \over 8 |E|}}    + 
{9k^2 \over 4(16-9k^2)} \left(-\mu_2 u_{02}+ \log  \sqrt{{|C_2| \over 8 |E|}}  \right)  \,, \\
\phi_0 & = {9k \over 16-9k^2}   \left(\mu_2 u_{02}+\log   \sqrt{{|C_1| \over  |C_2|}}  \right) \,. 
\eea
Then 
\begin{equation}
W = {9 \mu_1 \over 4(4+3k)} e^{-4 A_\infty}e^{{4 \mu_1 \over 4 +3k } u } =  {9 \mu_1 \over 4(4+3k)} e^{-4 A_\infty+ {4\over 3} \phi_\infty} 
e^{- {4 \over 3}\phi} \,.
\end{equation}
In this case there is no cancellation and the coefficient depends on the parameters of the solution, as expected. 

The analysis of \cite{GKMN} then leads to the conclusion that all the flows that end in the singular IR solution are not acceptable because they cannot be regularized by a small horizon. Indeed we see that when we turn on the temperature, we have no regular flow that 
ends in the vicinity of $X=1$. The only regular vacuum flow is the one that ends at $X=0$, shown in Fig. \ref{fig:Xcomparison} {\bf C}, which ends at a different extremum of the potential; we do not need to analyse it in detail since the asymptotics are those of AdS.

\end{document}